\documentclass[english]{article}
\pdfoutput=1

\usepackage{lmodern}
\usepackage[T1]{fontenc}
\usepackage[latin9]{inputenc}
\usepackage{amsmath}
\usepackage{amssymb}
\usepackage{bm}
\usepackage{graphicx}
\usepackage{esint}
\usepackage{xcolor}
\usepackage{babel}
\usepackage{fullpage}
\usepackage{hyperref}
\usepackage{caption}
\usepackage{subcaption}
\usepackage{float}
\usepackage{cite}
\usepackage{datetime}

\definecolor{blus}{cmyk}{1,1,0,0.6}
\definecolor{verdes}{cmyk}{0.99,0,0.59,0.65}
\definecolor{rossos}{cmyk}{0,1,1,0.55}
\definecolor{redy}{cmyk}{0,1,1,0.7}
\definecolor{greeny}{cmyk}{0.99,0,0.59,0.98}
\definecolor{green-go}{cmyk}{0.79,0,0.59,0.5}

\hypersetup{colorlinks,bookmarksopen,bookmarksnumbered,
citecolor=blue,
linkcolor=black,
pdfstartview=green,
urlcolor=purple}

\def\be{\begin{equation}}
\def\ee{\end{equation}}

\def\bea{\begin{eqnarray}}
\def\eea{\end{eqnarray}}

\def\hhref#1{\href{http://arxiv.org/abs/#1}{arXiv:#1}} 

\numberwithin{equation}{section}
\begin{document}

\begin{titlepage}

\rightline{IFT-UAM/CSIC-14-118}

\rule{0pt}{2.1cm}

%%%%%%%%%%%%%%%%%

\centerline{
\Large \bf  
On shape dependence of holographic mutual information in AdS$_4$
}

\vspace{1.4truecm}

 \centerline{\large
Piermarco Fonda$^{a,}$\footnote[1]{piermarco.fonda@sissa.it},  
Luca Giomi$^{b,}$\footnote[2]{giomi@lorentz.leidenuniv.nl},  
Alberto Salvio$^{c \, d,}$\footnote[3]{alberto.salvio@uam.es}
 and Erik Tonni$^{a,}$\footnote[4]{erik.tonni@sissa.it}
}

\vspace{1cm}

\centerline{$^a${\it  SISSA and INFN, via Bonomea 265, 34136, Trieste, Italy}}
\vspace{.3cm}
\centerline{$^b${\it  Instituut-Lorentz, Universiteit Leiden, P.O. Box 9506, 2300 RA Leiden, The Netherlands}}
\vspace{.3cm}
\centerline{$^c${\it Instituto de F\'isica Te\'orica IFT-UAM/CSIC, Universidad Aut\'onoma de Madrid, Madrid 28049, Spain }}
\vspace{.3cm}
\centerline{$^d${\it  Departamento de F\'isica Te\'orica, Universidad Aut\'onoma de Madrid, Madrid, Spain}}

\vspace{2truecm}

%%%%%%%%%%%%%%%%%
\centerline{\bf Abstract}

\vspace{.5truecm}

\noindent
We study the holographic mutual information in AdS$_4$ of disjoint spatial domains in the boundary which are delimited by smooth closed curves. 
A numerical method which approximates a local minimum of the area functional through triangulated surfaces is employed.
After some checks of the method against existing analytic results for the holographic entanglement entropy, we compute the holographic mutual information of equal domains delimited by ellipses, superellipses or the boundaries of two dimensional spherocylinders, finding also the corresponding transition curves along which the holographic mutual information vanishes.

\end{titlepage}

\tableofcontents

\section{Introduction}
\label{sec intro}

Entanglement entropy has been extensively studied during the last decade and its important role in quantum gravity, quantum field theory and condensed matter physics is widely recognized.  

Given a quantum system in its ground state $| \psi \rangle$ and assuming that its Hilbert space can be decomposed as $\mathcal{H} = \mathcal{H}_A \otimes \mathcal{H}_B$, one can introduce the reduced density matrix $\rho_A \equiv \textrm{Tr}_B \rho$ by tracing over $\mathcal{H}_B$  the density matrix $\rho = | \psi \rangle \langle \psi |$ of the whole system. Here we focus on a bipartition of the Hilbert space associated with a separation of a spatial slice into two complementary regions.
The entanglement entropy is the Von Neumann entropy associated with $\rho_A$, namely $S_A \equiv - \textrm{Tr}_A(\rho_A \log \rho_A)$, and it measures the entanglement between $A$ and $B$.
In the same way, one can introduce $\rho_B \equiv \textrm{Tr}_A \rho$ and $S_B$. Since $\rho$ is a pure state, we have that $S_B = S_A$. 
Understanding the dependence of $S_A$ on the geometry of the region $A$ is an important task.

Let us consider a conformal field theory in $D+1$ dimensions at zero temperature in its ground state. The entanglement entropy $S_A$ between a $D$ dimensional spatial region $A$ and its complement $B$ can be written as an expansion in the ultraviolet cutoff $\varepsilon$, where the leading divergence is $S_A \propto \textrm{Area}(\partial A)/\varepsilon^{D-1} + \dots$  \cite{Bombelli:1986rw, Srednicki:1993im}. 
This behaviour is known as the {\em area law} for the entanglement entropy  
and $\partial A$ is sometimes called entangling surface. When $D=1$ and the domain $A$ is an interval, $\partial A$ is made by its two endpoints and the area law is violated because the leading divergence is logarithmic. In particular, $S_A = (c/3) \log(\ell/\varepsilon)+\textrm{const}$, where $c$ is the central charge of the model \cite{Holzhey:1994we, Calabrese:2004eu}.

By virtue of the holographic correspondence \cite{Maldacena:1997re, Witten:1998qj, Gubser:1998bc} (see \cite{Aharony:1999ti} for a review), the entanglement entropy $S_A$ of a conformal field theory in a $D+1$ dimensional Minkowski spacetime can be also calculated from its dual gravitational model defined in a $D+2$ dimensional asymptotically anti-De Sitter (AdS) spacetime whose boundary is the spacetime of the original conformal field theory. In the regime where it is enough to consider only classical gravity, the holographic prescription to compute the entanglement entropy is \cite{Ryu:2006bv, Ryu:2006ef}
\be
\label{RT formula}
S_A = \frac{\mathcal{A}_A}{4 G_N}\,,
\ee
where $G_N$ is the $D+2$ dimensional Newton constant and $\mathcal{A}_A$ is the area of the codimension two minimal area spacelike surface $\tilde{\gamma}_A$ at some fixed time slice such that $\partial \tilde{\gamma}_A= \partial A$. Since $\tilde{\gamma}_A$ reaches the boundary of the asymptotically AdS$_{D+2}$ spacetime, its area $\mathcal{A}_A$ is divergent and therefore it must be regularized through the introduction of a cutoff $\varepsilon$ in the holographic direction, which corresponds to the ultraviolet cutoff of the dual conformal field theory. The leading divergence of (\ref{RT formula}) as $\varepsilon \to 0$ provides the area law of the entanglement entropy.  The covariant generalization of (\ref{RT formula}) has been proposed in \cite{Hubeny:2007xt} and it has been extensively  employed to study holographic models of thermalization. Recent reviews on entanglement entropy in quantum field theory and holography are \cite{Calabrese:2009qy, Casini:2009sr, Takayanagi:2012kg}.

The minimal area surfaces anchored on a given curve defined on the boundary of AdS$_{D+2}$ occur also in the holographic dual of the expectation values of the Wilson loops \cite{Maldacena:1998im, Rey:1998ik}. Nevertheless, while the bulk surfaces for the Wilson loops are always two dimensional, for the holographic entanglement entropy they have codimension two. Thus, when $D=2$ the minimal surfaces to compute for the holographic entanglement entropy (\ref{RT formula}) are the same ones occurring in the gravitational counterpart of the correlators of spacelike Wilson loops. 

As for the dependence of $\mathcal{A}_A$ on the geometry of $\partial A$, analytic results have been found for the infinite strip and for the sphere when $D$ is generic \cite{Ryu:2006bv, Ryu:2006ef}.
Spherical domains play a particular role because their reduced density matrix can be related to a thermal one \cite{Casini:2011kv}. 
When $D=2$, the $O(1)$ term in the expansion of $S_A$ as $\varepsilon \to 0$ for circular domains provides the quantity $F$, which decreases along any renormalization group flow \cite{Myers:2010tj, Klebanov:2011gs, Casini:2012ei}.
Some interesting results have been found about $\mathcal{A}_A$ for an entangling surface $\partial A$ with a generic shape \cite{Solodukhin:2008dh, Hubeny:2012ry, Klebanov:2012yf, Myers:2013lva, Papadimitriou:2010as, Hung:2011ta, Astaneh:2014uba, Allais:2014ata}, but a complete understanding is still lacking.

When $A= A_1 \cup A_2$ is made by two disjoint spatial regions, an important quantity to study is the mutual information
\be
\label{MI def}
I_{A_1, A_2} \equiv S_{A_1} + S_{A_2} - S_{A_1 \cup A_2}\,.
\ee
It is worth remarking  that $S_{A_1 \cup A_2}  $ provides the entanglement between $A_1 \cup A_2$ and the remaining part of the spatial slice. 
In particular, it does not quantify the entanglement between $A_1$ and $A_2$, which is measured by other quantities, such as the logarithmic negativity \cite{Vidal:2002zz, Calabrese:2012ew, Calabrese:2012nk, Calabrese:2014yza}.
In the combination (\ref{MI def}), the area law divergent terms cancel and the subadditivity of the entanglement entropy guarantees that $I_{A_1, A_2}  \geqslant 0$. 
For two dimensional conformal field theories, the mutual information depends on the full operator content of the model \cite{Caraglio:2008pk, Furukawa:2008uk, Calabrese:2009ez, Calabrese:2010he}. When $D\geqslant 2$, the computation of (\ref{MI def}) is more difficult because non local operators must be introduced along $\partial A$ \cite{Cardy:2013nua, Casini:2008wt, Hung:2014npa}.

The holographic mutual information is (\ref{MI def}) with $S_A$ given by (\ref{RT formula}). The crucial term to evaluate is $S_{A_1 \cup A_2}$, which depends on the geometric features of the  entangling surface $\partial A= \partial A_1 \cup \partial A_2$, including also the distance between $A_1$ and $A_2$ and their relative orientation, being $\partial A$ made by two disjoint components.
It is well known that, keeping the geometry of $A_1$ and $A_2$ fixed while their distance increases, the holographic mutual information has a kind of phase transition with discontinuous first derivative, such that $I_{A_1, A_2} =0$ when the two regions are distant enough.
This is due to the competition between two minima corresponding to a connected configuration and to a disconnected one.
While the former is minimal at small distances, the latter is favoured for large distances, where the holographic mutual information therefore vanishes \cite{Hubeny:2007re,Headrick:2010zt, Tonni:2010pv}. 
This phenomenon has been also studied much earlier in the context of the gravitational counterpart of the expectation values of circular spacelike Wilson loops \cite{Gross:1998gk, Zarembo:1999bu, Olesen:2000ji, Kim:2001td}.
The transition of the holographic mutual information is a peculiar prediction of (\ref{RT formula}) and it does not occur if the quantum corrections are taken into account \cite{Faulkner:2013ana}.
A similar transition due to the competition of two local minima of the area functional occurs also for the holographic entanglement entropy of a single region at finite temperature \cite{Headrick:2007km, Azeyanagi:2007bj, Hubeny:2013gta}.

In this paper we focus on $D=2$ and we study the shape dependence of the holographic entanglement entropy and of the holographic mutual information (\ref{RT formula}) in AdS$_4$, which is dual to the zero temperature vacuum state of the three dimensional conformal field theory on the boundary. This reduces to finding the minimal area surface $\tilde{\gamma}_{A}$ spanning a given boundary curve $\partial A$ (the entangling curve) defined in some spatial slice of the boundary of AdS$_4$. 
The entangling curve $\partial A$ could be made by many disconnected components.
When $\partial A$ consists of one or two circles, the problem is analytically tractable \cite{Maldacena:1998im, Drukker:2005cu, Ryu:2006bv, Ryu:2006ef, Hirata:2006jx, Dekel:2013kwa, Krtous:2013vha, Krtous:2014pva}. 
However, for an entangling curve having a generic shape (and possibly many components), finding analytic solutions becomes a formidable task. 
In order to make some progress, we tackle the problem numerically with the help of Surface Evolver \cite{evolverpaper, evolverlink}, a widely used open source software for the modelling of liquid surfaces shaped by various forces and constraints. 
A section at constant time of AdS$_4$ gives the Euclidean hyperbolic space $\mathbb{H}_{3}$.
Once the curve embedded in $\mathbb{H}_{3}$ is chosen, this software constructs a triangular mesh which approximates the surface spanning such curve which is a local minimum of the area functional, computing also the corresponding finite area. The number of vertices $V$, edges $E$ and faces $E$ of the mesh are related via the Euler formula, namely $V-E+F = \chi$, being $\chi=2-2g-b$ the Euler characteristic of the surface, where $g$ is its genus  and $b$ the number of its boundaries. 
In this paper we deal with surfaces of genus $g=0$ with one or more boundaries. 

The paper is organized as follows. 
In \S \ref{sec:minimal_surfaces} we state the problem, introduce the basic notation and review some properties of the minimal surfaces occurring in our computations.
In \S \ref{sec simply connected} we address the case of surfaces spanning simply connected curves. First we review two analytically tractable examples, the circle and the infinite strip; then we address the case of some elongated curves (i.e. ellipse, superellipse and the boundary of the two dimensional spherocylinder) and polygons. Star shaped and non convex domains are also briefly discussed. 
In \S \ref{sec 2 disjoint} we consider $\partial A$ made by two disjoint curves. 
The minimal surface spanning such disconnected curve can be either connected or disconnected, depending on the geometrical features of the boundary, including the distance between them and their relative orientation. 
The cases of surfaces spanning two disjoint circles, ellipses, superellipses and the boundaries of two dimensional spherocylinders are quantitatively investigated for a particular relative orientation.
Further discussions and technical details are reported in the appendices.

%%%%%%%%%%%%%%%%%%%%%%%%%%%%%%%%%

\section{Minimal surfaces in AdS$_4$}
\label{sec:minimal_surfaces}

Finding the minimal area surface spanning a curve is a classic problem in geometry and physics. In $\mathbb{R}^{3}$ this is known as Plateau's problem. A physical realization of the problem is obtained by dipping a stiff wire frame of some given shape in soapy water and then removing it: as the energy of the film is proportional to the area of the water/air interface, the lowest energy configuration consists of a surface of minimal area. In this mundane setting, the requirement of minimal area results into a well known equation
\begin{equation}\label{eq:minimal_surface_in_R3}
H = 0\,,	
\end{equation}
where $H=k_{i}^{i}/2$ is the mean-curvature given by the trace of the extrinsic curvature tensor $k_{ij}=\bm{e}_{i,j}\cdot\bm{N}$, with $\bm{N}$ the surface normal vector, $\bm{e}_{i}$ a generic tangent vector, such that the surface metric tensor is $h_{ij} = \bm{e}_{i}\cdot\bm{e}_{j}$, and $(\,\cdot\,)_{,i}=\partial_{i}(\,\cdot\,)$.

The metric of AdS$_4$ in Poincar\'e coordinates reads
\be
\label{ads metric xy}
ds^2 = \frac{-dt^{2}+dx^{2}+dy^{2}+dz^{2}}{z^2}\,,
\ee
where the AdS radius has been set to one for simplicity. 
The spatial slice $t=\textrm{const}$ provides the Euclidean hyperbolic space $\mathbb{H}_3$ and the region $A$ is defined in the $z=0$ plane. According to the prescription of \cite{Ryu:2006bv, Ryu:2006ef}, to compute the holographic entanglement entropy, first we have to restrict ourselves to a $t=\textrm{const}$ slice and then we have to find, among all the surfaces $\gamma_{A}$ spanning the curve $\partial A$, the one minimizing the area functional
\begin{equation}
\label{area fun general}
\mathcal{A}[\gamma_{A}] = 
\int_{\gamma_{A}} d\mathcal{A} = 
\int_{U_{A}}\frac{\sqrt{h}\, du^{1}du^{2}}{z^{2}}\,,
\end{equation}
where $U_{A}$ is a coordinate patch associated with the coordinates $(u^{1},u^{2})$ and $h=\det(h_{ij})$. We denote by $\tilde{\gamma}_{A}$ the area minimizing surface, so that $\mathcal{A}[\tilde{\gamma}_A] \equiv \mathcal{A}_A $ provides the holographic entanglement entropy through the Ryu-Takayanagi formula (\ref{RT formula}). Since all the surfaces $\gamma_A$ reach the boundary of AdS$_4$, their area is divergent and therefore one needs to introduce a cutoff in the holographic direction to regularize it, namely $z\geqslant \varepsilon >0$, where $\varepsilon$ is an infinitesimal parameter. The holographic dictionary tells us that this cutoff corresponds to the ultraviolet cutoff in the dual three dimensional conformal field theory. Considering $z\geqslant \varepsilon >0$, the area $\mathcal{A}[\gamma_A] $ and therefore $\mathcal{A}_A$ as well become $\varepsilon$ dependent quantities which diverge when $\varepsilon \to 0$. 
Important insights can be found by writing $\mathcal{A}_A$ as an expansion for $\varepsilon \rightarrow 0$. 
When $\partial A$ is a smooth curve, this expansion reads
\be
\label{SA no corners}
\mathcal{A}_A = \frac{P_A}{\varepsilon} - F_A  + o(1)\,,
\ee
where $P_A = \textrm{length}(\partial A)$ is the perimeter of the entangling curve and $o(1)$ indicates vanishing terms when $\varepsilon \to 0$.
When the entangling curve curve $\partial A$ contains a finite number of vertices, also a logarithmic divergence occurs, namely
\be
\label{SA corners}
\mathcal{A}_A  =  \frac{P_A}{\varepsilon} -  B_A \log(P_A/\varepsilon) - W_A + o(1)\,.
\ee
The functions $F_A$, $B_A$ and $W_A$ are defined through (\ref{SA no corners}) and (\ref{SA corners}). They depend on the geometry of $\partial A$ in a very non trivial way. 
We remark that the section of $\tilde{\gamma}_A$ at $z=\varepsilon$ provides a curve which does not coincide with $\partial A$ because of the non 
trivial profile of $\tilde{\gamma}_A$ in the bulk.

As the area element in AdS$_{4}$ is factorized in the form $d\mathcal{A}=du^{1}du^{2}\sqrt{h}/z^{2}$, a surface in AdS$_{4}$ is equivalent to a surface in $\mathbb{R}^{3}$ endowed with a potential energy density of the form $1/z^{2}$. By using the standard machinery of surface geometry (see \S\ref{app math}), one can find an analog of \eqref{eq:minimal_surface_in_R3} in the form
\be
\label{eq:minimal_surface_in_AdS4}
H+\frac{\hat{\boldsymbol{z}}\cdot\boldsymbol{N}}{z} = 0\,,	
\ee
where $\hat{\bm{z}}$ is a unit vector in the $z$ direction. The relation \eqref{eq:minimal_surface_in_AdS4} implies that, in order for the mean curvature to be finite, the surface must be orthogonal to the $(x,y)$ plane at $z=0$: i.e. $\hat{\boldsymbol{z}}\cdot\boldsymbol{N}=0$ at $z=0$. As a consequence of the latter property, the boundary is also a geodesic of $\tilde{\gamma}_{A}$ (see \S\ref{app math}).

%%%%%%%%%%%%%%%%%%%%%%%%%%%%%%%%%%%%%%

\section{Simply connected regions}
\label{sec simply connected}

In this section we consider cases in which the region $A$ is a simply connected domain. 
We first review the simple examples of the disk and of the infinite strip, which can be solved analytically \cite{Ryu:2006bv, Ryu:2006ef}. In \S\ref{sec superellipse} we numerically analyze the case in which $A$ is an elongated region delimited by either an ellipse, a superellipse or the boundary of a two dimensional spherocylinder, while in \S\ref{sec corners} we address the case in which $\partial A$ is a regular polygon.
In \S\ref{sec nonconvex}, star shaped and non convex domains are briefly discussed. 

If $A$ is a disk of radius $R$, the minimal area surface $\tilde{\gamma}_A$ is a hemisphere, as it can be easily proved from a direct substitution in  \eqref{eq:minimal_surface_in_AdS4}.  Taking $\bm{N}=\bm{r}/|\bm{r}|$, with $\bm{r}=(x,y,z)$ and $|\bm{r}|=R$, one finds $\hat{\bm{z}}\cdot\bm{N}=z/R$, hence $H=-1/R$, which is the mean curvature of a sphere whose normal is outward directed. The area of the part of the hemisphere such that $\varepsilon \leqslant z \leqslant R$ is 
\be
\label{Ahemisphere ads4}
\mathcal{A}_A  =  \frac{2\pi R}{\varepsilon} -2\pi \,.
\ee
Comparing this expression with (\ref{SA no corners}), one finds that $F_A = 2\pi$ in this case. It is worth remarking , as peculiar feature of the disk, that in (\ref{Ahemisphere ads4}) $o(1)$ terms do not occur. 

A special case of  \eqref{eq:minimal_surface_in_AdS4} is obtained when the surface is fully described by a function $z=z(x,y)$ representing the height of the surface above the $(x,y)$ plane at $z=0$. 
In this case
\be
\label{area z(x,y)}
\mathcal{A}[\gamma_{A}] = 
\int_{\gamma_{A}} \frac{1}{z^{2}}
 \,\sqrt{1+z_{,x}^2 + z_{,y}^2}
\; dx dy \,,
\ee
and \eqref{eq:minimal_surface_in_AdS4} becomes the following second order non linear partial differential equation for $z$ (see \S\ref{app math} for some details on this derivation)
\be
\label{eq:cartesian_equation}
z_{,xx}(1+z_{,y}^{2})+z_{,yy}(1+z_{,x}^{2})-2z_{,xy}z_{,x}z_{,y}+\frac{2}{z}(1+z_{,x}^{2}+z_{,y}^{2})=0\,,
\ee
with the boundary condition that $z=0$ when $(x,y) \in \partial A$.
The partial differential equation (\ref{eq:cartesian_equation}) is very difficult to solve analytically for a generic curve $\partial A$; but for some domains $A$ it reduces to an ordinary differential equation. 
Apart from the simple hemispherical case previously discussed, this happens also for an infinite strip $A = \{(x,y) \in \mathbb{R}^2, |y| \leqslant R_2 \}$, whose width is $2R_2$. The corresponding minimal surface is invariant along the $x$ axis and therefore it is fully characterized by the profile $z=z(y)$ for $ |y| \leqslant R_2$. Taking $z_{,x}=0$ in  \eqref{eq:cartesian_equation} yields
\be
\label{eom z(y)}
z_{,yy}+\frac{2}{z}(1+z_{,y}^{2})=0\,.
\ee
Equivalently, the infinite strip case can be studied by considering the one dimensional problem obtained substituting $z=z(y)$ directly in (\ref{area z(x,y)})
\cite{Aharony:1999ti,Ryu:2006bv, Ryu:2006ef}.
Since the resulting effective Lagrangian does not depend on $y$ explicitly, one easily finds that $z^2 \sqrt{1+z_{,y}^2}$ is independent of $y$. Taking the derivative with respect to $y$ of this conservation law, (\ref{eom z(y)}) is recovered, as expected. The constant value can be found by considering $y=0$, where $z(0) \equiv z_\ast$ and $z_{,y}(0)=0$. Notice that $z_\ast$ is the maximal height attained by the curve along the $z$ direction.
Integrating the conservation law, one gets
\be
\label{profile for strip}
y(z) = 
\frac{\sqrt{\pi}\; \Gamma(3/4)}{\Gamma(1/4)}\,z_\ast 
-\frac{z^3}{3z_\ast^2}\;
_2F_1 \bigg( 
\frac{1}{2} ,  \frac{3}{4} ;  \frac{7}{4} ; \frac{z^4}{z_\ast^4}
\bigg)\,,
\qquad
z_\ast = \frac{\Gamma(1/4)}{\sqrt{\pi}\;\Gamma(3/4)}\, R_2\,,
\ee
where $\Gamma$ is the Euler gamma and $_2F_1$ is the hypergeometric function. 
Thus, the minimal surface $\tilde{\gamma}_{A}$ consists of a tunnel of infinite length along the $x$ direction, finite width $R_{2}$ along the $y$ direction and whose shape in the $(y,z)$ plane is described by  \eqref{profile for strip}. 
Considering a finite piece of this surface which extends for $R_{1}\gg R_{2}$ in the $x$ direction, whose projection on the $(x,y)$ plane is delimited by the dashed lines in the bottom panel of Fig.\,\ref{fig:squircle min surfs}, its area is given by
\cite{Maldacena:1998im, Rey:1998ik, Ryu:2006bv, Ryu:2006ef}
\be
\label{area infinite strip}
\mathcal{A}_A  = 
\frac{4R_1}{\varepsilon} -  \frac{R_1 s_\infty}{R_2}+o(1)\,,
\qquad
s_\infty \equiv \frac{8\pi^3}{\Gamma(1/4)^4}\,,
\ee
where $\varepsilon \leqslant z \leqslant z_\ast$. 
Comparing (\ref{SA no corners}) with $P_A =4R_1$ and (\ref{area infinite strip}), one concludes that $F_A = s_\infty R_1/ R_2$.

In order to compare \eqref{area infinite strip} with our numerical results, we find it useful to construct an auxiliary surface by closing this long tunnel segment with two planar ``caps'' placed at $x=\pm R_{1}$, whose profile is described in the $(y,z)$ plane by  \eqref{profile for strip}, with a cutoff at $z=\varepsilon$. These regions are identical by construction and their area (see \S\ref{app caps}) is given by $\mathcal{A}_{\rm cap} = 2R_2 / \varepsilon-\pi/2+ o(1)$.
Thus, the total area of the auxiliary surface reads
\be
\label{area RT strip capped}
\mathcal{A}_A  + 2 \mathcal{A}_{\rm cap}  = 
\frac{4(R_1+R_2)}{\varepsilon} - \frac{R_1 s_\infty}{R_2} - \pi + o(1)\,,
\ee
where the coefficient of the leading divergence is the perimeter of the rectangle in the boundary (dashed curve in Fig.\,\ref{fig:squircle min surfs}). It is worth remarking  that this surface is not the minimal area surface anchored on the dashed rectangle in Fig.\,\ref{fig:squircle min surfs}. Indeed, in this case an additional logarithmic divergence occurs (see \S\ref{sec corners}).

Since in the following we will compute numerically $\mathcal{A}_A$ for various domains keeping $\varepsilon$ fixed, let us introduce
\be
\label{FAtilde def}
\widetilde{F}_A \equiv -\left(\mathcal{A}_A - \frac{P_A}{\varepsilon}\right) .
\ee
From (\ref{SA no corners}) one easily observes that $\widetilde{F}_A = F_A + o(1)$ when $\varepsilon \to 0$.
Notice that for the disk we have $\widetilde{F}_A = F_A$.

In Fig.\,\ref{fig:data squircles} the values of $\widetilde{F}_A$ for the surfaces discussed above are represented together with other ones coming from different curves that will be introduced in \S\ref{sec superellipse}:
the black dot corresponds to the disk (see (\ref{Ahemisphere ads4})),
the dotted horizontal line is obtained from (\ref{area infinite strip}) for the infinite strip, while the dashed line is found from the area (\ref{area RT strip capped}) of the auxiliary surface.
\\

\subsection{Superellipse and two dimensional spherocylinder}
\label{sec superellipse}

\begin{figure}[t] 
\vspace{-.9cm}
\hspace{-.0cm}
\begin{center}
\includegraphics[width=.75\textwidth]{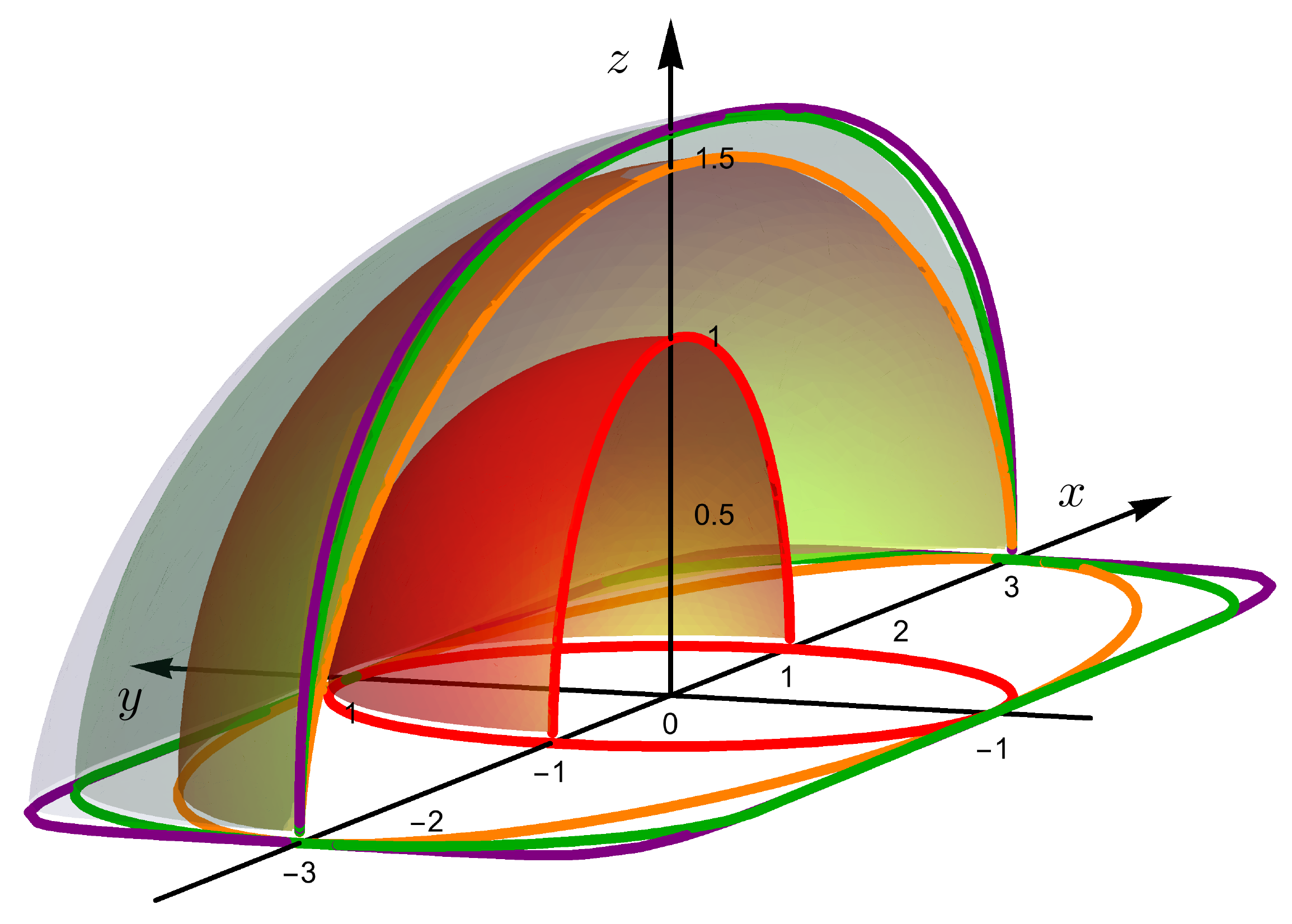}
\\
\rule{0pt}{5.7cm}
\hspace{.7cm}
\includegraphics[width=.85\textwidth]{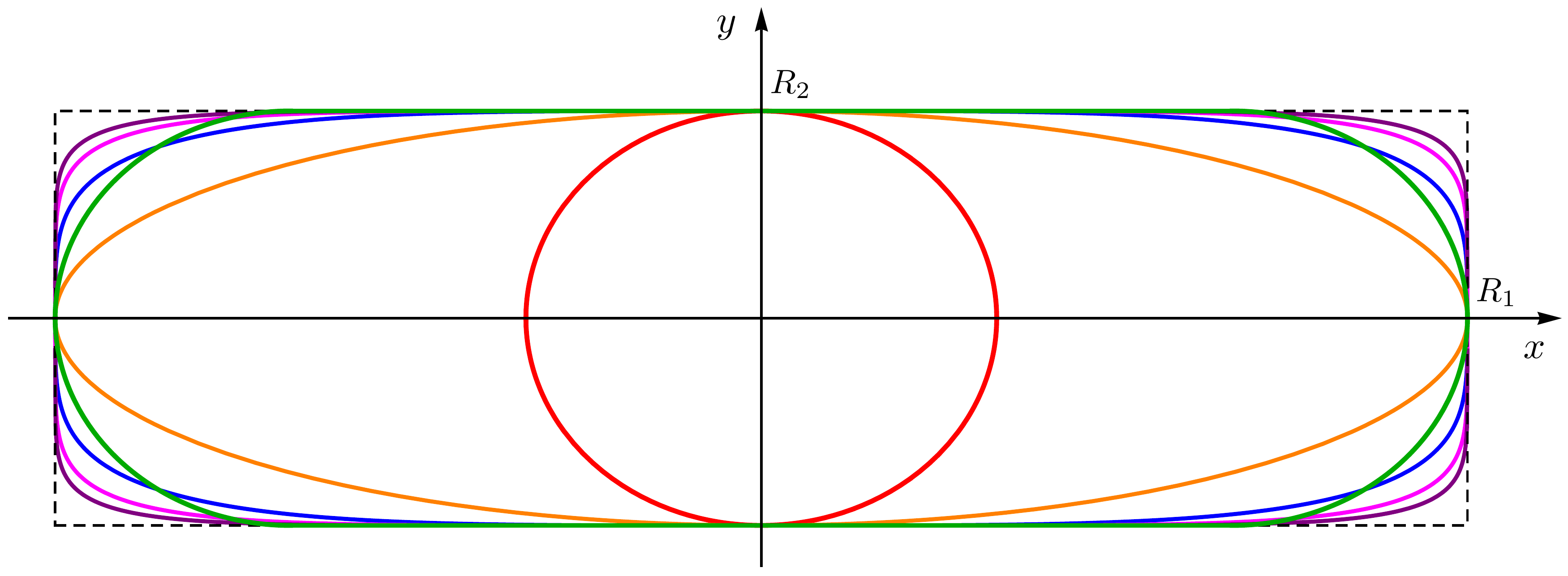}
\\
\end{center}
\vspace{.3cm}
\caption{\label{fig:squircle min surfs}
Top panel:
Minimal surfaces constructed by using Surface Evolver where the entangling curve $\partial A$ is a circle with radius $R=1$ (red), an ellipse (orange), a superellipse (\ref{eq superellipse}) with $n=8$ (purple) and the boundary of a spherocylinder (green) with $R_1=3R_2$. 
The cutoff is $\varepsilon = 0.03$ and only the $y \geqslant 0$ part of the minimal surfaces has been depicted to highlight the curves provided by the section $y=0$.
Bottom panel: In the $(x,y)$ plane, we show
the superellipses with $R_1=3R_2$ with $n=2$ (orange), $n=4$ (blue), $n=6$ (magenta) and $n=8$ (purple), the circle with radius $R_1$ (red curve) and the rectangle circumscribing the superellipses (dashed lines).
The green curve is the boundary of the two dimensional spherocylinder with $R_2=3R_1$. 
}
\end{figure}

\begin{figure}[t] 
\vspace{-.5cm}
\hspace{-.7cm}
\includegraphics[width=1.03\textwidth]{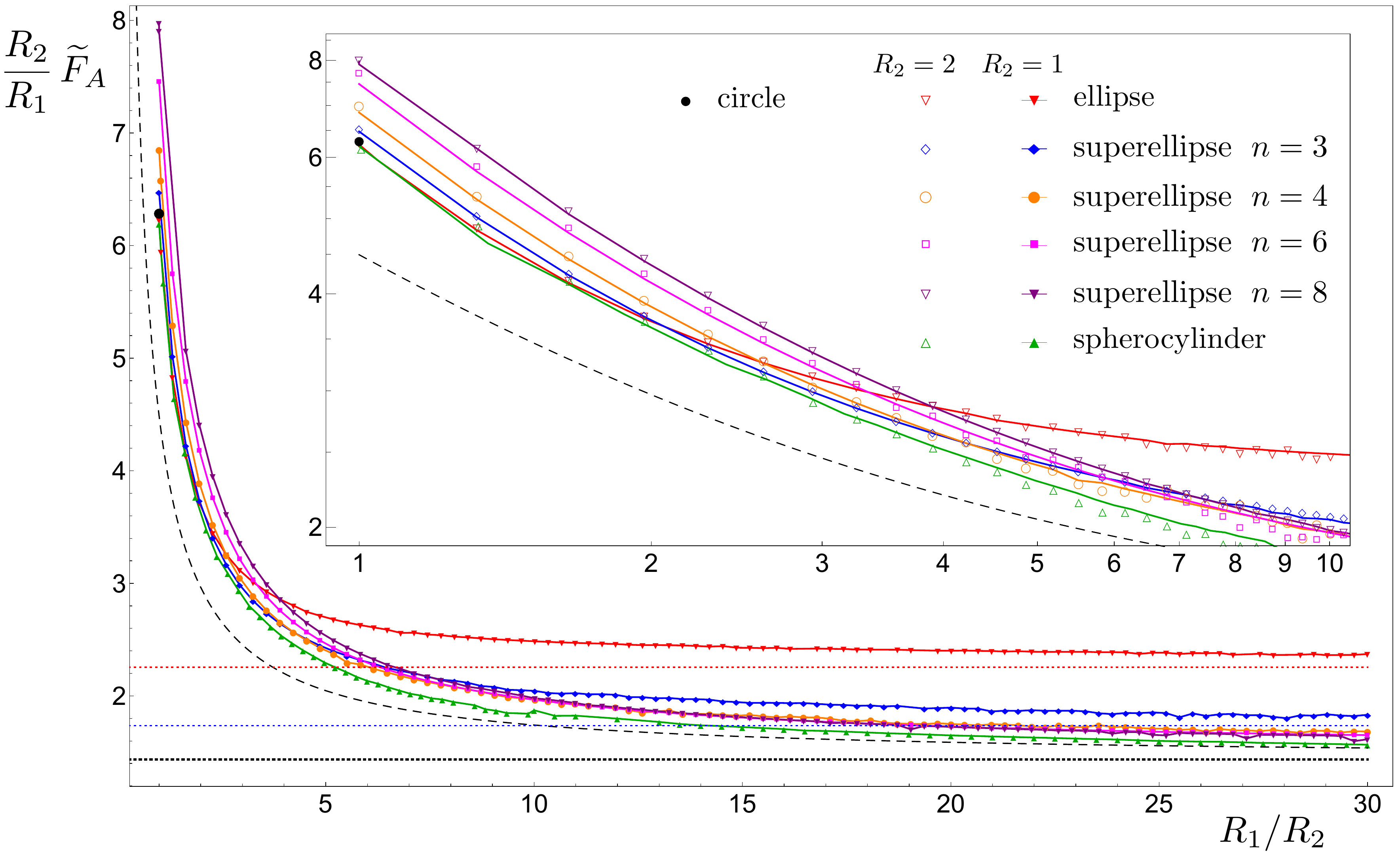}
\vspace{-.1cm}
\caption{\label{fig:data squircles}
Numerical data for $\widetilde{F}_A$, defined in (\ref{FAtilde def}), corresponding to domains $A$ which are two dimensional spherocylinders or delimited by superellipses.
Here $\varepsilon = 0.03$. In the main plot $R_2 =1$, while in the inset, which shows a zoom of the initial part of the main plot in logarithmic scale on both the axes, we have also reported data with $R_2 =2$.
The horizontal dotted black line corresponds to the infinite strip (\ref{area infinite strip})
and the dashed one to the auxiliary surface where the sections at $x=\pm R_1$ have been added (see (\ref{area RT strip capped})).
The red and blue dotted horizontal lines come from the asymptotic result (\ref{Fa star expanded}) evaluated for $n=2$ and $n=3$ respectively.
}
\end{figure}

The first examples of entangling curves $\partial A$ we consider for which analytic expressions of the corresponding minimal surfaces are not known are the superellipse and the boundary of the two dimensional spherocylinder, whose geometries depend on two parameters. The two dimensional spherocylinder nicely interpolates between the circle and the infinite strip. 

In Cartesian coordinates, a superellipse centered in the origin with axes parallel to the coordinate axes is described by the equation
\be
\label{eq superellipse}
 \frac{|x|^n}{R_1^n}  + \frac{|y|^n}{R_2^n}  = 1\,,
\qquad
\hspace{.4cm} R_1 \geqslant R_2 > 0\,,
\qquad
\hspace{.4cm} n \geqslant 2\,,
\ee
where $R_1$, $R_2$ and $n$ are real and positive parameters.
The curve (\ref{eq superellipse}) is also known as Lam\'e curve and here we consider only integers $n \geqslant 2$ for simplicity.
The special case $n=2$ in (\ref{eq superellipse}) is the ellipse with semi-major and semi-minor axes given by $R_1$ and $R_2$ respectively.
As the positive integer $n$ increases, the superellipse approximates the rectangle with sides $2R_1$ and $2R_2$. 
When $R_1=R_2$, the curves (\ref{eq superellipse}) for various $n$ are known as {\it squircles} because they have intermediate properties between the ones of a circle ($n=2$) and the ones of a square ($n \to \infty$).
In the bottom panel of Fig.\,\ref{fig:squircle min surfs}, we show some superellipses with $R_1 =3 R_2$, the circle with radius $R_1$ included in all the superellipses and  the rectangle circumscribing them.

In order to study the interpolation between the circle and the infinite strip, a useful domain to consider is the two dimensional spherocylinder. The spherocylinder (also called capsule) is a three dimensional volume consisting of a cylinder with hemispherical ends. Here we are interested in its two dimensional version, which is a rectangle with semicircular caps.
In particular, the two dimensional spherocylinder circumscribed by the rectangle with sides $2R_1$ and $2R_2$ is defined as the set $\mathcal{S} \equiv \mathcal{D}  \cup \mathcal{C}_+ \cup \mathcal{C}_-$,
where the rectangle $\mathcal{D} $ and the disks $\mathcal{C}_\pm$ are 
\be
\label{eq spherocylinder parts}
\mathcal{D} \equiv \big\{ (x,y)\,, \, |y| \leqslant R_2 \,, \, |x| \leqslant R_1 -R_2 \big\}\,,
\qquad
\mathcal{C}_\pm \equiv  \big\{ (x,y)\,, \, \big[x \pm (R_1-R_2)\big]^2+y^2 \leqslant R_2 \big\}\,.
\ee
The perimeter of this domain is $P_A = 2\pi R_2 +4(R_1-R_2)$ and an explicit example of  $\partial \mathcal{S}$ with $R_2=3R_1$ is given by the green curve in the bottom panel of Fig.\,\ref{fig:squircle min surfs}.
When $R_1= R_2$,  the curve $\partial \mathcal{S}$ becomes a circle, while for $R_1 \gg R_2$ it provides a kind of regularization of the infinite strip. 
Indeed, when $R_1 \to \infty$ at fixed $R_2$ the two dimensional spherocylinder $\mathcal{S}$ becomes the infinite strip with width $2R_2$.
Let us remark that the curvature of $\partial \mathcal{S}$ is discontinuous while the curvature of the superellipse (\ref{eq superellipse}) is continuous. 
Moreover, the choice to regularize the infinite strip through the circles $\mathcal{C}_\pm$ in (\ref{eq spherocylinder parts}) is arbitrary; other domains can be chosen (e.g. regions bounded by superellipses) without introducing vertices in the entangling curve. 
A straightforward numerical analysis allows to observe that a superellipses with $n>2$ intersects once the curve $\partial \mathcal{S}$ in the first quadrant outside the Cartesian axes.

In Fig.\,\ref{fig:data squircles} we show the numerical data for $\widetilde{F}_A$, defined in (\ref{FAtilde def}), when $A$ is given by the domains discussed above: disk, infinite strip, two dimensional spherocylinder and two dimensional regions delimited by superellipses.
In particular, referring to the bottom panel of Fig.\,\ref{fig:squircle min surfs}, we fixed $R_2$ and increased $R_1$.
For the two dimensional spherocylinder, this provides an interpolation between the circle and the infinite strip. 
Surface Evolver has been employed to compute the area $\mathcal{A}_A$ and for the cutoff in the holographic direction we choose $\varepsilon = 0.03$. Below this value, the convergence of the local minimization algorithm employed by Surface Evolver becomes problematic, as well as for too large domains $A$, as discussed in \S\ref{app technical details}. 

When $R_1=R_2$, we observe that $\widetilde{F}_A$ for the squircles with different $n>2$ increases with $n$. 
For large $R_1/R_2$, the limits of $\widetilde{F}_A/(R_1/R_2)$ for the domains we address are finite and positive. The values of these limits associated with the superellipses are ordered in the opposite way in $n$ with respect to the starting point at $R_1=R_2$ and therefore they cross each other as $R_1/R_2$ increases.
We remark that the curve corresponding to the two dimensional spherocylinder stays below the ones associated with the superellipses for the whole range of $R_1/R_2$ that we considered. 
In Fig.\,\ref{fig:data squircles} the horizontal black dotted line corresponds to the infinite strip (see (\ref{area infinite strip})) while the dashed curve is obtained from the auxiliary surface described above (see (\ref{area RT strip capped})). The latter one is our best analytic approximation of the data corresponding to the two dimensional spherocylinder.

Focussing on the regime of large $R_1/R_2$, from Fig.\,\ref{fig:data squircles}  we observe that the asymptotic value of $\widetilde{F}_A/(R_1/R_2)$ for the two dimensional spherocylinder is very close to the one of the auxiliary surface obtained from (\ref{area RT strip capped}) and therefore it is our best approximation of the result corresponding to the infinite strip. 
This is reasonable because the two dimensional spherocylinder is a way to regularize the infinite strip without introducing vertices in the entangling curve, as already remarked above. 
As for the minimal surfaces spanning a superellipse with a given $n\geqslant 2$, in \S\ref{app bounds} an asymptotic lower bound is obtained (see (\ref{Fa star expanded})), generalizing the construction of \cite{Allais:2014ata}. 
In Fig.\,\ref{fig:data squircles} this bound is shown explicitly for $n=2$ and $n=3$ (red and blue dotted horizontal lines respectively). 
Since this value is strictly larger than the corresponding one associated with the infinite strip (see (\ref{area infinite strip})), we can conclude that $\widetilde{F}_A/(R_1/R_2)$ for the superellipse at fixed $n$ does not converge to the value $s_\infty$ associated with the infinite strip.

%%%%%%%%%%%%%%%%%%%%%%%%%%%%%%%%%%

\subsection{Polygons}
\label{sec corners}

\begin{figure}[t] 
\vspace{-.2cm}
\hspace{-.1cm}
\includegraphics[width=1.0\textwidth]{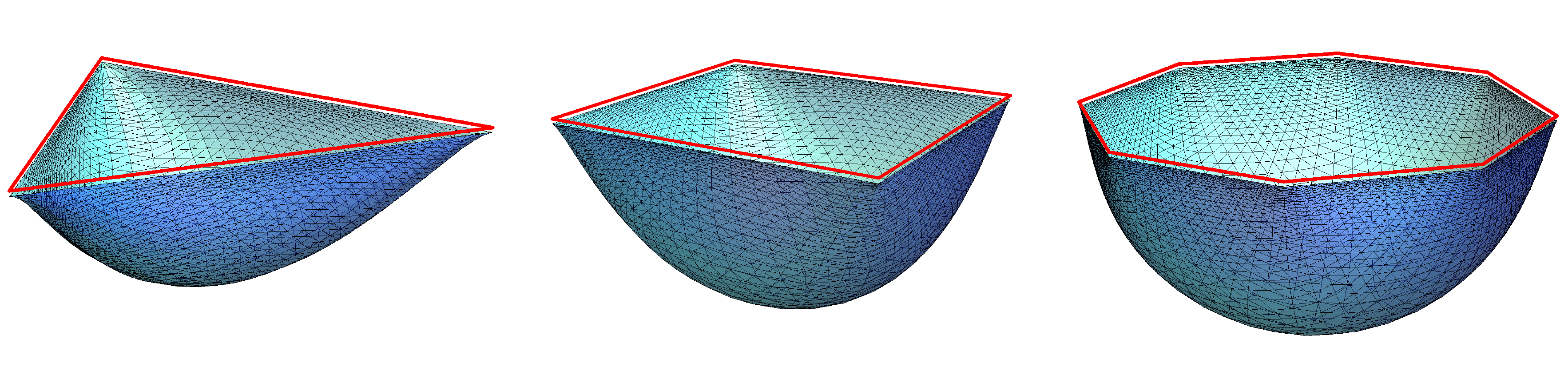}
\vspace{-.55cm}
\caption{\label{fig:polygons min surf}
Minimal area surfaces constructed with Surface Evolver whose $\partial A$ is a polygon with three (left), four (middle) and eight (right) sides. 
The red polygons $\partial A$ lie in the plane at $z=0$ and the $z$ axis points downward but, according to our regularization, the triangulated surfaces are anchored to the same polygons at $z=\varepsilon$.
The pair $(V,F)$ giving the number of vertices $V$ and the number of faces $F$ for these surfaces is $(1585,3072)$ (left), $(2113,4096)$ (middle) and $(4225,8192)$ (right). The number of edges can be found from the Euler formula with vanishing genus and one boundary.
}
\end{figure}

\begin{figure}[t] 
\vspace{.1cm}
\hspace{-.3cm}
\includegraphics[width=1.01\textwidth]{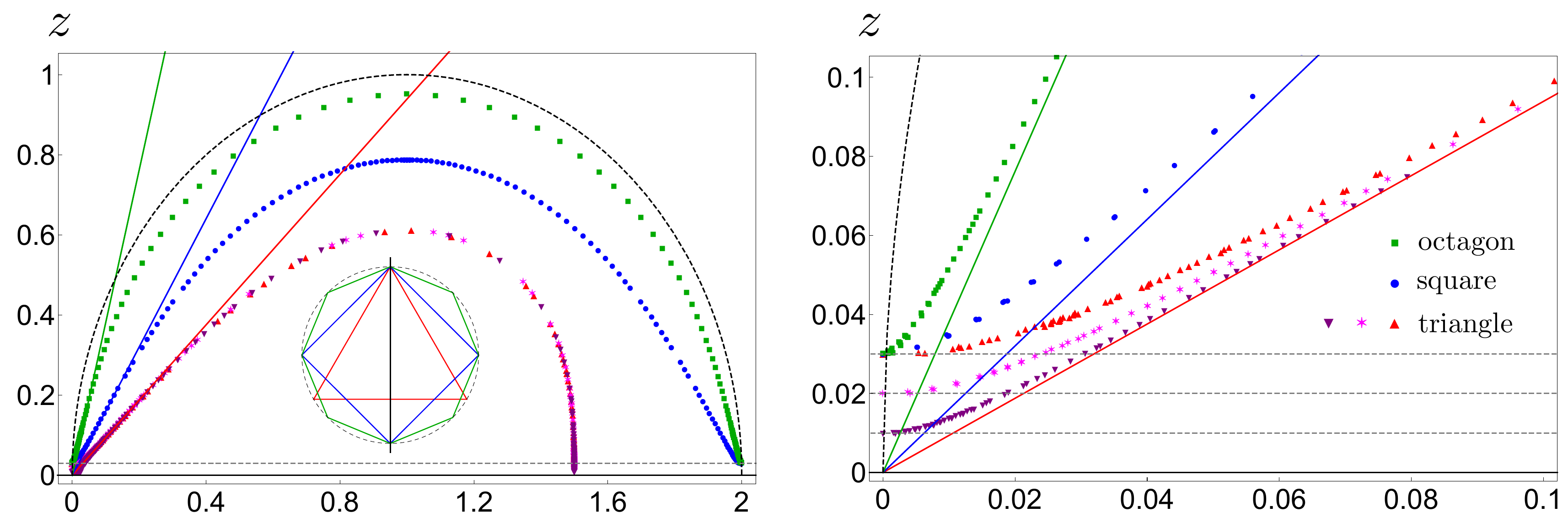}
\vspace{-.3cm}
\caption{\label{fig:polygons profiles}
Left: Section of the minimal surfaces anchored to an equilateral triangle (red, magenta and purple points), a square (blue points) or an octagon (green points) inscribed in a circle, as indicated in the inset by the black line. 
The continuos lines are $z=\rho/f_0(\alpha_N)$, where $f_0(\alpha)$ is found from (\ref{integ phi 2}) with $N=3$ (red), $N=4$ (blue) or $N=8$ (green).
The dashed black curve is the hemisphere corresponding to the circle circumscribing the polygons at $z=0$ (dashed in the inset), while the dashed grey horizontal line corresponds to the cutoff $\varepsilon = 0.03$.
Right: A zoom of the left panel around the origin, placed in the common vertex of the polygons.
For the triangle, three different values of $\varepsilon \in \{0.03, 0.02, 0.01\}$ has been considered to highlight how the agreement with the analytic result improves as $\varepsilon \to 0$.
}
\end{figure}

In this section we consider the minimal area surfaces associated with simply connected regions $A$ whose boundary is a convex polygon with $N$ sides. These are prototypical examples of minimal surfaces spanning entangling curves with geometric singularities. For quantum field theory results about the entanglement entropy of domains delimited by such curves, see e.g. \cite{Fradkin:2006mb, Casini:2006hu, Kallin:2014oka}.

The main feature to observe about the area $\mathcal{A}_A$  of the minimal surface is the occurrence of a logarithmic divergence, besides the leading one associated with the area law,  in its expansion as $\varepsilon \to 0$.
We find it convenient to introduce
\be
\label{BAtilde def}
\widetilde{B}_A \,\equiv \,
\frac{1}{\log(\varepsilon/P_A)}\left(\mathcal{A}_A - \frac{P_A}{\varepsilon}\right) .
\ee
Since (\ref{SA corners}) holds in this case, we have that $\widetilde{B}_A = B_A+ o(1)$.

When $\partial A$ is a convex polygon with $N$ sides, denoting by $\alpha_i < \pi$ its internal angle at the $i$-th vertex, for the coefficient of the logarithmic term in (\ref{SA corners}) we can write
\be
\label{BA sum}
B_A \equiv  2\sum_{i=1}^N b(\alpha_i)\,.
\ee
The function $b(\alpha)$ has been first found in \cite{Drukker:1999zq}, where the holographic duals of the correlators of Wilson loops with cusps have been studied, by considering the minimal surface near a cusp whose opening angle is $\alpha$.
Notice that (\ref{BA sum}) does not depend on the lengths of the edges but only on the convex angles of the polygon. 
Further interesting results have been obtained in the context of the holographic entanglement entropy \cite{Hirata:2006jx, Myers:2012vs}.

\begin{figure}[t] 
\vspace{-.7cm}
\hspace{1.cm}
\includegraphics[width=.87\textwidth]{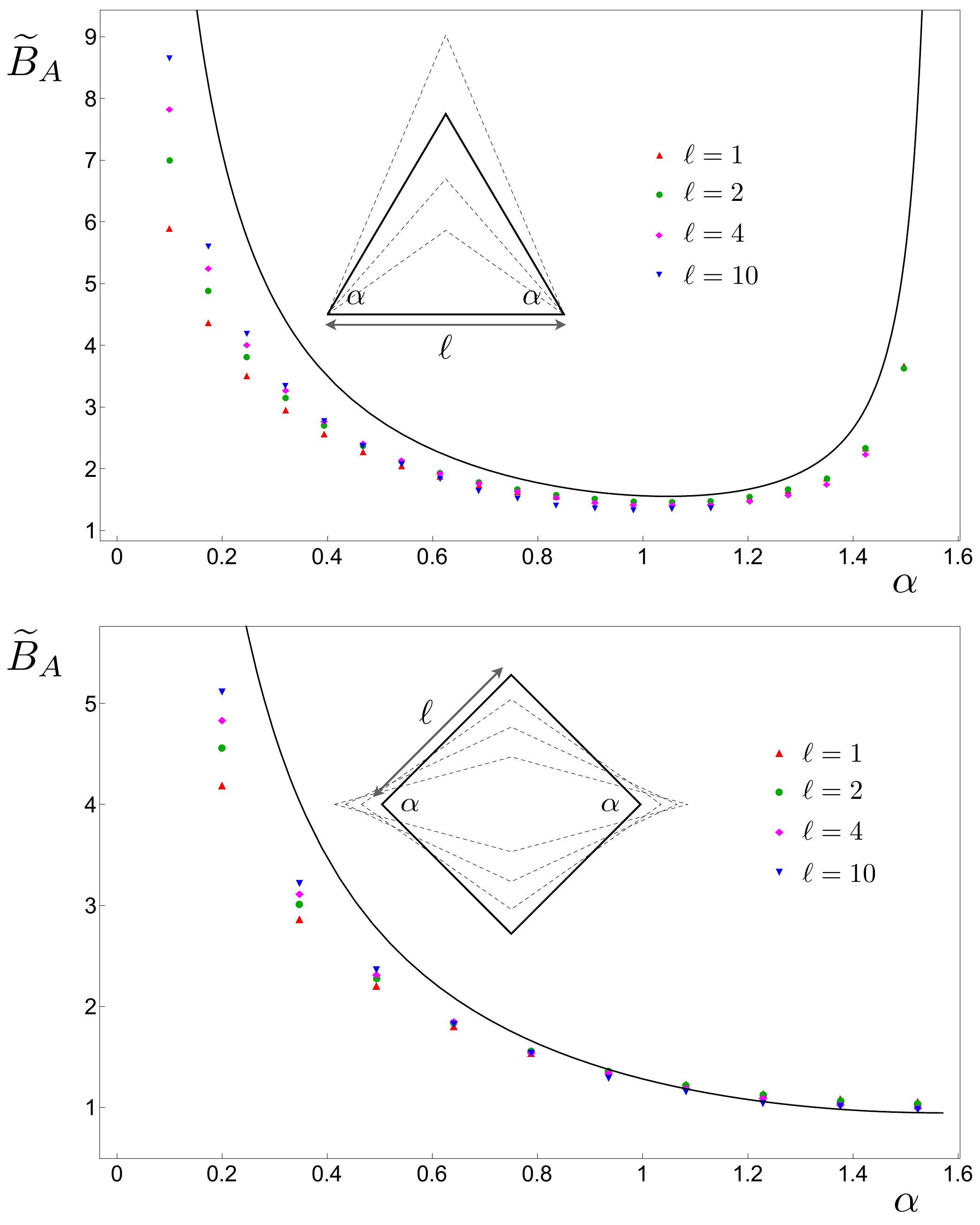}
\vspace{-.0cm}
\caption{\label{fig:data polygons 34}
The quantity $\widetilde{B}_A$ in (\ref{BAtilde def}) with $\mathcal{A}_A$ evaluated with Surface Evolver  when the entangling curve $\partial A$ is either an isosceles triangle whose basis has length $\ell$ (top panel) or a rhombus whose side length is $\ell$ (bottom panel). Here $\varepsilon =0.03$.
The black continuous curves are obtained from (\ref{BA sum}) and (\ref{corner b-function def}). 
}
\end{figure}

\begin{figure}[t] 
\vspace{-.5cm}
\hspace{.6cm}
\includegraphics[width=.88\textwidth]{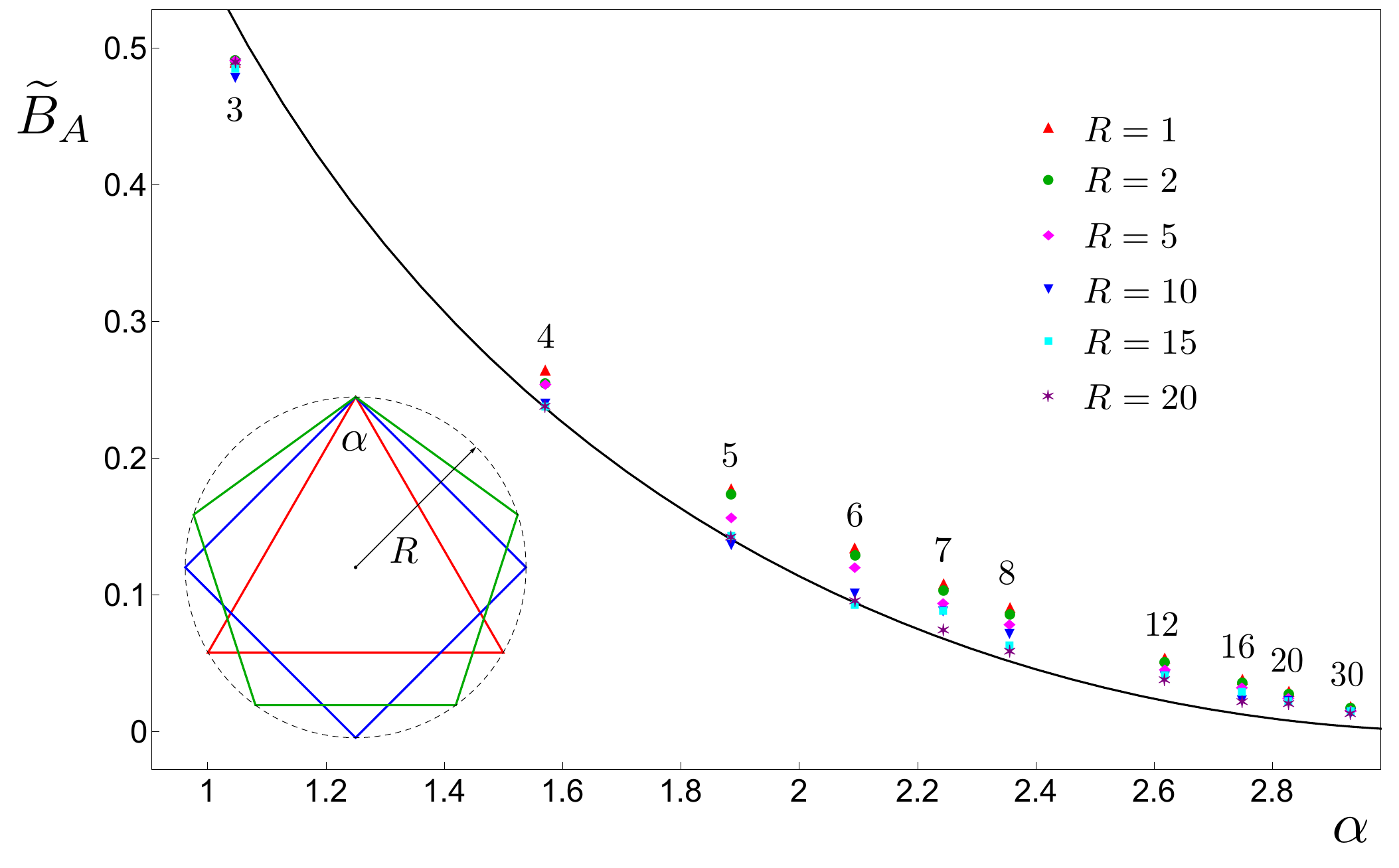}
\vspace{-.2cm}
\caption{\label{fig:data polygons N}
The quantity $\widetilde{B}_A$ in (\ref{BAtilde def})  corresponding to $\partial A$ given by polygons with $N$ equal sides circumscribed by a circle with radius $R$.
The cutoff is $\varepsilon =0.03$ and the values of $N$ are indicated above the corresponding series of data points. 
The black curve is given by (\ref{BA sum}) and (\ref{corner b-function def}). 
}
\end{figure}

Introducing the polar coordinates $(\rho, \phi)$ in the $z=0$ plane, one considers the domain $\{ |\phi| \leqslant \alpha/2\,, \rho < L\}$, where $L\gg 1$. 
By employing scale invariance, one introduces the following ansatz \cite{Drukker:1999zq}
\be
\label{wedge prof ansatz}
z = \frac{\rho}{f(\phi)}\,,
\ee
in terms of a positive function $f(\phi)$, which is even in the domain $|\phi| \leqslant \alpha/2$ and $f \to +\infty$ for $|\phi|  \to \alpha/2$.
Plugging (\ref{wedge prof ansatz}) into the area functional, the problem becomes one dimensional, similarly to the case of the infinite strip slightly discussed in \S\ref{sec simply connected}.
Since the resulting integrand does not depend explicitly on $\phi$, the corresponding conservation law tells us that $(f^4+f^2)/\sqrt{(f')^2+f^4 + f^2}$ is independent of $\phi$.
Thus, the profile for $0\leqslant \phi < \alpha/2$ (the part of the surface with $-\alpha/2 < \phi \leqslant 0$ is obtained by symmetry) is given by
\be
\label{integ phi}
\phi \,=\,
\int_{f_0}^f
\frac{1}{\zeta}
\left[
(\zeta^2+1)
\left(\,
\frac{\zeta^2(\zeta^2+1)}{f_0^2(f_0^2+1)}-1
\right)
\right]^{-\frac{1}{2}}
d\zeta\,,
\ee
being $f_0\equiv f(0)$.
When $f \to \infty$, we require that the l.h.s. of (\ref{integ phi}) becomes $\alpha/2$ and,
by inverting the resulting relation, one finds $f_0=f_0(\alpha)$.
In this limit the integral in (\ref{integ phi}) can be evaluated analytically in terms of elliptic integrals $\Pi$ and $\mathbb{K}$ (see \S\ref{sec:elliptic integrals} for their definitions) as follows
\be
\label{integ phi 2}
\alpha(f_0) =
2 \tilde{f}_0
\sqrt{ \frac{1-2\tilde{f}_0^2}{1-\tilde{f}_0^2}}
\,
\Big[
\Pi\big(1-\tilde{f}_0^2,\tilde{f}_0^2\big)-\mathbb{K}\big(\tilde{f}_0^2\big)
\Big]\,,
\qquad
\tilde{f}_0^2 \equiv \frac{f_0^2}{1+2f_0^2} \in [0,1/2]\,.
\ee
Notice that when $f_0 \to 0$ we have $\alpha\to  \pi$, which means absence of the corner, while $\alpha\to 0$ for $f_0 \to \infty$.

As for the area of the minimal surface given by (\ref{wedge prof ansatz}), one finds that 
\be
\label{corner b-function def}
b(\alpha) 
\equiv 
\int_0^\infty \left(1-\sqrt{\frac{\zeta^2+f_0^2+1}{\zeta^2+2f_0^2+1}}\,\right) d\zeta
\,=\,
\frac{\mathbb{E} \big(\tilde{f}_0^2\big) - \big(1-\tilde{f}_0^2\big) \mathbb{K}\big(\tilde{f}_0^2\big)}{\sqrt{1-2\tilde{f}_0^2}}\,,
\ee
where $f_0 = f_0(\alpha)$ can be found by inverting numerically (\ref{integ phi 2}).
The function (\ref{corner b-function def}) has a pole when $\alpha \to 0$ (in particular, $b(\alpha) = \Gamma(\tfrac{3}{4})^4/(\pi \alpha)+\dots$) while $b(\pi)=0$, which is expected because $\alpha=\pi$ means no cusp and the logarithmic divergence does not occur for smooth entangling curves.

An interesting family of curves to study is the one made by the convex regular polygons. 
They are equilateral, equiangular and all vertices lie on a circle.
For instance, a rhombus does not belong to this family.
Denoting by $R$ the radius of the circumscribed circle and by $N$ the number of sides, the length of each side is $\ell = 2R \sin(\pi/N)$ and all the internal angles are $\alpha_N\equiv\tfrac{N-2}{N}\,\pi$.
When $N \rightarrow \infty$ we have that $\alpha_N \rightarrow \pi$ and the polygon becomes a circle.
Thus, the area of the minimal surface spanning these regular polygons is (\ref{SA corners}) with $P_A=N \ell$ and $B_A = 2N b(\alpha_N)$.

It is interesting to compare the analytic results presented above with the corresponding numerical ones obtained with Surface Evolver. 
Some examples of minimal surfaces anchored on curves $\partial A$ given by a polygon are given in  Fig.\,\ref{fig:polygons min surf}, where the triangulations are explicitly shown.
In Fig.\,\ref{fig:polygons profiles} we take as $\partial A$ an equilateral triangle, a square and an octagon which share a vertex and consider the section of the corresponding minimal surfaces through a vertical plane which bisects the angles associated with the common vertex, as shown in the inset of the left panel. 
Focussing on the part of the curves near the common vertex, we find that the numerical  results are in good agreement with the analytic expression $z=\rho/f_0$, where $f_0 = f_0(\alpha_N)$ is obtained from (\ref{integ phi}). 
It would be interesting to find analytic results for the profiles shown in the left panel of Fig.\,\ref{fig:polygons profiles}.

By employing Surface Evolver, we can also consider entangling curves given by polygons which are not regular, as done in Fig.\,\ref{fig:data polygons 34}, where we have reported the data for $\widetilde{B}_A$ (defined in (\ref{BAtilde def})) corresponding to the area of the minimal surfaces $\tilde{\gamma}_A$ when $\partial A$ is either an isosceles triangle (top panel) or a rhombus with side $\ell$ (bottom panel).
These examples allow us to consider also cusps with small opening angles.  
The size of the isosceles triangles has been changed by varying the angles $\alpha$ adjacent to the basis. Thus, the limiting regimes are the segment ($\alpha =0$) and the semi infinite strip ($\alpha=\pi$). 
As for the rhombus, denoting by $\alpha$ the angle indicated in the inset, its limiting regimes are the segment ($\alpha =0$) and the square ($\alpha =\pi$).
The cutoff in the holographic direction has been fixed to $\varepsilon =0.03$ (see the discussion in \S\ref{app technical details}).
Increasing the size of the polygon improves the agreement with the curve given by (\ref{BA sum}) and (\ref{corner b-function def}), as expected, because $\varepsilon/P_A$ gets closer to zero.  
Moreover, the agreement between the numerical data and the analytic curve gets worse as $\alpha$ becomes very small.

In Fig.\,\ref{fig:data polygons N} we report the data for $\widetilde{B}_A$ found with Surface Evolver for regular polygons with various number $N$ of edges. The agreement with the curve given by (\ref{BA sum}) and (\ref{corner b-function def}) is quite good and it improves for larger domains.

It is worth emphasizing that, for entangling surfaces $\partial A$ containing corners, the way we have employed to construct the minimal surfaces with Surface Evolver (i.e. by defining $\partial A$ at $z=\varepsilon$) influences the term $W_A$ in the expansion (\ref{SA corners}) for the area, as already remarked in \cite{Drukker:1999zq}.

It could be helpful to compute the length $P_\varepsilon$ of the curve defined as the section at $z=\varepsilon$ of the minimal surface anchored on the long segments of a large wedge with opening angle $\alpha$, which has been introduced above. From (\ref{wedge prof ansatz}) we find that, in terms of polar coordinates whose center is the projection of the vertex at $z=\varepsilon$, this curve is given by $\rho = \varepsilon f(\phi)$. Being $L\gg 1$, we find that $P_\varepsilon$ reads
\be
\label{P_eps def}
P_\varepsilon =
2 \int_0^{\alpha_\varepsilon/2} \sqrt{\rho^2 +(\partial_\phi \rho)^2} \, d\phi
=
2 \varepsilon \int_0^{\alpha_\varepsilon/2} \sqrt{f^2 +(\partial_\phi f)^2} \,d\phi
=
2  \varepsilon \int_{f_0}^{L/\varepsilon} \sqrt{1 +f^2 (\partial_f \phi)^2} \,df 
=
2L- 2  f_0 \varepsilon  +\dots ,
\ee
where $\alpha_\varepsilon \simeq \alpha$  is defined by the relation $L= \varepsilon f(\alpha_\varepsilon/2) $ and in the last step a change of variable has been performed. 
It is easy to observe that $\alpha_\varepsilon < \alpha$.
Considering the integral in the intermediate step of (\ref{P_eps def}), one notices that it diverges because of its upper limit of integration (see the text below (\ref{wedge prof ansatz})), while the lower limit of integration gives a finite result, providing a contribution $O(\varepsilon)$ to $P_\varepsilon$.
The expression of $\partial_f \phi$ can be read from the integrand of (\ref{integ phi}), finding that $f^2 (\partial_f \phi)^2=O(1/f^6)$ when $f \to +\infty$.
Since $L/\varepsilon \gg 1$, by expanding the integrand in (\ref{P_eps def}) for large $f$, we obtain that this integral diverges like $L/\varepsilon - f_0 + \dots$, where the finite term has been found numerically.
As a cross check of the finite term, we observe that $f_0=0$ when $\alpha=\pi$ (see below (\ref{integ phi 2})), as expected. 
Thus, we can conclude that $P_\varepsilon = 2L + O(\varepsilon)$, being $P_A =2L$ the length of the boundary of the wedge at $z=0$.
Notice that, performing this computation for the minimal surface anchored on a circle of radius $R$, which is a hemisphere, one finds that $P_\varepsilon = 2\pi R + O(\varepsilon^2)$.

Let us remark that $P_\varepsilon$ is not related to the regularization we adopt in our numerical analysis, as it can be realized from the right panel of Fig.\,\ref{fig:polygons profiles}. Indeed, in order to analytically the profiles given by the numerical data in the 
right panel of Fig.\,\ref{fig:polygons profiles} the ansatz (\ref{wedge prof ansatz}) cannot be employed and a partial differential equation must be solved.

\begin{figure}[t] 
\vspace{-.5cm}
\hspace{-.0cm}
\begin{center}
\includegraphics[width=.85\textwidth]{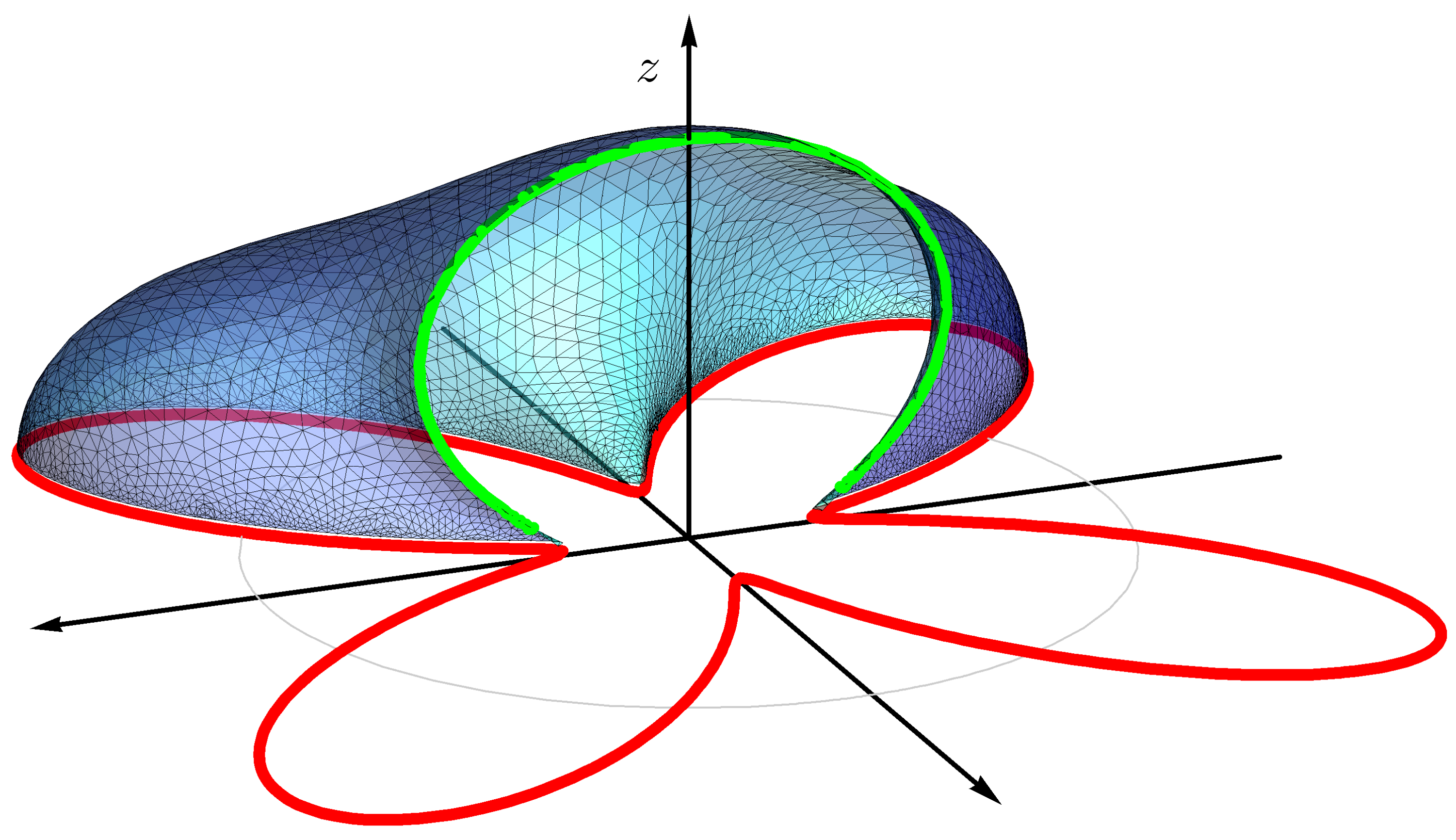}
\end{center}
\vspace{.1cm}
\caption{\label{fig:nonconvex min surfs star}
Minimal surface constructed with Surface Evolver corresponding to a star convex domain delimited by the red curve given by  $r(\phi) = R_0 +a_0 \cos( k \phi)$ in polar coordinates in the $z=0$ plane, with $R_0=1$, $a_0=0.7$ and $k=4$. 
Here the cutoff is $\varepsilon = 0.03$ and $(V,F)=(6145,11776)$. Only half of the minimal surface is shown in order to highlight the section given by the green curve. 
}
\end{figure}

\begin{figure}[t] 
\vspace{-.5cm}
\hspace{-.0cm}
\begin{center}
\includegraphics[width=1\textwidth]{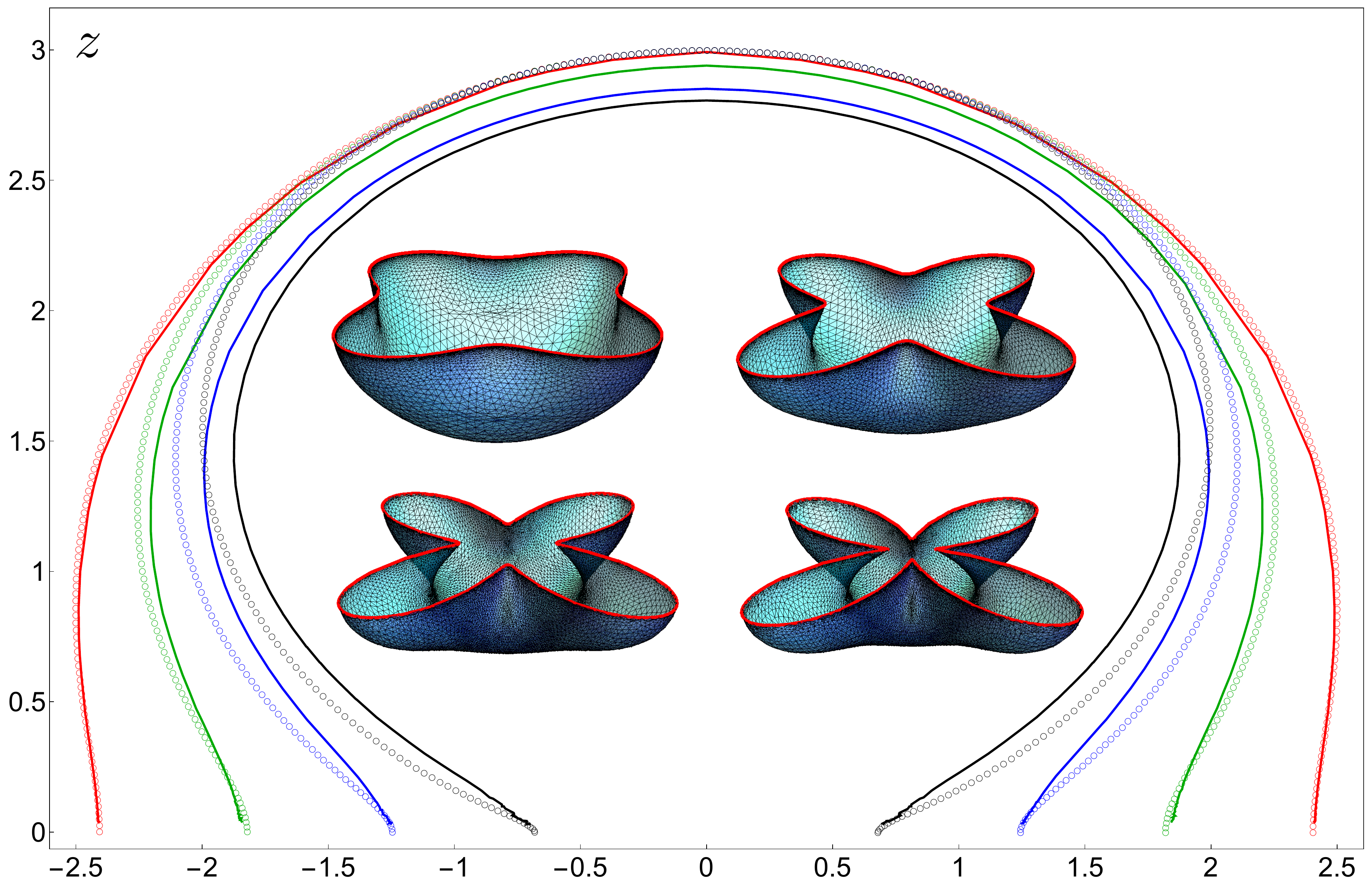}
\end{center}
\vspace{-.1cm}
\caption{\label{fig:section star}
Minimal surfaces corresponding to entangling curves $\partial A$ at $z=0$ given by (\ref{R z=0 order 2}) with $R=3$, $k =4$, $\mu = 0$ and for different values of the parameter $a$,
which delimit star shaped domains (red curves in the inset).
In the inset, where the $z$ direction points downward, we show the minimal surfaces constructed through Surface Evolver with $\varepsilon =0.03$. 
In the main plot, the solid curves are their sections of the minimal surfaces of the inset at $\phi=\pi/4$ (like the green curve in Fig.\,\ref{fig:nonconvex min surfs star}), while the curves made by the empty small circles are obtained from the linearized solution of \cite{Hubeny:2012ry}. 
The colors in the main plot correspond to different values of $a\in \{ 0.2, 0.4, 0.6, 0.8\}$  (red, green, blue and black respectively), while in the inset $a$ increases
starting from the top left surface and going to the top right, bottom left and bottom right ones.
}
\end{figure}

\subsection{Star shaped and non convex regions}
\label{sec nonconvex}

\begin{figure}[t] 
\vspace{-.5cm}
\hspace{-.0cm}
\begin{center}
\includegraphics[width=.85\textwidth]{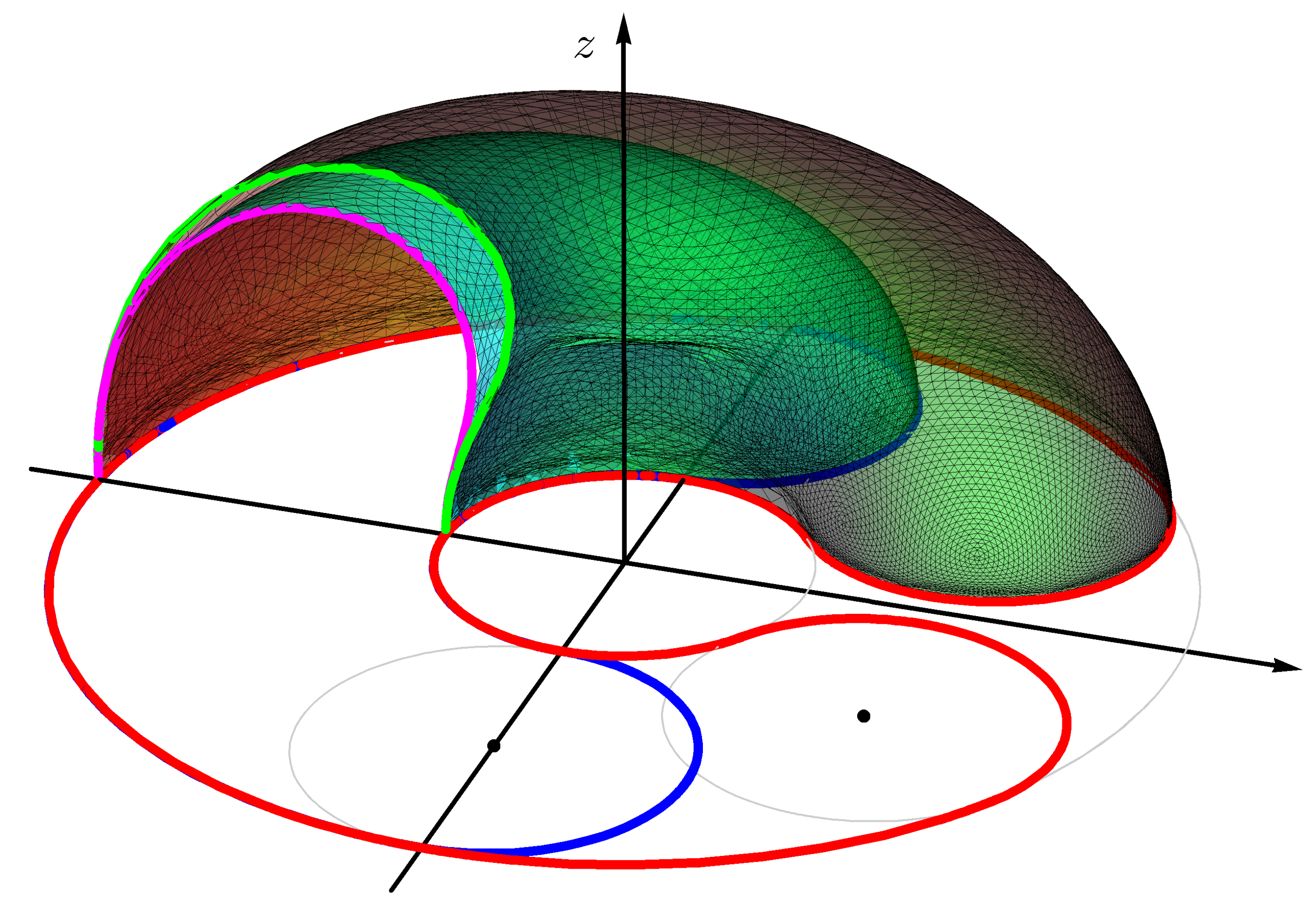}
\end{center}
\vspace{.1cm}
\caption{\label{fig:nonconvex min surfs}
Minimal surfaces constructed with Surface Evolver corresponding to non convex domains at $z=0$ delimited by the red and blue curves, which are made by arcs of circle centered either in the origin or in the points identified by the black dots. 
The green and magenta curves are sections of the minimal surfaces anchored on the red and the blue curves respectively. 
}
\end{figure}

The crucial assumption throughout the above discussions is that the minimal surface $\tilde{\gamma}_A$ can be fully described by $z=z(x,y)$, where $(x,y) \in A$. 
Nevertheless, there are many domains $A$ for which this parameterization cannot be employed because there are pairs of different points belonging to the minimal surfaces $\tilde{\gamma}_A$ with the same projection $(x,y) \notin A$ in the $z=0$ plane.
In these cases, being the analytic approach quite difficult in general, one can employ our numerical method  to find the minimal surfaces and to compute their area.
The numerical data obtained with Surface Evolver would be an important benchmark for analytic results that could be found in the future. 

An interesting class of two dimensional regions to consider is given by the star shaped domains.
A region $A$ at $z=0$ belongs to this set of domains if a point $P_0 \in A$ exists such that the segment connecting any other point of the region to $P_0$ entirely belongs to $A$.
As for the minimal surface anchored on a star shaped domain $A$, by introducing a spherical polar coordinates system $(r,\phi, \theta)$ centered in $P_0$ (the angular ranges are $\phi \in [0,2\pi)$ and $\theta \in [0,\pi/2]$), one can parameterize the entire minimal surface. 
Thus, we have $\rho = r \sin \theta$ and $z=r \cos \theta$, being $(\rho,\phi)$ the polar coordinates of the $z=0$ plane.
Some interesting analytic results about these domains have been already found. In particular, \cite{Hubeny:2012ry} considered minimal surfaces obtained as smooth perturbations around the hemisphere and in \cite{Klebanov:2012yf} the behaviour in the IR regime for gapped backgrounds \cite{Klebanov:2007ws} has been studied. 
Our numerical method allows a more complete analysis because, within our approximations, we can find (numerically) the area of the corresponding minimal surface without restrictions. 

In Fig.\,\ref{fig:nonconvex min surfs star} we show a star convex domain $A$ delimited by the red curve at $z=0$, which does not contain vertices, and the corresponding minimal surface $\tilde{\gamma}_A$ anchored on it. Notice that there are pairs of points belonging to $\tilde{\gamma}_A$ having the same projection $(x,y) \notin A$ on the $z=0$ plane. It is worth recalling that in our regularization the numerical construction of the minimal surface with Surface Evolver has been done by defining the entangling curve $\partial A$ at $z=\varepsilon$.

In order to give a further check of our numerical method, we find it useful to compare our numerical results against the analytic ones obtained in \cite{Hubeny:2012ry}, where the equation of motion coming from (\ref{area fun general}) written in polar coordinates $(r,\phi, \theta)$ has been linearized to second order around the hemisphere solution with radius $R$, finding
\be
\label{soln a-expansion}
r(\theta,\phi) = R + a \,r_1(\theta,\phi) + a^2 r_2(\theta,\phi) +O(a^3)\,,
\ee
where the $r_1(\theta,\phi)$ and $r_2(\theta,\phi)$ are given by \cite{Hubeny:2012ry}
\bea
\label{r1 linear}
r_1(\theta,\phi)  &=& [\tan (\theta/2)]^k (1+ k \cos \theta) \cos(k \phi)\,,
\\
\label{r2 linear}
\rule{0pt}{.7cm}
r_2(\theta,\phi)  &=& 
\frac{[\tan (\theta/2)]^{2 k}}{4R}\,
\Big\{
(1+ k \cos \theta)^2
+\big[
\mu \,(1+2 k \cos \theta) + k^2 \cos^2\theta
\big] \cos(2k \phi)
\Big\}\,,
\eea
being $k \in \mathbb{N}$ and $\mu \in \mathbb{R}$ two parameters of the linearized solution. 
The minimal surface equation coming from (\ref{area fun general}) is satisfied by (\ref{soln a-expansion}) at $O(a^2)$. 
Notice that $r_1(\theta=0,\phi) = r_2(\theta=0,\phi) = 0$, which means that the maximum value reached by the linearized solution along the $z$ direction is $R$, like for the hemisphere. 
Neglecting the $O(a^3)$ terms in (\ref{soln a-expansion}), one has a surface spanning the curve $r(\pi/2,\phi)\equiv R_2(\phi)$ at $z=0$, which reads
\be
\label{R z=0 order 2}
R_2(\phi) \equiv 
R + a \cos(k \phi) +\frac{a^2}{4R} \,\big[1+\mu \cos(2k \phi)\big]\, .
\ee
In Fig.\,\ref{fig:section star} we construct the minimal surfaces providing the holographic entanglement entropy of some examples of star shaped regions $A$ delimited by (\ref{R z=0 order 2}) where $R$ and $\mu$ are kept fixed while $a$ takes different values, taking the $\phi=\pi/4$ section of these surfaces (see also the green curve in Fig.\,\ref{fig:nonconvex min surfs star}). 
Compare the resulting curves (the solid ones in the main plot of Fig.\,\ref{fig:section star}) with the corresponding  ones obtained from the second order linearized solution (\ref{soln a-expansion}) (made by the empty circles), we observe that the agreement is very good for small values of $a/R$ and it gets worse as $a/R$ increases, as expected.

Our numerical method is interesting because it does not rely on any particular parameterization of the surface and this allows us to study the most generic non convex domain. 
In Fig.\,\ref{fig:nonconvex min surfs} we show two examples of non convex domains $A$ which are not star shaped: one is delimited by the red curve and the other one by the blue curve. 
We could see these domains as two two dimensional spherocylinders which have been bended in a particular way.
Constructing the minimal surfaces $\tilde{\gamma}_A$ anchored on their boundaries and considering their sections given by the green and magenta curves, one can clearly observe that some pair of points belonging to the minimal surfaces have the same projection $(x,y) \notin A$ on the $z=0$ plane, as already remarked above. 
An analytic description of these surfaces is more difficult with respect to the minimal surfaces anchored on the boundary of star shaped domains because it would require more patches.

%%%%%%%%%%%%%%%%%%%%%%%%%%%%%%%%%

\section{Two disjoint regions}
\label{sec 2 disjoint}

\begin{figure}[t!] 
\vspace{-.5cm}
\hspace{.65cm}
\includegraphics[width=.92\textwidth]{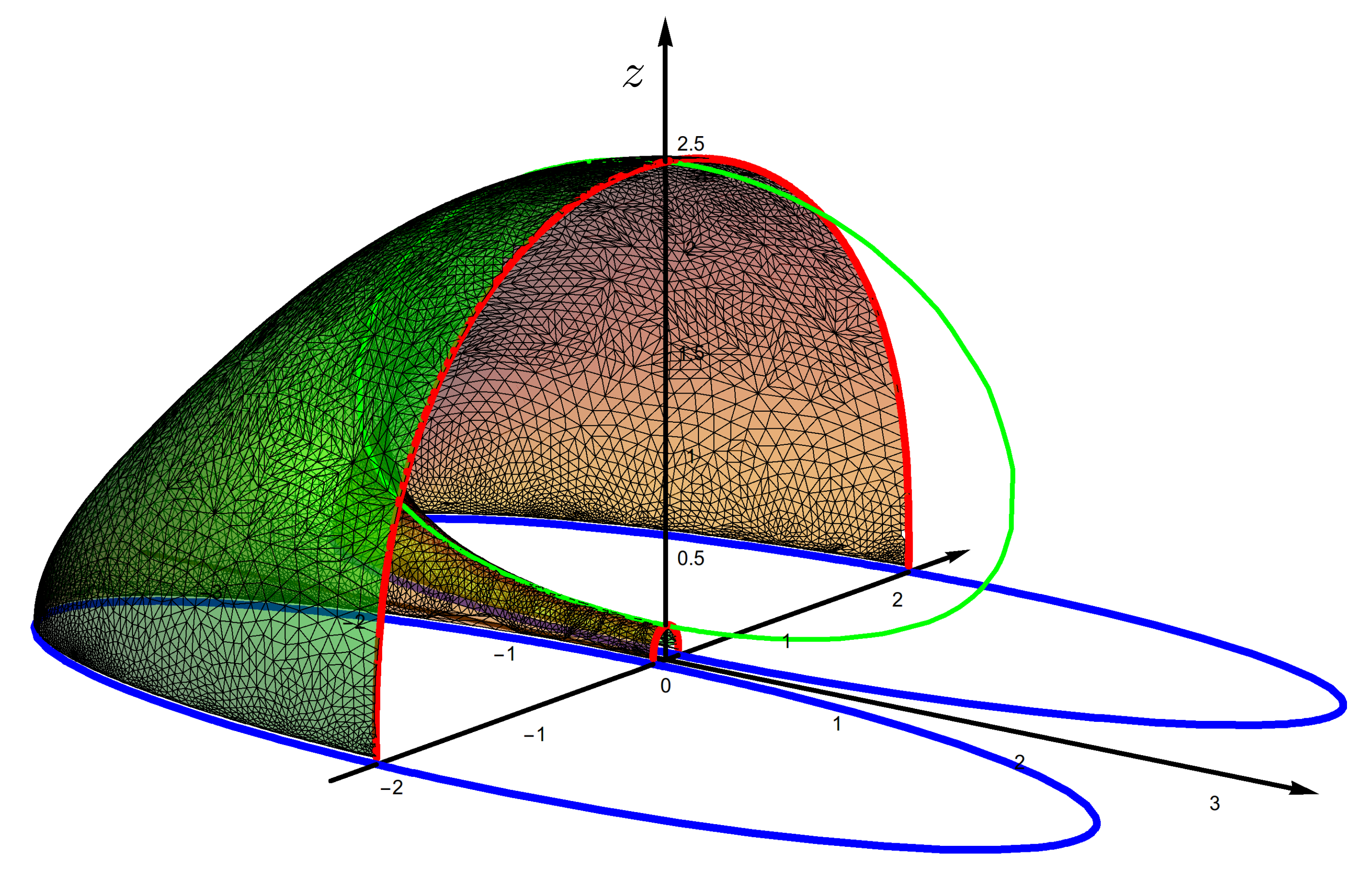}
\vspace{.3cm}
\caption{\label{fig:2disjoint example}
Minimal surface constructed with Surface Evolver for a domain $A=A_1 \cup A_2$ delimited by two disjoint and equal ellipses at $z=0$ (blue curves). 
Here $\varepsilon =0.03$ and the minimal surface is anchored on $\partial A$ defined at $z=\varepsilon $, according to our regularization prescription. 
The minimal surface has $(V,F)=(18936,37616)$ (the number of edges $E$ can be found from the Euler formula with vanishing genus and two boundaries). Only half surface is shown in order to highlight the curves given by the two sections suggested by the symmetry of the surface.
}
\end{figure}

In this section we discuss the main result of this paper, which is the numerical study of the holographic mutual information of disjoint equal domains delimited by some of the smooth curves introduced in \S\ref{sec superellipse}. 
For two equal disjoint ellipses, an explicit example of the minimal surface whose area determines the corresponding holographic mutual information is shown in Fig.\,\ref{fig:2disjoint example}.

Let us consider two dimensional domains $A=A_1 \cup A_2$ made by two disjoint components $A_1$ and $A_2$, where each component is a simply connected domain delimited by a smooth curve. 
The boundary is $\partial A= \partial A_1 \cup \partial A_2$ and the shapes of $\partial A_1$ and $\partial A_2$ could be arbitrary, but we will focus on the geometries discussed in \S\ref{sec simply connected}.
Since the area law holds also for $S_{A_1 \cup A_2}$ and $P_A = P_{A_1} + P_{A_2}$, the leading divergence $O(1/\varepsilon)$ cancels in the combination (\ref{MI def}), which is therefore finite when $\varepsilon \to 0$.

Considering the mutual information (\ref{MI def}) with the entanglement entropy computed through the holographic formula  (\ref{RT formula}), we find it convenient to introduce $\mathcal{I}_{A_1, A_2} $ as follows
\be
\label{MI holog}
I_{A_1, A_2} \equiv  \frac{\mathcal{I}_{A_1, A_2} }{4G_N} \,,
\ee
where $G_N$ is the four dimensional Newton constant.
Since $\partial A_1$ and $\partial A_2$ are smooth curves, from (\ref{SA no corners}) and (\ref{FAtilde def}) we have
\be
\mathcal{I}_{A_1, A_2}  
= \widetilde{F}_{A_1 \cup A_2}  - \widetilde{F}_{A_1} - \widetilde{F}_{A_2} 
= F_{A_1 \cup A_2}  - F_{A_1} - F_{A_2} + o(1)\,.
\ee
In the following we study $\mathcal{I}_{A_1, A_2}$ when $\partial A$ is made either by two circles (\S\ref{sec 2 disjoint circles}) or by two superellipses or by the boundaries of two two dimensional spherocylinders.
Once $A_1$, $A_2$ and their relative orientation have been fixed, we can only move their relative distance. A generic feature of the holographic mutual information is that it diverges when $A_1$ and $A_2$ become tangent, while it vanishes when the distance between $A_1$ and $A_2$ is large enough.

\subsection{Circular boundaries}
\label{sec 2 circular bdy}

In this section we consider domains $A$ whose boundary $\partial A$ is made by two disjoint circles. 
The corresponding disks can be either overlapping (in this case $A$ is an annulus) \cite{Drukker:2005cu, Hirata:2006jx, Dekel:2013kwa} or disjoint \cite{Krtous:2013vha, Krtous:2014pva}.

\subsubsection{Annular regions}
\label{sec annulus}

Let us consider the annular region $A$  bounded by two concentric circles with radii $R_{\textrm{\tiny in}} < R_{\textrm{\tiny out}} $. 
The complementary domain $B$ is made by two disjoint regions and, since we are in the vacuum,  $S_A = S_B$. The minimal surfaces associated with this case have been already studied in \cite{Drukker:2005cu, Dekel:2013kwa} as the gravitational counterpart of the correlators of spatial Wilson loops and in \cite{Hirata:2006jx} from the holographic entanglement entropy perspective.

In \S\ref{app annulus} we discuss the construction of the analytic solution in $D$ dimensions for completeness, but here we are interested in the $D=2$ case.
Because of the axial symmetry, it is convenient to introduce polar coordinates $(\rho,\phi)$ at $z=0$. Then, the profile of the minimal surface is completely specified by a curve in the plane $(\rho, z)$.

A configuration providing a local minimum of the area functional is made by the disjoint hemispheres anchored on the circles with radii $R_{\textrm{\tiny in}}$ and $R_{\textrm{\tiny out}} $.
In the plane $(\rho, z)$, they are described by two arcs centered in the origin with an opening angle of $\pi/2$ (see the dashed curve in Fig.\,\ref{fig:appannulusprof}).
Another surface anchored on $\partial A$ that could give a local minimum of the area functional is the connected one having the same topology of a half torus.
This solution is fully specified by its profile curve in the plane $(\rho, z)$, which connects the points $(R_{\textrm{\tiny in}},0 )$ and $(R_{\textrm{\tiny out}} ,0)$.
Thus, we have two qualitatively different surfaces which are local minima of the area functional and we have to establish which is the global minimum in order to compute the holographic entanglement entropy.
Changing the annulus $A$, a transition occurs between these two types of surfaces, as we explain below. This is the first case that we encounter of a competition between two saddle points of the area functional.

\begin{figure}[t] 
\vspace{-.2cm}
\hspace{-.5cm}
\includegraphics[width=1.035\textwidth]{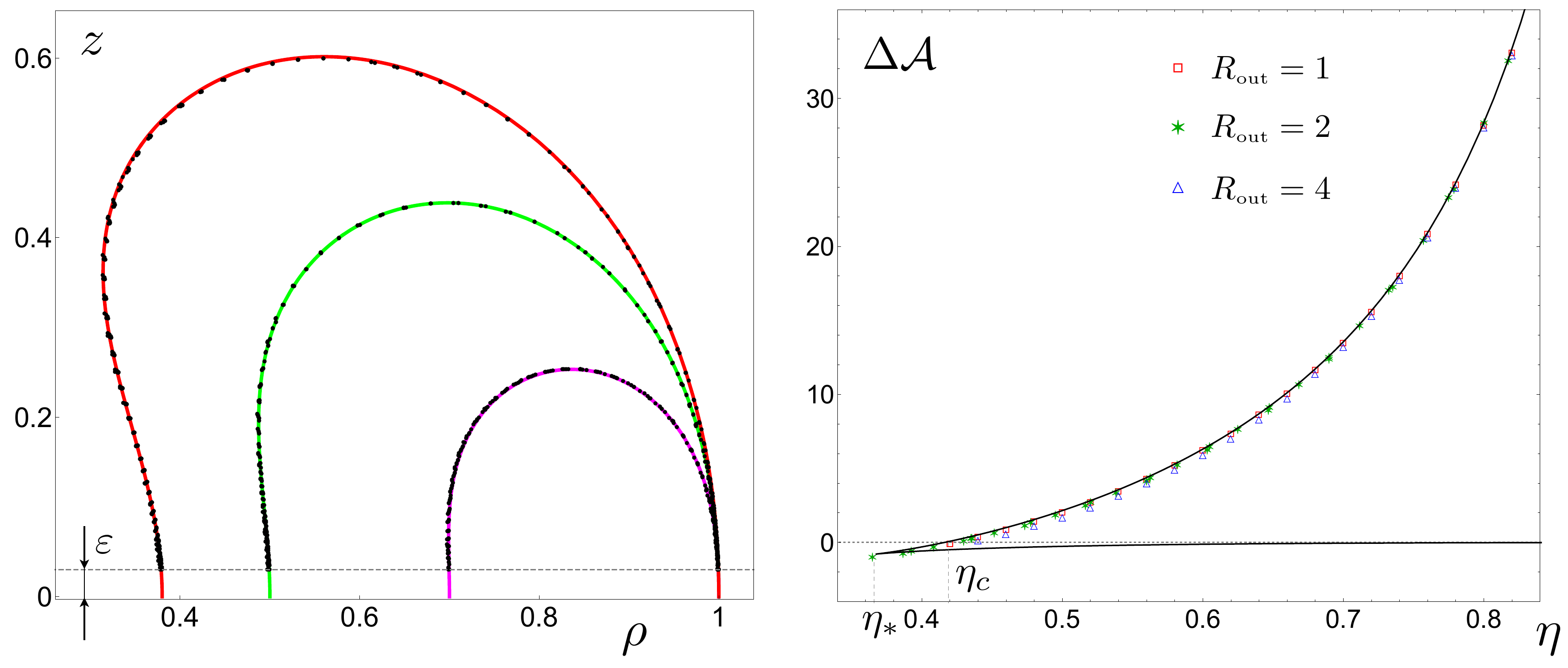}
\vspace{-.3cm}
\caption{\label{fig:2disjoint data annulus}
Left panel: 
Radial profiles of the connected surfaces anchored on the boundary of an annulus $A$ which are local minima of the area functional.
Comparison between the section of the surfaces constructed with Surface Evolver (black dots) and the analytic expressions reported in \S\ref{sec annulus}.
While the external radius is kept fixed to $R_{\textrm{\tiny out}}= 1$, for the internal one the values $R_{\textrm{\tiny in}}= 0.38$ (red), $0.5$ (green) and $0.7$ (magenta) have been chosen. 
The cutoff is $\varepsilon=0.03$ and, according to our regularization prescription, $\partial A$ has been defined at $z=\varepsilon$ in the numerical construction.
Right panel: The sign of $\Delta \mathcal{A} $ establishes the minimal area surface between the 
connected surface and the two disjoint hemispheres. 
The black curve is obtained from (\ref{Delta A d=2}) by varying $K >0$ and it is made by two branches joining at $\eta=\eta_\ast$, where the lower one corresponds to the connected solution which is not the minimal one between the two connected ones.
The data points have been found with Surface Evolver for various annular domains. 
Notice that in the left panel $\eta<\eta_c$ only for the red curve.
}
\end{figure}

The existence of the connected solution depends on the ratio $\eta \equiv R_{\textrm{\tiny in}} / R_{\textrm{\tiny out}} <1$.
As discussed in \S\ref{app annulus}, a minimal value $\eta_\ast$ can be found such that  for $0< \eta < \eta_\ast$ only the disconnected configuration of two hemispheres exists, while for $\eta_\ast < \eta <1$, besides the disconnected configuration, there are two connected configurations which are local minima of the area functional (see Fig.\,\ref{fig:appannulusprof}). 
 In the latter case, one has to find which of these two connected surfaces has the lowest area and then compare it with the area of the two disconnected hemispheres.
This comparison provides a critical value $\eta_c > \eta_\ast$ such that when $\eta\in (\eta_c,1)$ the minimal surface is given by the connected configuration, while for $\eta\in (0,\eta_c)$ the minimal area configuration is the one made by the two disjoint hemispheres.

Let us give explicit formulas about these surfaces by specifying to $D=2$ the results found in \S\ref{app annulus} (in order to simplify the notation adopted in \S\ref{app annulus}, in the following we report some formulas from that appendix omitting the index $D$).
The profile of the radial section of the connected minimal surface in the plane $(\rho,z)$ is given by the following two branches
\be
\label{branches R1R2 D=2}
\Bigg\{\begin{array}{l}
\rho = R_{\textrm{\tiny in}} \, e^{-f_{-, K}(z/\rho)}\,,
\\
\rule{0pt}{.5cm}
\rho = R_{\textrm{\tiny out}} \, e^{-f_{+, K}(z/\rho)}\,,
\end{array}
\ee
where, by introducing $\tilde{z} \equiv z/\rho$, the functions $f_{\pm, K}(\tilde{z})$ are defined as follows (from (\ref{fD def}))
\be
\label{fD=2 def}
f_{\pm, K}(\tilde{z})
\equiv
\int_0^{\tilde{z}} 
\frac{\lambda}{1+\lambda^2} 
\left(1\pm \frac{\lambda}{\sqrt{K \,(1+\lambda^2)-\lambda^4}}\right)
d\lambda\,,
\hspace{.3cm}
\qquad
0 \leqslant \tilde{z} \leqslant \tilde{z}_m\,,
\qquad
\tilde{z}_m^2 = \frac{K+\sqrt{K(K+4)}}{2}\,.
\ee
The integral occurring in $f_{\pm, K}$ can be computed in terms of the incomplete elliptic integrals of the first and third kind (see \S\ref{sec:elliptic integrals}), finding
\be
f_{\pm,K} (\tilde{z}) =
\frac{1}{2} \log (1+\tilde{z}^2)
\pm
\kappa \, 
\sqrt{\frac{1-2\kappa^2}{\kappa^2-1}}\,
\Big[\,
\mathbb{F}\big(\omega(\tilde{z}) | \kappa^2\big)
-
\Pi \big(1-\kappa^2, \omega(\tilde{z}) | \kappa^2 \big)
\Big]\,,
\ee
where we have introduced
\be
\omega(\tilde{z})
\,\equiv\, 
\arcsin \left(
\frac{\tilde{z}/\tilde{z}_m}{\sqrt{1+\kappa^2({\tilde{z}/\tilde{z}_m-1)}}}
\right), 
\qquad 
\kappa \equiv
\sqrt{\frac{1+\tilde{z}_m^2}{2+\tilde{z}_m^2}}\,.
\ee
The matching condition of the two branches (\ref{branches R1R2 D=2}) provides a relation  between $\eta \geqslant \eta_\ast$ and the constant $K$, 
namely (from (\ref{integ ratio}))
\be
\label{integ ratio D=2}
 \log(\eta)
=
-\int_0^{\tilde{z}_m} 
 \frac{2\,\lambda^2}{(1+\lambda^2)\sqrt{K (1+\lambda^2)-\lambda^4}} \, 
d\lambda 
=
2 \kappa \, 
\sqrt{\frac{1-2\kappa^2}{\kappa^2-1}}
\Big(
\mathbb{K}\big(\kappa^2\big)
-
\Pi\big(1-\kappa^2 , \kappa^2\big)
\Big)\,,
\ee
where $\mathbb{K}(m)$ and $\Pi(n,m)$ are the complete elliptic integrals of the first and third kind respectively.  

The relation (\ref{integ ratio D=2}) tells us $\eta = \eta(K)$ and $\kappa \in [1/\sqrt{2},1]$. 
As discussed in \S\ref{app annulus}, where also related figures are given, plotting this function one gets a curve whose global minimum tells us that $\eta_\ast = 0.367$.
From this curve it is straightforward to observe that, for any given $\eta \in (\eta_\ast,1)$, there are two values of $K$ fulfilling the matching condition (\ref{integ ratio D=2}). 
This means that, correspondingly, there are two connected surfaces anchored on the same pair of concentric circles on the boundary which are both local minima of the area functional. 
We have to compute their area in order to establish which one has to be compared with the configuration of disjoint hemispheres to find the global minimum.

Performing the following integral up to an additive constant (from (\ref{Rcon}) for $D=2$)
\be
\int 
\frac{d\tilde{z}}{\tilde{z}^2 \sqrt{1+\tilde{z}^2-\tilde{z}^{4}/K}}
=
\frac{
\sqrt{(\tilde{z}_m^2-\tilde{z}^2)(\tilde{z}_m^2+\tilde{z}^2 \tilde{z}_m^2+\tilde{z}^2)}
}
{\tilde{z}\, \tilde{z}_m^3} 
+
\frac{\mathbb{E}\big(\arcsin(\tilde{z}/\tilde{z}_m)|\kappa^2\big)
+(\kappa^2-1)
\mathbb{F}\big(\arcsin (\tilde{z}/\tilde{z}_m) |\kappa^2\big)}{\sqrt{2\kappa^2-1}}\,,
\ee
one obtains the area of the connected surface \cite{Drukker:1999zq, Dekel:2013kwa}
\bea
\label{area connected annulus d=2}
\mathcal{A}_{\textrm{\tiny  con}}  
&=& 
2\pi
\left(
\int_{\varepsilon/R_{\textrm{\tiny out}}}^{\tilde{z}_m}
\frac{d\tilde{z}}{\tilde{z}^2\sqrt{1+\tilde{z}^2-\tilde{z}^4/K}}
+
\int_{\varepsilon/R_{\textrm{\tiny in}}}^{\tilde{z}_m}
\frac{d\tilde{z}}{\tilde{z}^2\sqrt{1+\tilde{z}^2-\tilde{z}^4/K}}
\right)
\\
\label{area connected annulus d=2 bis}
\rule{0pt}{.7cm}
& = &
\frac{2\pi (R_{\textrm{\tiny in}}+ R_{\textrm{\tiny out}})}{\varepsilon} 
- \frac{4\pi}{\sqrt{2\kappa^2-1}}\,
\Big( \mathbb{E}\big(\kappa^2\big)-(1-\kappa^2)\, \mathbb{K}\big(\kappa^2\big) \Big)
+O(\varepsilon)\,.
\eea
Plotting the $O(1)$ term of this expression in terms of $K$, it is straightforward to realize that the minimal area surface between the two connected configurations corresponds to the smallest value of $K$.

As for the area of the configuration made by two disconnected hemispheres, from (\ref{Rsphere}) 
one gets
\be
\label{area concentric spheres d=2}
\mathcal{A}_{\textrm{\tiny  dis}}  
= 2\pi
\left(
\int^{\infty}_{\varepsilon/R_{\textrm{\tiny in}} } \frac{d\tilde{z}}{\tilde{z}^2\, \sqrt{1+\tilde{z}^2}}
+
\int^{\infty}_{\varepsilon/R_{\textrm{\tiny out}} } \frac{d\tilde{z}}{\tilde{z}^2\, \sqrt{1+\tilde{z}^2}}
\right)
= 
\frac{2\pi (R_{\textrm{\tiny in}}+ R_{\textrm{\tiny out}})}{\varepsilon} - 4\pi +O(\varepsilon)\,.
\ee
We find it convenient to introduce $\Delta \mathcal{A} \equiv \mathcal{A}_{\textrm{\tiny  dis}} - \mathcal{A}_{\textrm{\tiny  con}}$, which is finite when $\varepsilon \to 0$.
In particular, $\Delta \mathcal{A} \to  2\pi \Delta \mathcal{R}$ as $\varepsilon \to 0$, 
where $\Delta \mathcal{R}$ is (\ref{delta R finite}) evaluated at $D=2$.
From (\ref{area connected annulus d=2 bis}) and (\ref{area concentric spheres d=2}), we have
\be
\label{Delta A d=2}
\lim_{\varepsilon \to 0} \Delta \mathcal{A}  
=
4\pi
\left(
\frac{\mathbb{E}\big(\kappa^2\big)-(1-\kappa^2)\,\mathbb{K}\big(\kappa^2\big)}{\sqrt{2\kappa^2-1}}
-1
\right) .
\ee
Considering as the connected surface the one with minimal area, the sign of $\Delta \mathcal{A} $ determines the minimal surface between the disconnected configuration and the connected one and therefore the global minimum of the area functional.
The root $\eta_c$ of $\Delta \mathcal{A} $ can be found numerically and one gets $\eta_c=0.419$
\cite{Olesen:2000ji, Drukker:2005cu}. 
Thus, the connected configuration is minimal for $\eta\in (\eta_c, 1)$, while for $\eta\in (0,\eta_c)$ the minimal area configuration is the one made by the disjoint hemispheres. 

By employing Surface Evolver, we can construct the surface anchored on the boundary of the annulus at $z=0$ which is a local minimum, compute its area and compare it with the analytic results discussed above. This is another important benchmark of our numerical method. 

In the left panel of Fig.\,\ref{fig:2disjoint data annulus} we consider the profile of the connected configuration in the plane $(\rho,z)$. The black dots correspond to the radial section of the  surface obtained with Surface Evolver, while the solid line is obtained from the analytic expressions discussed above.
Let us recall that the triangulated surface is numerically constructed by requiring that it is anchored to the two concentric circles with radii $R_{\textrm{\tiny in}} < R_{\textrm{\tiny out}}$ at $z=\varepsilon$ and not at $z=0$, as it should. 
Despite this regularization, the agreement between the analytic results and the numerical ones is very good for our choices of the parameters. 
It is worth remarking  that, when $\eta \geqslant \eta_\ast$ and therefore two connected solutions exist for a given $\eta$, Surface Evolver finds the minimal area one between
them.
Nevertheless, it is not able to establish whether it is the global minimum.
Indeed, for example, the red curve in the left panel of Fig.\,\ref{fig:2disjoint data annulus} has $\eta_\ast < \eta < 1$ and therefore the corresponding surface is minimal but it is not the global minimum. 
Instead, considering an annulus with $\eta< \eta_\ast$, even if one begins with a rough triangulation of a connected surface, Surface Evolver converges towards the  configuration made by the two disconnected hemispheres.

In the right panel of Fig.\,\ref{fig:2disjoint data annulus} we compare the values of $\Delta \mathcal{A}$ obtained with Surface Evolver with the analytic curve from (\ref{Delta A d=2}), finding a very good agreement. 
Numerical points having $\eta_\ast < \eta < \eta_c$ are also found, for the reason just explained.

\subsubsection{Two disjoint disks}
\label{sec 2 disjoint circles}

In this section we consider domains $A$ made by two disjoint disks  by employing the analytic results for the annulus reviewed in \S\ref{sec annulus} and some isometries of $\mathbb{H}_3$.
This method has been used in \cite{Berenstein:1998ij} for the case of a circle, while the case of two disjoint circles has been recently studied in \cite{Krtous:2013vha, Krtous:2014pva}.
The analytic results found in this way provide another important benchmark for the numerical data obtained with Surface Evolver.

Let us consider the following reparameterizations of $\mathbb{H}_3$, which correspond to the special conformal transformations on the boundary \cite{Berenstein:1998ij}
\be
\label{ads rep}
\tilde{x} =
\frac{x+b_x (|\boldsymbol{v}|^2 +z^2)}{
1+2\boldsymbol{b} \cdot \boldsymbol{v}
+ |\boldsymbol{b}|^2 (|\boldsymbol{v}|^2 +z^2)}\,,
\qquad
\tilde{y} =
\frac{y+b_y (|\boldsymbol{v}|^2 +z^2)}{
1+2\boldsymbol{b} \cdot \boldsymbol{v}
+ |\boldsymbol{b}|^2 (|\boldsymbol{v}|^2 +z^2)}\,,
\qquad
\tilde{z} =
\frac{z}{
1+2\boldsymbol{b} \cdot \boldsymbol{v}
+ |\boldsymbol{b}|^2 (|\boldsymbol{v}|^2 +z^2)}\,,
\ee
being $\boldsymbol{b} \equiv (b_x, b_y)$ a vector in $\mathbb{R}^2$ and $\boldsymbol{v} \equiv (x, y)$.

\begin{figure}[t] 
\vspace{-.5cm}
\hspace{-.0cm}
\begin{center}
\includegraphics[width=.98\textwidth]{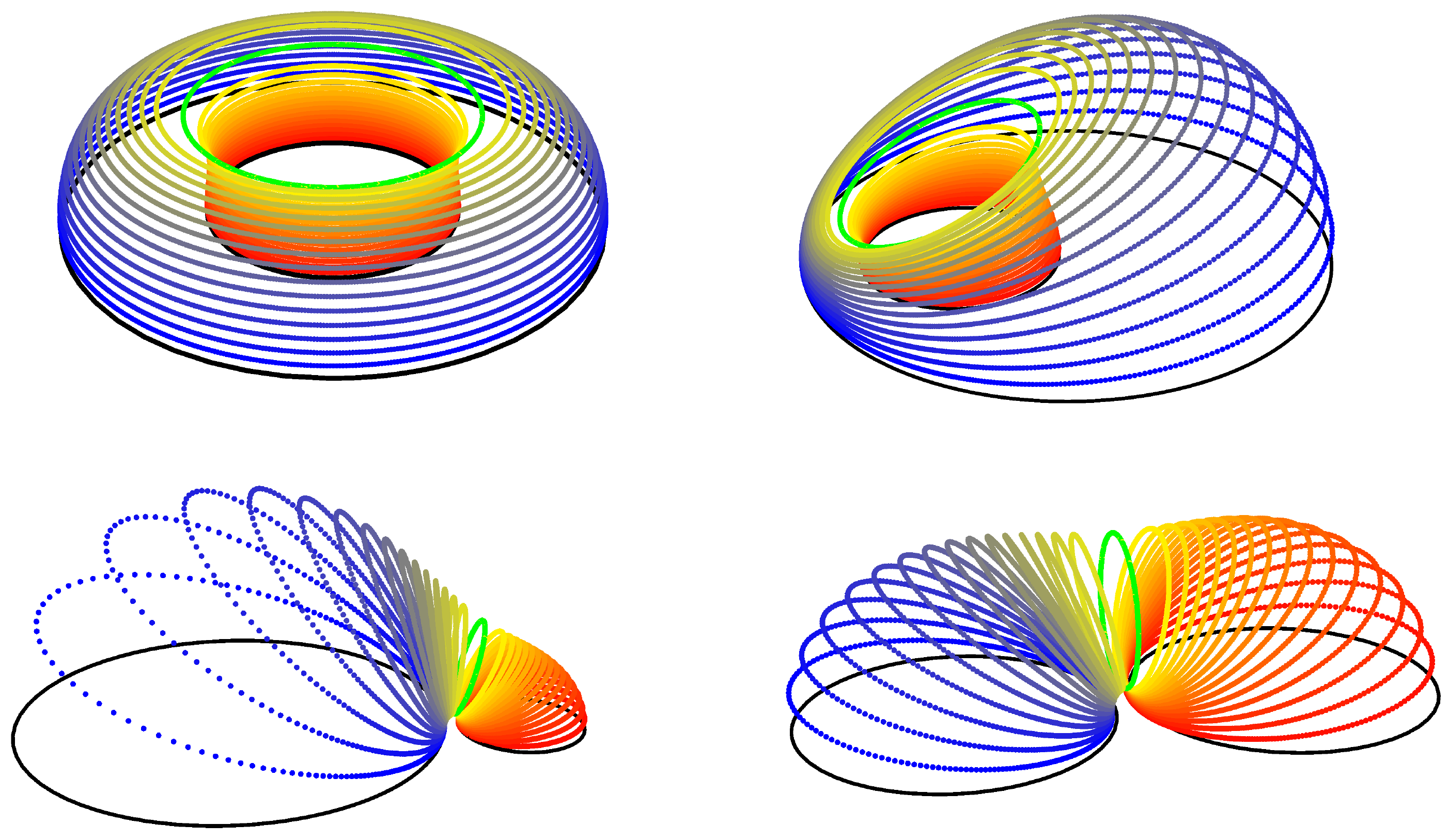}
\end{center}
\vspace{.1cm}
\caption{\label{fig:annulus to bridge}
The connected surface anchored on the boundary of an annulus at $z=0$ (top left panel), which is a local minimum of the area functional, can be mapped through (\ref{ads rep}) into one of the connected surfaces anchored on the configurations of circles at $z=0$ shown in the remaining panels, depending on the value of the parameter of the transformation (\ref{ads rep}), as discussed in \S\ref{sec 2 disjoint circles}. 
The mapping preserves the color code.  
The green circle in the top left panel corresponds to the matching of the two branches given by (\ref{branches R1R2 D=2}) and (\ref{integ ratio D=2}) (see the point $P_m$ in Fig.\,\ref{fig:appannulusprof}) and it is mapped into the vertical circle in the bottom right panel. 
}
\end{figure}

\begin{figure}[t] 
\vspace{-.6cm}
\hspace{-.0cm}
\begin{center}
\includegraphics[width=.7\textwidth]{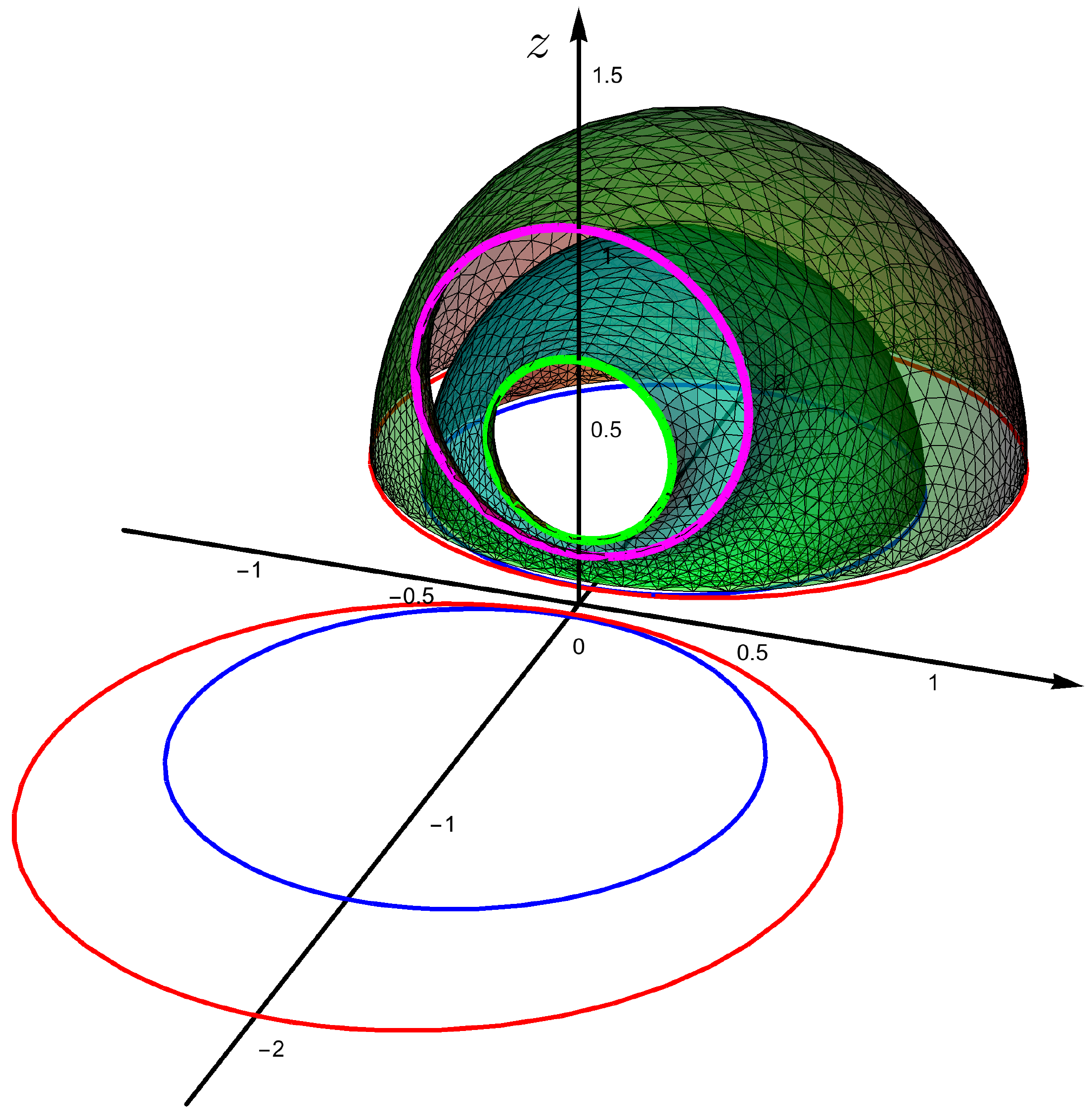}
\end{center}
\vspace{-.2cm}
\caption{\label{fig:2circles section}
Two examples of minimal surfaces (constructed with Surface Evolver) corresponding to $A$ made by two disjoint and equal disks ($\partial A$ is given by the red and blue circles). Only half of the surfaces is shown in order to highlight their section through a plane orthogonal to $z=0$ and to the segment connecting the centers. 
This section provides a circle whose radius and center are given in (\ref{vertical circle}). In this figure $\varepsilon =0.03$, the red circles have radius $R=1$ and the distance between their centers is $d=2.16$, while for the blue ones $R=0.75$ and $d=1.68$.
}
\end{figure}

When $z=0$ in (\ref{ads rep}), the maps $(x,y) \to (\tilde{x} , \tilde{y} )$ are the special conformal transformations of the Euclidean conformal group in two dimensions.
These transformations in the $z=0$ plane send a circle $\mathcal{C}$ with center $\boldsymbol{c} =(c_x, c_y) $ and radius $R$ into another circle $\widetilde{\mathcal{C}}$ with center $\tilde{\boldsymbol{c}}=(\tilde{c}_x, \tilde{c}_y) $ and radius $\widetilde{R}$ which are given by 
\be
\label{sct circ}
\tilde{c}_i = 
\frac{c_i +b_i (|\boldsymbol{c}|^2 -R^2)}{
1+2\boldsymbol{b} \cdot \boldsymbol{c}
+ |\boldsymbol{b}|^2 (|\boldsymbol{c}|^2 -R^2)}
\hspace{.5cm} i \in \{x,y\}\,,
\hspace{1cm} 
\widetilde{R} = 
\frac{R}{
\big|1+2\boldsymbol{b} \cdot \boldsymbol{c}
+ |\boldsymbol{b}|^2 (|\boldsymbol{c}|^2 -R^2)\big|}\,.
\ee
Notice that the center $\tilde{\boldsymbol{c}}$ is not the image of the center $\boldsymbol{c}$ under (\ref{ads rep}) with $z=0$.
Moreover, when $\boldsymbol{c}$ is such that the denominator in (\ref{sct circ}) vanishes, the circle is mapped into a straight line \cite{Berenstein:1998ij}.

Considering two concentric circles at $z=0$ with radii $R_{\textrm{\tiny in}} < R_{\textrm{\tiny out}} $,
their images are two different circles at $z=0$ which do not intersect.  
In order to deal with simpler expressions for the mapping, let us place the center of the concentric circles in the origin, i.e. $\boldsymbol{c}=(0,0)$.
By introducing $\eta \equiv R_{\textrm{\tiny in}}/ R_{\textrm{\tiny out}} <1$ for the initial configuration of concentric circles centered in the origin and
denoting by $\widetilde{R}_1 \equiv R_{\textrm{\tiny in}}/|1-|\boldsymbol{b}|^2 R_{\textrm{\tiny in}}^2|$ and $\widetilde{R}_2 \equiv R_{\textrm{\tiny out}}/|1-|\boldsymbol{b}|^2 R_{\textrm{\tiny out}}^2|$ the radii of the circles after the mapping,
 the distance between the two centers reads
\be
\label{distance centers after map}
d
=
\frac{(1-\eta^2) \beta}{ | (1-\beta^2)(\beta^2- \eta^2)|}\, R_{\textrm{\tiny in}}
=
\frac{(1-\eta^2) \beta}{ | \beta^2- \eta^2|}\, \widetilde{R}_1
\,,
\ee
where $\beta^2 \equiv |\boldsymbol{b}|^2  R_{\textrm{\tiny in}}^2$.
Thus, $\eta$ and $\beta$ fix the value of the ratio $\tilde{\delta}\equiv d/\widetilde{R}_1$.
The final disks are either disjoint or fully overlapping, depending on the sign of the expression within the absolute value in the denominator of (\ref{distance centers after map}).
In particular, when $\beta^2 \in (\eta^2,1) $ the two disks are disjoint, while when 
$\beta^2 \in (0,\eta^2) \cup  (1, +\infty ) $ they overlap.
As for their ratio $\tilde{\eta}  \equiv \widetilde{R}_1/  \widetilde{R}_2$, we find
\be
\label{eta tilde ris}
\tilde{\eta} = 
\left\{\begin{array}{lll}
\displaystyle     \frac{\beta^2 - \eta^2}{\eta (\beta^2-1)}
\hspace{.5cm} &
\beta^2 \in (0,\eta^2) \cup  (1, \infty ) 
\hspace{.5cm} &
\textrm{overlapping disks\,,}
\\
\rule{0pt}{.8cm}
\displaystyle      \frac{\beta^2 - \eta^2}{\eta (1-\beta^2)}
&
\beta^2 \in (\eta^2,1) 
&
\textrm{disjoint disks\,.}
\end{array}\right.
\ee
Notice that  $\tilde{\eta} \to 1/\eta>1$ for $\beta^2\to \infty$.
Thus, given $\eta$ and $\beta$, the equations (\ref{distance centers after map}) and (\ref{eta tilde ris}) provide $\tilde{\delta}$ and $\tilde{\eta}$.
By inverting them, one can write  $\eta$ and $\beta$ in terms of $\tilde{\delta}$ and $\tilde{\eta}$.
The system is made by two quadratic equations and some care is required to distinguish the various regimes.

When the disks after the mapping are disjoint, i.e. $\eta^2 < \beta^2 <1$, an interesting special case to discuss is $\widetilde{R}_1 = \widetilde{R}_2$, namely when the disjoint disks have the same radius $\widetilde{R} = R_{\textrm{\tiny in}} /(1-\eta)= R_{\textrm{\tiny out}} /(\eta^{-1}-1 ) $, being $R_{\textrm{\tiny in}} < R_{\textrm{\tiny out}} $ the radii of the two concentric circles at $z=0$ centered in the origin.
Setting $\tilde{\eta} =1$ in (\ref{eta tilde ris}), one finds that it happens for $\beta^2=\eta$, i.e. $|\boldsymbol{b}|^2=1/(R_{\textrm{\tiny in}} R_{\textrm{\tiny out}} )$. 
The distance corresponding to this value of $\beta$ can be found  from (\ref{distance centers after map}) and it is given by $d/R_{\textrm{\tiny in}} = (1+\eta)/\big[\sqrt{\eta} (1-\eta)\big]$ or, equivalently,  by $\tilde{\delta} = (1+\eta)/\sqrt{\eta}$.
By inverting this relation, one finds $\eta(\tilde{\delta}) = \big\{ \tilde{\delta}^2-2-\big[(\tilde{\delta}^2-2)^2 -4\big]^{1/2}\big\}/2$, where the root $\eta(\tilde{\delta}) < 1$ has been selected and $\tilde{\delta} >2$ must be imposed in order to avoid the intersection of the two equal disks.

\begin{figure}[t] 
\vspace{-.2cm}
\hspace{-.5cm}
\includegraphics[width=1.035\textwidth]{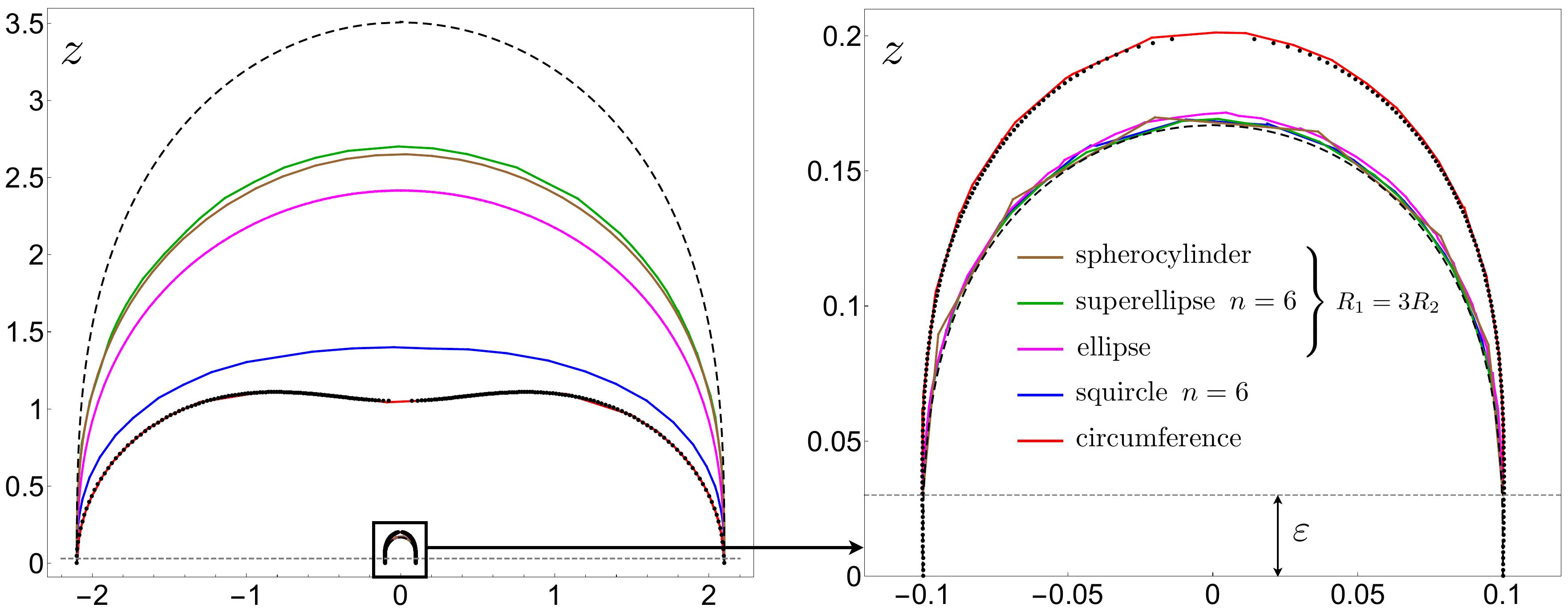}
\vspace{-.2cm}
\caption{\label{fig:2disjoint data sections}
Left: Sections of minimal surfaces when $A$ is made by two equal disjoint domains with smooth boundaries, like the red curves in Fig.\,\ref{fig:2disjoint example}.
The coloured solid lines are the numerical results  found with Surface Evolver for the shapes indicated in the common legend in the right panel. Here $R_2=1$ and $\varepsilon=0.03$.
The black dots (notice that they reach $z=0$) correspond to the minimal surface for two disjoint circles and they have been found by mapping the connected minimal surface for the annulus  through the transformations (\ref{ads rep}) (see \S\ref{sec 2 disjoint circles} and Fig.\,\ref{fig:annulus to bridge}).
The dashed curve corresponds to two infinite strips.
Right: Zoom of the part of the left panel enclosed by the black rectangle.
}
\end{figure}

Once the vector $\boldsymbol{b} = (b_x, b_y) = |\boldsymbol{b}| (\cos \phi_b, \sin \phi_b)$ is chosen by fixing the initial and final configurations of circles at $z=0$, the transformations (\ref{ads rep}) for the points in the bulk are fixed as well and they can be used to map the points belonging to the minimal surfaces spanning the initial configuration of circles. 
In particular, let us consider a circle given by $(R_\star \cos\phi,R_\star \sin\phi, z_\star)$ for $\phi \in [0,2\pi)$, lying in a plane at $z=z_\star$ parallel to the boundary. 
This circle is mapped through (\ref{ads rep}) into another circle $\widehat{\mathcal{C}}$ whose radius is given by
\be
\label{Rhat def}
\widehat{R} = 
\frac{R_\star}{\sqrt{1+2 |\boldsymbol{b}|^2 (z_\star^2-R_\star^2)+|\boldsymbol{b}|^4 (z_\star^2+R_\star^2)^2}}\,,
\ee
and whose center $\hat{\boldsymbol{c}} \equiv (\hat{c}_x , \hat{c}_y, \hat{c}_z)$ has coordinates
\be
\label{hat c def}
\hat{c}_i = 
\frac{|\boldsymbol{b}|^2 (R_\star^2+z_\star^2)^2+ z_\star^2-R_\star^2 
}{1+2 |\boldsymbol{b}|^2 (z_\star^2-R_\star^2)+|\boldsymbol{b}|^4 (z_\star^2+R_\star^2)^2} \, b_i 
\hspace{.5cm} i \in \{x,y\}\,,
\qquad
\hat{c}_z = 
\frac{[1+|\boldsymbol{b}|^2 (R_\star^2+z_\star^2)]\, z_\star}{1+2 |\boldsymbol{b}|^2 (z_\star^2-R_\star^2)+|\boldsymbol{b}|^4 (z_\star^2+R_\star^2)^2} \;.
\ee
Setting $z_\star =0$, $R_\star =  R$ and $\widehat{R} = \widetilde{R}$ in (\ref{Rhat def}) and (\ref{hat c def}), the expressions in (\ref{sct circ}) with $\boldsymbol{c}=(0,0)$ are recovered.
The circle $\widehat{\mathcal{C}}$ lies in a plane orthogonal to the following unit vector
\be
\boldsymbol{v}_\perp = 
(-\cos\phi_b \sin \theta_\perp,-\sin \phi_b \sin \theta_\perp, \,\cos \theta_\perp)\,, 
\qquad
\theta_\perp \equiv \arcsin (2 z_\star  | \boldsymbol{b} | \widehat{R}/R_\star )\,,
\ee
where $2 z_\star  | \boldsymbol{b} | \widehat{R}/R_\star <1$, as can be easily observed from (\ref{Rhat def}).

In the top left panel of Fig.\,\ref{fig:annulus to bridge} we consider as initial configuration the annulus at $z=0$ for some given value of $\eta$ and the corresponding connected minimal surface in the bulk anchored on its boundary, which has been discussed in \S\ref{sec annulus}. The transformation (\ref{ads rep}) with $\beta = \sqrt{\eta}$ maps this surface into the connected surface anchored on two equal and disjoint circles (bottom right panel in Fig.\,\ref{fig:annulus to bridge}). 
It is interesting to follow the evolution of the former surface into the latter one as $\beta \in [0, \sqrt{\eta}]$ increases: in Fig.\,\ref{fig:annulus to bridge} we show two intermediate steps where the surfaces are qualitatively different and they correspond to different regimes of $\beta$ separated by $\beta =\eta$. 
For $0<\beta < \eta$ the disks at $z=0$ are still overlapping but they are not concentric (top right panel of Fig.\,\ref{fig:annulus to bridge}). 
Within this range of $\beta$, the radius of the largest disk, which is $R_{\textrm{\tiny out}}/|1-\beta^2/\eta^2|$, increases with $\beta$ and it diverges when as $\beta \to \eta$. 
When $\eta<\beta \leqslant \sqrt{\eta}$, instead, the disks at $z=0$ are disjoint and the images of the initial surface through (\ref{ads rep}) are shown in the bottom panels of Fig.\,\ref{fig:annulus to bridge}, where the surface on the left has $\eta<\beta <\sqrt{\eta}$, while the one on the right corresponds to the final stage of disjoint equal disks ($\beta = \sqrt{\eta}$).
In Fig.\,\ref{fig:annulus to bridge} the mapping preserves the color code and we have highlighted the green circle because in the top left panel it corresponds to the circle at $z=z_m$ along which the two branches given by (\ref{branches R1R2 D=2}) match, as imposed by the condition (\ref{integ ratio D=2}). When $\beta = \sqrt{\eta}$, this matching circle is mapped into the vertical one shown in the bottom right panel, 
whose radius $\widetilde{R}_v$ and whose coordinate $z_v > \widetilde{R}_v$ of its center along the holographic direction are given respectively by
\be
\label{vertical circle}
\widetilde{R}_v = \frac{1-\eta}{2\tilde{z}_m\sqrt{\eta}}\, \widetilde{R}\,,
\qquad
z_v= \frac{(1-\eta) \sqrt{1+\tilde{z}_m^2}}{2\tilde{z}_m\sqrt{\eta}}\, \widetilde{R} \,,
\ee
where $ \widetilde{R}$ is the radius of the two equal disjoint disks written above and $\tilde{z}_m$ is a function of $\eta$ (see (\ref{fD=2 def}) and (\ref{integ ratio D=2})).
In Fig.\,\ref{fig:2circles section} we show two examples of minimal surfaces constructed with Surface Evolver which provide the holographic mutual information of two equal disjoint disks. 
Considering the section of these surfaces through a vertical plane which is orthogonal to the boundary and to the line passing through the centers of the disks, we find a good agreement with (\ref{vertical circle}).

As for the finite part of the area, once $\eta$ and $\beta$ have been written in terms of $\tilde{\eta}$ and $\tilde{\delta}$ by inverting (\ref{distance centers after map}) and (\ref{eta tilde ris}), the limit $\varepsilon \to 0$ of either $\Delta \mathcal{A}  $ or $\mathcal{I}_{A_1, A_2}$ (depending on whether the final disks are either overlapping or disjoint respectively) is given by the r.h.s. of (\ref{Delta A d=2}), where $\kappa = \kappa(\eta)$ is obtained through the numerical inversion of (\ref{integ ratio D=2}), being $\eta = \eta(\tilde{\delta},\tilde{\eta})$ found above.

The special case of two equal disjoint disks corresponds to $\tilde{\eta} =1$ and $\tilde{\delta} = (1+\eta)/\sqrt{\eta}$, and therefore the limit $\varepsilon \to 0$ of $\mathcal{I}_{A_1, A_2}$ depends only on the parameter $\tilde{\delta}$, as expected. The relation $\tilde{\delta} = (1+\eta)/\sqrt{\eta}$ can be used to find the critical distance $d_c$ between the centers beyond which the holographic mutual information vanishes and also the distance $d_\ast > d_c$ beyond which the connected surface does not exist anymore. They correspond to $\eta_c$ and $\eta_\ast$ respectively and, in particular, one gets $\tilde{\delta}_c = 2.192$ and $\tilde{\delta}_\ast= 2.256$.

\begin{figure}[t!] 
\vspace{-.5cm}
\hspace{.3cm}
\includegraphics[width=.92\textwidth]{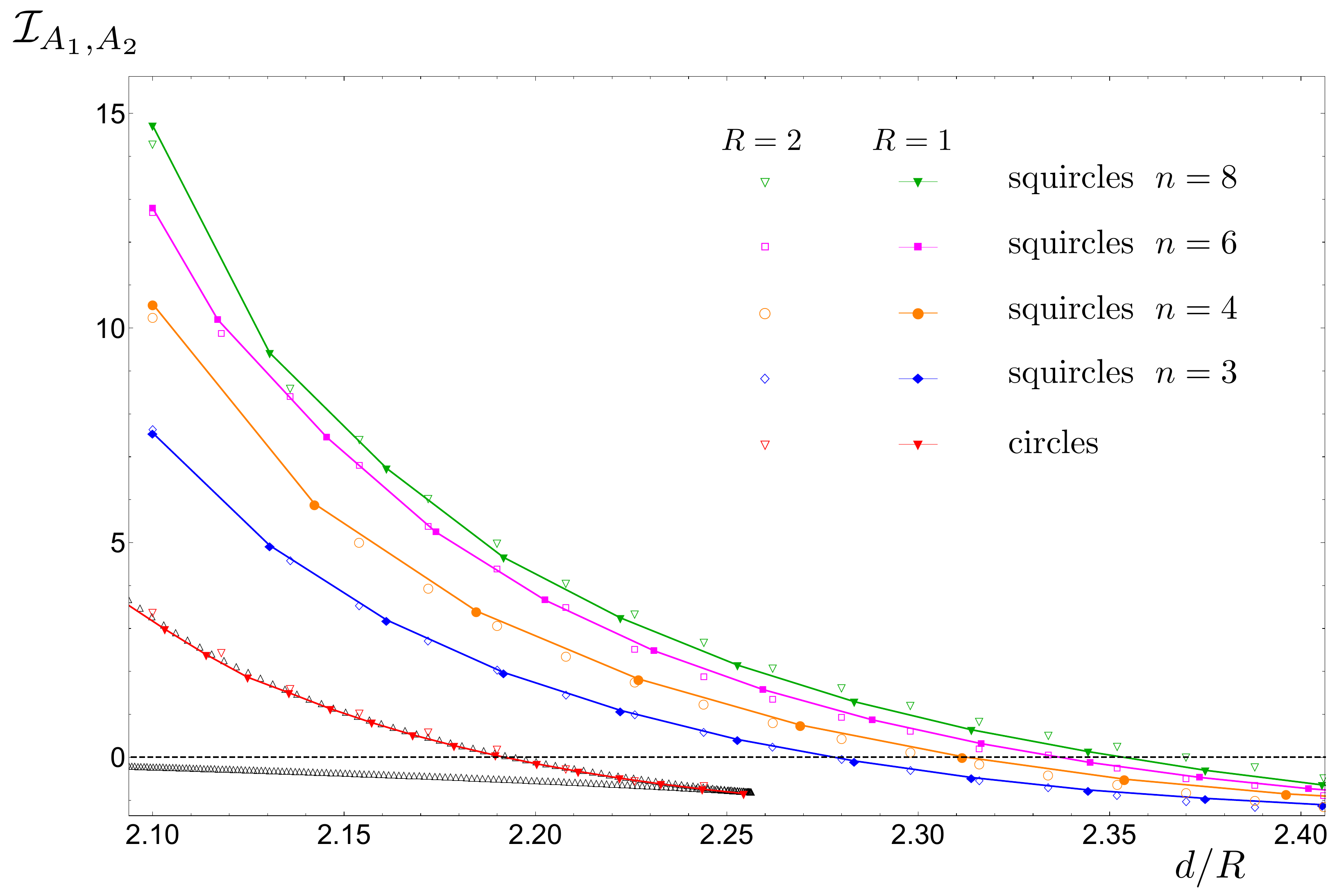}
\vspace{.0cm}
\caption{\label{fig:transquircles}
Holographic mutual information of two disjoint and equal domains delimited by squircles for various $n$.
The coloured points are the numerical data obtained with Surface Evolver, while
the black triangles correspond to the solid black curve of Fig.\,\ref{fig:2disjoint data annulus} (right panel) mapped through the transformation (\ref{eta tilde ris}) with $\beta^2=\eta$. The transition between the connected surface and the configuration of disconnected surfaces occurs at the zero of each curve. A point having $\mathcal{I}_{A_1, A_2} <0$ corresponds to a connected surface which is a local minimum of the area functional but it is not the global minimum for the corresponding entangling curve.
}
\end{figure}

In order to check that the surfaces obtained through (\ref{ads rep})  are local minima of the area functional, one can compare the analytic results found as explained above against the corresponding surfaces constructed by Surface Evolver. 
In Fig.\,\ref{fig:2disjoint data sections} we have performed this check for a section profile: the black dots come from the surface obtained as in the bottom right panel of Fig.\,\ref{fig:annulus to bridge} (notice that the black dots reach $z=0$), while the red curve is the section of the corresponding surface constructed by Surface Evolver (see also the red curves in Fig.\,\ref{fig:2disjoint example} for a similar construction with different $A$).
In Fig.\,\ref{fig:transquircles} we have performed another comparison between the analytic expressions and the numerical data of Surface Evolver by computing the holographic mutual information of a domain $A$ made by two equal disjoint disks.
The black triangles have been found by mapping the black curve for the annulus in the right panel of Fig.\,\ref{fig:2disjoint data annulus} (which is given by the r.h.s. of (\ref{Delta A d=2})) through $\eta=\eta(\tilde{\delta})$ found above.
The agreement with the corresponding data obtained with Surface Evolver (red curve) is very good. 
Notice that, as already observed for the annulus in \S\ref{sec annulus}, also in this case Surface Evolver finds a surface which is a local minimum of the area functional, even if it is not the global minimum. 
Let us conclude by emphasizing that, while this numerical method is very efficient in finding surfaces which are local minima for the area functional when they exist, it is not suitable for studying the existence of a surface with a given topology.

\begin{figure}[t!] 
\vspace{-.2cm}
\hspace{-1cm}
\includegraphics[width=1.05\textwidth]{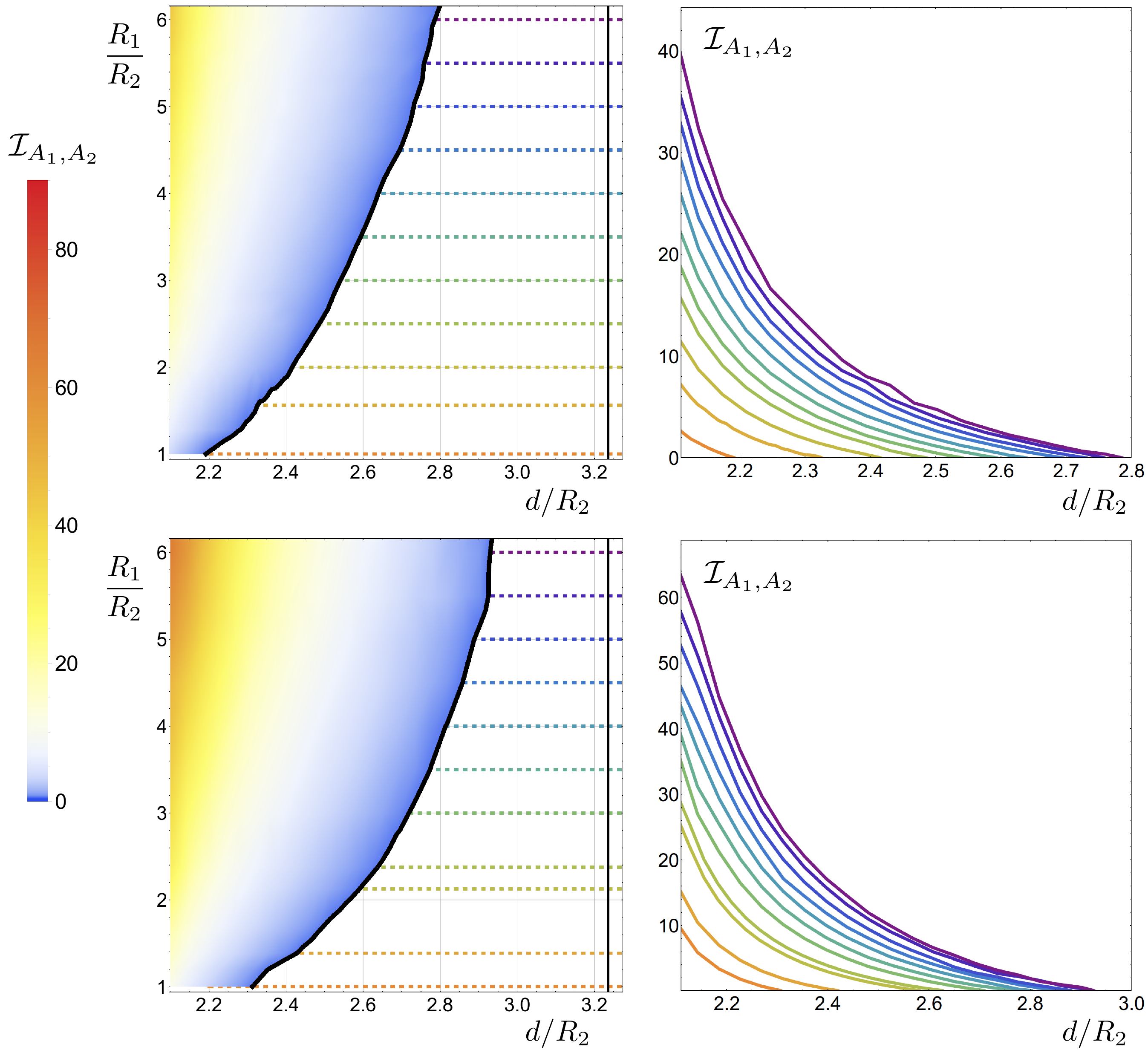}
\vspace{.2cm}
\caption{\label{fig:densitysuper}
Holographic mutual information of two equal and disjoint domains delimited by ellipses (top panels) or superellipses with $n=4$ (bottom panels), which are defined by $R_1$ and $R_2$ (see the bottom  panel of Fig.\,\ref{fig:squircle min surfs} and (\ref{eq superellipse})), while $d$ is the distance between their centers. 
The relative orientation is like in Fig.\,\ref{fig:2disjoint example}.
Left panels: Density plots for $\mathcal{I}_{A_1, A_2}$ whose zero provides the corresponding transition curve (solid black line) in the plane $(d/R_2, R_1/R_2)$. 
The straight vertical line indicates the transition when $A$ is made by two equal and disjoint infinite strips whose width is $2R_2$ and the distance between their central lines is $d$. 
Right panels:  $\mathcal{I}_{A_1, A_2}$ in terms of $d/R_2$ for various fixed values of $R_1/R_2$ indicated by the horizontal dashed lines in the corresponding left panel, with the same color code.
The lower curves (orange) in the right panels correspond to the squircles ($R_1=R_2$) with $n=2$ (top) and $n=4$ (bottom) and therefore they reproduce the red and orange curves in Fig.\,\ref{fig:transquircles} respectively.
The data reported here have been found with $R_2=1$ and some checks have been done also with $R_2=2$.
}
\end{figure}

\begin{figure}[t!] 
\vspace{-.2cm}
\hspace{-.1cm}
\includegraphics[width=.97\textwidth]{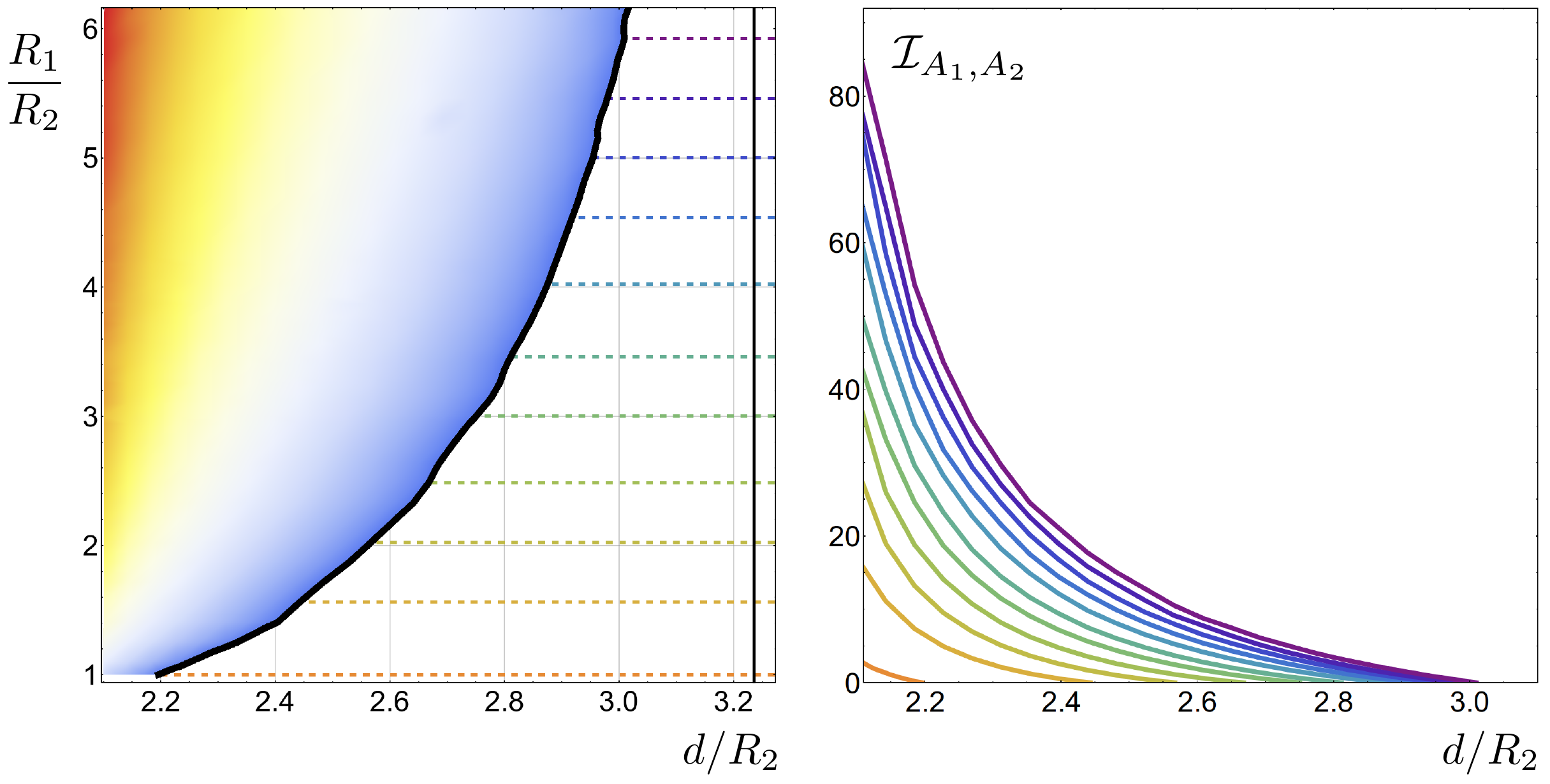}
\vspace{.1cm}
\caption{\label{fig:densitysphero}
Holographic mutual information of two equal and disjoint two dimensional spherocylinders oriented like the two ellipses in Fig.\,\ref{fig:2disjoint example}. The parameters $R_1$ and $R_2$ specify the domains (see the bottom  panel of Fig.\,\ref{fig:squircle min surfs} and (\ref{eq spherocylinder parts})) and $d$ is the distance between their centers.
The same notation and color coding of Fig.\,\ref{fig:densitysuper} has been adopted. 
}
\end{figure}

%%%%%%%%%%%%%%%%%%%%%%%%%%%%%%%%%%%%%%%%

\subsection{Other shapes}
\label{sec 2 disjoint shapes}

In \S\ref{sec 2 disjoint circles} we have considered the holographic mutual information of two disjoint circular domains, for which analytic results are available.
When $A= A_1 \cup A_2$ is not made by two disjoint disks, analytic results for the corresponding holographic mutual information are not known and therefore a numerical approach could be very useful.
Here we employ Surface Evolver to study $\mathcal{I}_{A_1, A_2} $ (defined in (\ref{MI holog})) of disjoint regions delimited by some of the smooth curves introduced in \S\ref{sec superellipse}.

The holographic mutual information of non circular domains depends on the geometries of their boundaries, on their distance and also on their relative orientation.
Independently of the shapes of $\partial A_1$ and $\partial A_2$, once the domains and their relative orientation have been fixed, the holographic mutual information vanishes when the distance between $A_1$ and $A_2$ is large enough. The critical distance $d_c$ beyond which $\mathcal{I}_{A_1, A_2} = 0$ depends on the configuration of the domains. This transition occurs because, for a generic distance $d$ between the centers of $A_1$ and $A_2$, the global minimal area surface comes from a competition between a connected surface anchored on $\partial A$ and a configuration made by two disconnected surfaces spanning $\partial A_1$ and $\partial A_2$, which are both local minima. Beyond the critical distance between the centers, the disconnected configuration becomes the global minimum and therefore $\mathcal{I}_{A_1, A_2}$ vanishes.

In Fig.\,\ref{fig:2disjoint example} we show an example of a connected surface constructed with Surface Evolver where $\partial A$ is made by two equal and disjoint ellipses at $z=0$. Let us recall that in our numerical analysis we have regularized the area by defining $\partial A$ at $z=\varepsilon$, as discussed in \S\ref{app technical details}.
In the figure, we have highlighted two sections of the surface suggested by the symmetry of this configuration of domains, which are given by the red curves and by the green one. 

We have constructed minimal area connected surfaces also for configurations of equal disjoint domains with other shapes and in Fig.\,\ref{fig:2disjoint data sections} we have reported the corresponding curves obtained from the section giving the red curves in Fig.\,\ref{fig:2disjoint example}.
The red curves in Fig.\,\ref{fig:2disjoint data sections} are associated with circular domains and they can be recovered analytically (black dots), as explained in \S\ref{sec 2 disjoint circles}. Instead, for the remaining curves analytic expressions are not available and therefore they provide a useful benchmark for analytic results that could be found in the future. 

Besides the profiles for various sections, Surface Evolver computes also the area of the surfaces that it constructs. 
Considering a configuration of disjoint domains with given shapes and relative orientation, we can compute $\mathcal{I}_{A_1, A_2}$ while the distance $d$ between their centers changes. In Fig.\,\ref{fig:transquircles} we show the results of this analysis when $\partial A_1$ and $\partial A_2$ are squircles (i.e. (\ref{eq superellipse}) with $R_1=R_2\equiv R$). 
As for their relative orientation, drawing the squares that circumscribe $\partial A_1$ and $\partial A_2$, their edges are parallel.
Since $\mathcal{I}_{A_1, A_2} \geqslant 0$, the critical distance $d_c$ corresponds to the zero of the various curves and $\mathcal{I}_{A_1, A_2}$ vanishes for $d \geqslant d_c$. Thus, $\mathcal{I}_{A_1, A_2}$ is continuos with a discontinuous first derivative at $d=d_c$. 
The points found numerically which have $\mathcal{I}_{A_1, A_2}<0$ correspond to connected surfaces that  Surface Evolver constructs but they are not the global minimum for the area functional because the disconnected configuration is favoured for that distance.

Once the relative orientation has been chosen, a configuration of two equal and disjoint squircles is completely determined by two parameters: the distance $d$ between the centers and the size $R$ of the squircles.
Instead, when $A_1$ and $A_2$ are two equal two dimensional spherocylinders or equal domains delimited by two disjoint superellipses and the relative orientation has been chosen, we have three parameters to play with: 
the distance $d$ between the centers and the parameters $R_1$ and $R_2$ which specify the two equal domains (see the bottom panel of Fig.\,\ref{fig:squircle min surfs}).
In Fig.\,\ref{fig:densitysuper} we show $\mathcal{I}_{A_1, A_2}$ for two disjoint domains delimited by ellipses and superellipses with $n=4$, whose relative orientation is like in Fig.\,\ref{fig:2disjoint example}. In the left panels,
the black thick curve is the transition curve along which the holographic mutual information vanishes, while the continuos straight line identifies the transition value corresponding to two disjoint infinite strips \cite{Tonni:2010pv}.
Comparing the transition curve in the top left panel with the one in the bottom left panel, it is evident that the one associated with the superellipses having $n=4$ is closer to the value corresponding to the infinite strips than the one associated with the ellipses. 
In Fig.\,\ref{fig:densitysphero} we study $\mathcal{I}_{A_1, A_2}$ for a domain $A$ made by two equal and disjoint two dimensional spherocylinders.
In this case the transition curve is closer to the line corresponding to the transition for two infinite strips with respect to the transition curves of Fig.\,\ref{fig:densitysuper}. 
Nevertheless, from our data we cannot conclude that the transition curve for the two dimensional spherocylinders approaches the value corresponding to the infinite strips as $R_1/R_2 \to \infty$.
It would be interesting to have further data and some analytic argument to understand whether some bounds prevent the transition curves to approach the value associated with the infinite strips for $R_1/R_2 \to \infty$.
Let us remark that the lowest curves (orange) in the right panels of Figs.\,\ref{fig:densitysuper} and \ref{fig:densitysphero} 
correspond to disjoint squircles with $n=2$ (i.e. circles) or $n=4$ and therefore they reproduce the red and the orange curves of Fig.\,\ref{fig:transquircles}.
Configurations of domains having smaller values of $d$ than the ones shown in the plots provide unstable numerical results. 

\begin{figure}[t] 
\vspace{-.2cm}
\hspace{-.25cm}
\includegraphics[width=1.02\textwidth]{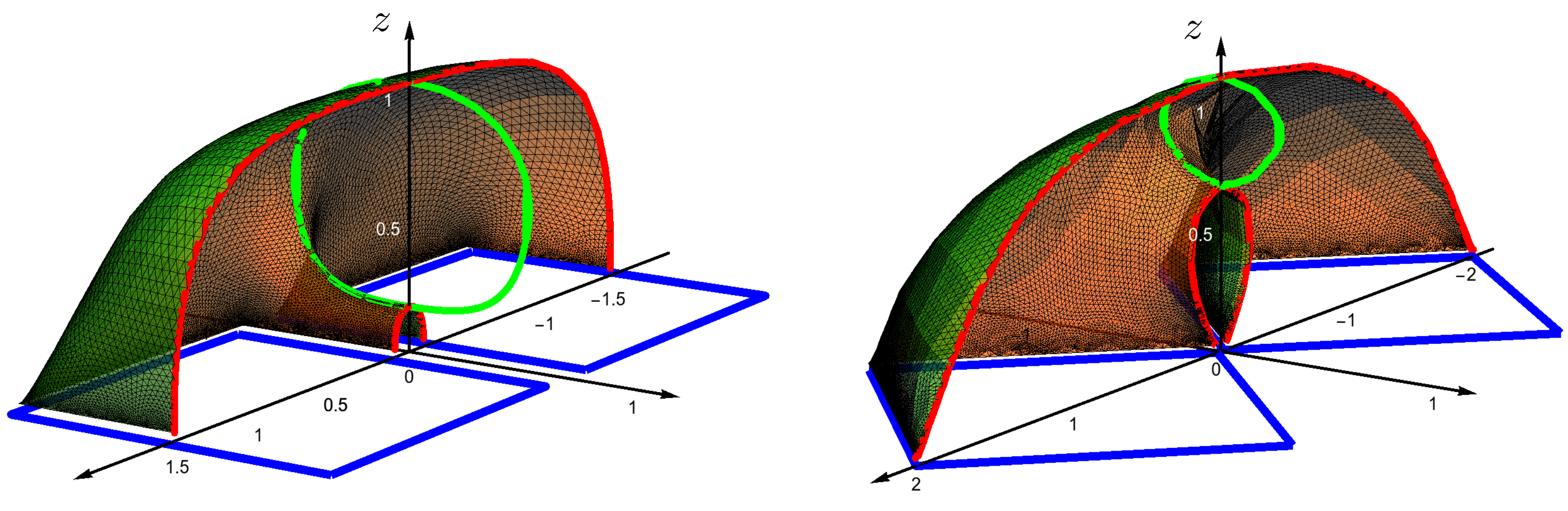}
\vspace{-.1cm}
\caption{\label{fig:2disjoint squares}
Minimal surfaces obtained with Surface Evolver for a domain $A=A_1\cup A_2$ made by the interior of two disjoint and equal squares. 
All the squares have the same size but the relative orientation of $A_1$ and $A_2$ is different in the two panels.
}
\end{figure}

By employing Surface Evolver, we could also study the holographic mutual information of disjoint domains whose boundaries contain corners.
In particular, one could take both $A_1$ and $A_2$ bounded by polygons, but also  $A_1$ bounded by a smooth curve and $A_2$ by a polygon.
In Fig.\,\ref{fig:2disjoint squares} we show the minimal area surfaces corresponding to $\partial A$ made by two equal and disjoint squares having different relative orientation. 
As discussed in \S\ref{sec corners}, when $\partial A$ has vertices a further logarithmic divergence occurs after the area law term in the $\varepsilon \to 0$ expansion (see (\ref{SA corners})).
If the coefficient of the logarithmic divergence in (\ref{SA corners}) is additive, i.e. $B_{A_1 \cup A_2}=B_{A_1}+B_{A_2}$ for two disjoint regions, then the holographic mutual information is finite. 
An expression like (\ref{BA sum}) with the sum extended over the vertices of both the components of $\partial A$ is additive, leading to a finite $\mathcal{I}_{A_1, A_2}$.
Also for these cases we could find plots similar to Figs.\,\ref{fig:densitysuper} and \ref{fig:densitysphero} but the curves would not be suitable for a comparison with an analytic formula because of the regularization procedure that we have adopted. Indeed, in our numerical computations $\partial A$ is defined at $z=\varepsilon$ and this regularization affects the $O(1)$ term in (\ref{SA corners}) \cite{Drukker:1999zq}, as already mentioned in the closing part of \S\ref{sec corners}.

%%%%%%%%%%%%%%%%%%%%%%%%%%%%%%%%%%%%

\section{Conclusions}
\label{sec conclusions}

In this paper we have studied the area of the minimal surfaces in AdS$_4$ occurring in the computation of the holographic entanglement entropy and of the holographic mutual information, focussing on their dependence on the shape of the entangling curve $\partial A$ in the boundary of AdS$_4$.

Our approach is numerical and the main tool we have employed is the program Surface Evolver, which allows to construct  triangulated surfaces approximating a surface anchored on a given curve $\partial A$ which is a local minimum of the area functional.
We have computed the holographic entanglement entropy and the holographic mutual information for entangling curves given by (or made by the union of) ellipses, superellipses or the boundaries of two dimensional spherocylinders, for which analytic expressions are not known.
We have also obtained the transition curves for the holographic mutual information of disjoint domains delimited by some of these smooth curves (see Figs.\,\ref{fig:transquircles}, \ref{fig:densitysuper} and \ref{fig:densitysphero}), providing a solid numerical benchmark for analytic expressions that could be found in future studies. 
We focused on these simple examples, but the method can be employed to address more complicated domains.

Besides the fact that the surfaces constructed by Surface Evolver are triangulated, a  source of approximation in our numerical analysis is the way employed to define the curve spanning the minimal surface.
Indeed, once the cutoff $\varepsilon >0$ in the holographic direction has been introduced to regularize the area of the surfaces, the numerical data have been found by defining $\partial A$ at $z=\varepsilon$. 
It would be interesting to understand better this regularization with respect to some other ones and also to decrease $\varepsilon$ in a stable and automatically controlled way in order to get numerical data which provide better approximations of the analytic results.

\begin{figure}[t] 
\vspace{-.2cm}
\hspace{-.0cm}
\begin{center}
\includegraphics[width=.85\textwidth]{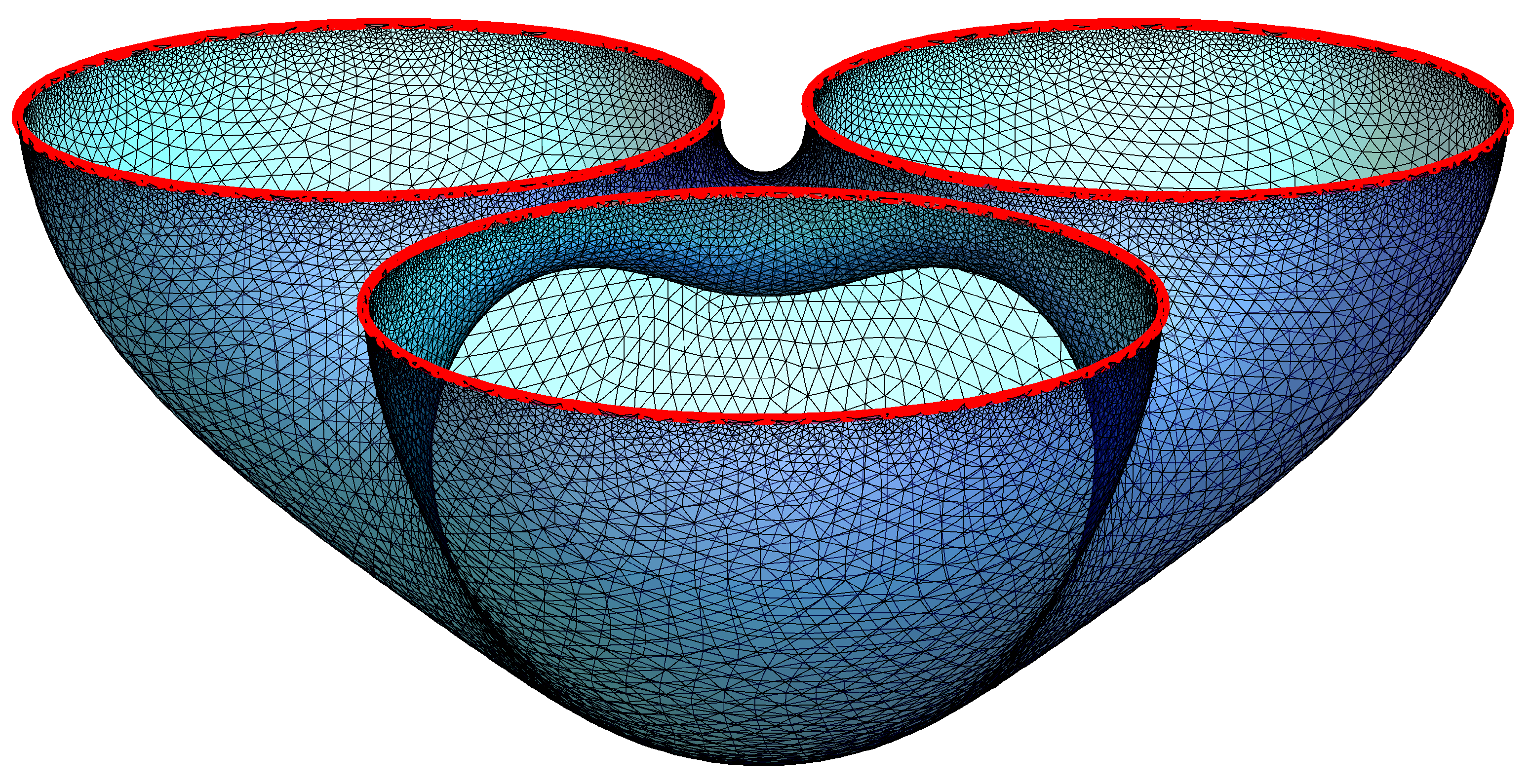}
\end{center}
\vspace{.2cm}
\caption{\label{fig:3 disjoint circles}
Minimal surface corresponding to three disjoint and equal red circles in the plane $z=0$ (the $z$ axis points downward).  
This surface has $13147$ vertices and $26624$ faces, while the number of edges is given by Euler formula with vanishing genus and 3 boundaries. 
This kind of surfaces occurs in the computation of the holographic tripartite information for the union of three disjoint disks.
}
\end{figure}

There are many possibilities to extend our work. 
The most important ones concern black hole geometries and higher dimensional generalizations.
An interesting extension involves domains $A$ made by three or more regions (see  \cite{Coser:2013qda} for some results in two dimensional conformal field theories and  \cite{Hayden:2011ag, Balasubramanian:2014hda, Headrick:2013zda} for a holographic viewpoint). 
In Fig.\,\ref{fig:3 disjoint circles} we show a minimal surface anchored to an entangling curve made by three disjoint circles. 
The area of this surface provides the holographic entanglement entropy between the union of the three disjoint disks and the rest of the plane, which is the most difficult term to evaluate in the computation of the holographic tripartite information \cite{Hayden:2011ag}. 
Another important application of the numerical method employed here involves time dependent backgrounds modelling the holographic thermalization \cite{Hubeny:2007xt, AbajoArrastia:2010yt, Balasubramanian:2011ur, Balasubramanian:2011at, Allais:2011ys, Callan:2012ip,  Liu:2013qca}.

Surface Evolver is a useful tool to get numerical results for the holographic entanglement entropy, 
which can be used to test analytic formulas that could be found in the future.

%%%%%%%%%%%%%%%%%%%%%%%%%%%%%%%%%%

\subsection*{Acknowledgments}

It is our pleasure to thank Ioannis Papadimitriou for collaboration in the initial part of this project and for many useful discussions during its development.
We wish to thank Hong Liu, Rob Myers, Mukund Rangamani, Domenico Seminara, Tadashi Takayanagi and in particular Mariarita de Luca, Veronika Hubeny, Alessandro Lucantonio for useful discussions.
We acknowledge Veronika Hubeny, Hong Liu, Mukund Rangamani, Tadashi Takayanagi and Larus Thorlacius for their comments on the draft. 
E.T. is grateful to Perimeter Institute and to the Center for Theoretical Physics at MIT for the warm hospitality during parts of this work. 
L.G. has been supported by The Netherlands Organization for Scientific Research (NWO/OCW).
A.S. has been supported by the Spanish Ministry of Economy and Competitiveness under grant FPA2012-32828, Consolider-CPAN (CSD2007-00042), the grant  SEV-2012-0249 of the ``Centro de Excelencia Severo Ochoa'' Programme and the grant  HEPHACOS-S2009/ESP1473 from the C.A. de Madrid.
E.T. has been supported by the ERC under  Starting Grant  279391 EDEQS.

%%%%%%%%%%%%%%%%%%%%%%%%%%%%%%%%%%%%%

\appendix

\section{Further details on minimal surfaces in $\mathbb{H}_3$}
\label{app math}

In this appendix we provide a derivation of  \eqref{eq:minimal_surface_in_AdS4} and describe some additional properties of minimal surfaces in AdS$_{4}$. Let us consider the area of a two dimensional surface $\gamma_{A}$ embedded in spatial slice $t={\rm const}$
\begin{equation}
\mathcal{A}[\gamma_{A}] 
= \int_{\gamma_{A}} d\mathcal{A} 
= \int_{U_{A}}\frac{\sqrt{h}\, du^{1}du^{2}}{z^{2}}\,,
\end{equation}
where $U_{A}$ is a coordinate patch. As mentioned in \S\ref{sec:minimal_surfaces}, $\mathcal{A}$ can be interpreted as the energy of a two dimensional interface immersed in $\mathbb{R}^{3}$ endowed with a potential energy of density $1/z^{2}$. To find the surface $\tilde{\gamma}_{A}$ minimizing $\mathcal{A}$ we consider a small displacement along the normal direction $\bm{N}$, parametrized as: $\bm{R}\rightarrow\bm{R}+w\bm{N}$, where $\bm{R}$ represents the position of a point on the surface and $w$ is a small normal displacement. The linear area variation can be straightforwardly calculated using classic differential geometry \cite{Kreyszig:1991}
\be
\delta \mathcal A[\gamma_{A}] 
=  
\int_{U_{A}} \delta\big(\sqrt{h}\,du^{1}du^{2} \big)\,\frac{1}{z^{2}}
+\int_{U_{A}} \delta\left(\frac{1}{z^{2}}\right) \sqrt{h}\,du^{1}du^{2} 
=
-\,2\int_{U_{A}} \frac{1}{z^{2}}\left(H+\frac{\bm{\hat{z}}\cdot\bm{N}}{z}\right)
w \, du^{1}du^{2}\,,
\ee
where $H$ is the surface mean curvature. Setting  $\delta \mathcal A[\gamma_{A}]$ to zero yields  \eqref{eq:minimal_surface_in_AdS4}. 

In a Monge patch $(u^{1},u^{2})=(x,y)$ and the surface can be represented as the graph of the function $z=z(x,y)$ representing the height of the surface above the $(x,y) $ plane. In this case the mean curvature reads
\begin{equation}\label{eq:cartesian_mean_curvature}
H = \frac{z_{,xx}(1+z_{,y}^{2})+z_{,yy}(1+z_{,x}^{2})-2z_{,xy}z_{,x}z_{,y}}{2(1+z_{,x}^{2}+z_{,y}^{2})^{3/2}}	\,,
\end{equation}
while the outward directed normal vector is given by
\be
\label{eq:cartesian_normal_vector}
\bm{N} = -\frac{z_{,x}\bm{\hat{x}}+z_{,y}\bm{\hat{y}}-\bm{\hat{z}}}{\sqrt{1+z_{,x}^{2}+z_{,y}^{2}}}\,.
\ee
Using Eqs. \eqref{eq:cartesian_mean_curvature} and \eqref{eq:cartesian_normal_vector} in  \eqref{eq:minimal_surface_in_AdS4} yields the Cartesian equation \eqref{eq:cartesian_equation}. 

In \S\ref{sec:minimal_surfaces} we argued that a surface described by  \eqref{eq:minimal_surface_in_AdS4} must be orthogonal to the $z=0$ plane. This orthogonality implies that the boundary curve $\partial\tilde{\gamma}_{A}$ is a geodesic of $\tilde{\gamma}_{A}$. To see this we can recall that the curvature $\kappa$ of a curve that lies on a surface can be decomposed as
\be
\label{eq:curvature_decomposition}
\kappa \, \bm{n} = \kappa_{n}\bm{N}+\kappa_{g}(\bm{N}\times\bm{t})\,,
\ee
with $\bm{t}$ the tangent vector of $\partial\tilde{\gamma}_{A}$, $\kappa\,\bm{n}=\bm{t}_{,s}$ (with $s$ the arc lenght) and $\kappa_{n}$ and $\kappa_{g}$ the normal and geodesic curvature respectively. 
Since $\partial\tilde{\gamma}_{A}$ lies on the $z=0$ plane and $\bm{\hat{z}}\cdot\bm{N}=0$ at $z=0$, then $\bm{N}=\pm\bm{n}$ where the choice of the sign is conventional. By virtue of  \eqref{eq:curvature_decomposition} this implies that $\kappa_{g}=0$. Thus $\partial\tilde{\gamma}_{A}$ is a geodesic over $\tilde{\gamma}_{A}$.

An interesting consequence of the previous statement is that the total Gaussian curvature of the surface is constant, regardless the shape of the boundary in the $z=0$ plane. The Gauss-Bonnet theorem tells us that
\be
\label{eq:gauss_bonnet}
\int_{\tilde{\gamma}_{A}}  K_{G} \, \sqrt{h}\, du^{1}du^{2}
+\oint_{\partial\tilde{\gamma}_{A}} \kappa_{g} \, ds 
\,=\, 2\pi\chi \,,	
\ee
where $K_{G}$ is the Gaussian curvature and $\chi$ is the Euler characteristic. 
Since $\kappa_{g}=0$ in our case, we have 
\be
\label{eq:total_curvature}
\int_{\tilde{\gamma}_{A}}  K_{G} \, \sqrt{h}\, du^{1}du^{2}
= 2\pi\chi	\,.
\ee
Let us recall that the Euler characteristic is $\chi = 2-2g -b$, where $g$ is the genus of the surface and $b$ is the number of its boundaries.

%%%%%%%%%%%%%%%%%%%%%%%%%%%%%%%%%%%
	
\section{Numerical Method}
\label{app technical details}	

\begin{figure}[t] 
\vspace{-.2cm}
\hspace{-.0cm}
\begin{center}
\includegraphics[width=1.0\textwidth]{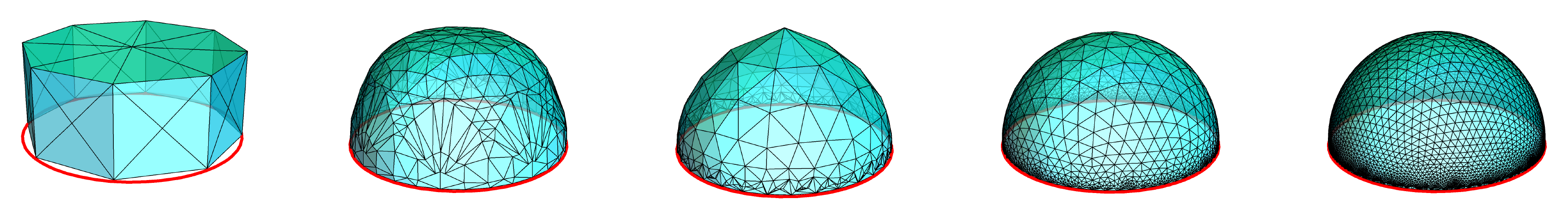}
\end{center}
\vspace{.0cm}
\caption{\label{fig:evolverexample}
Example of a typical \emph{evolution} obtained by Surface Evolver in the case of a circular boundary. The initial configuration consists of an octagonal prism composed of 40 triangles (left). The shape is then optimized and refined as described in \S\ref{app technical details}, finding the final configuration given by the rightmost surface, which consists of 10240 triangles and yields $\widetilde{F}_{A}=1.99843\pi$ whereas $F_{A}=2\pi$ is the exact value from the analytic result (\ref{Ahemisphere ads4}). In this example the radius of the circle is $R=1$ and $\varepsilon=0.03$.
}
\end{figure}

The numerical results presented in \S\ref{sec simply connected} and \S\ref{sec 2 disjoint} have been obtained with Surface Evolver \cite{evolverpaper,evolverlink}. This is a multipurpose shape optimization program created by Brakke \cite{evolverpaper} in the context of minimal surfaces and capillarity and then expanded to address generic problems on energy minimizing surfaces. A surface is implemented as a simplicial complex, i.e. a union of triangles. Given an initial configuration of the surface, the program evolves the surface toward a local energy minimum by a gradient descent method. The energy used in our calculations is the $\mathbb{H}_{3}$ area function given in  \eqref{area fun general}.

The initial configuration is preferably very simple and contains only the least number of triangles necessary to achieve a given surface topology (Fig.\,\ref{fig:evolverexample}). A typical {\em evolution} consists in a sequence of optimization and mesh-adjustment steps. During an optimization step, the coordinates of the vertices are updated by a local minimization algorithm (conjugate gradient in our case), resulting in a configuration of lower energy. The topology of the mesh (i.e. the number of vertices, faces and edges) is not altered during minimization. A mesh-adjustment step, on the other hand, consists of a set of operations whose purpose is to render the discretized surface smooth and uniform. These operations can be broadly divided in two class: mesh-refinements and mesh-repairs. In a mesh-refinement operation a finer grid is overlaid on the coarse one. This is obtained, for instance, by splitting a triangle in four smaller triangle obtained by joining the mid points of the original edges. In a mesh-repair operation, the triangles that are too distorted compared to the average are eliminated. This operation can change the topology of the mesh and possibly also the topology of the surface which can then breakup into two or more connected parts. This happens, for instance, in the case of the surfaces described in \S\ref{sec 2 disjoint}. As explained, the minimal surface spanning a disconnected boundary curve can be either connected or disconnected depending on the shape of the boundary. Evolving an initially connected surface in the regime of geometric parameters where the only stable solution is disconnected causes the surface to form narrow necks and eventually pinches off once the triangles around the necks become too squeezed. 

\begin{figure}[t]
\vspace{.0cm}
\hspace{1.3cm}
\includegraphics[width=.8\textwidth]{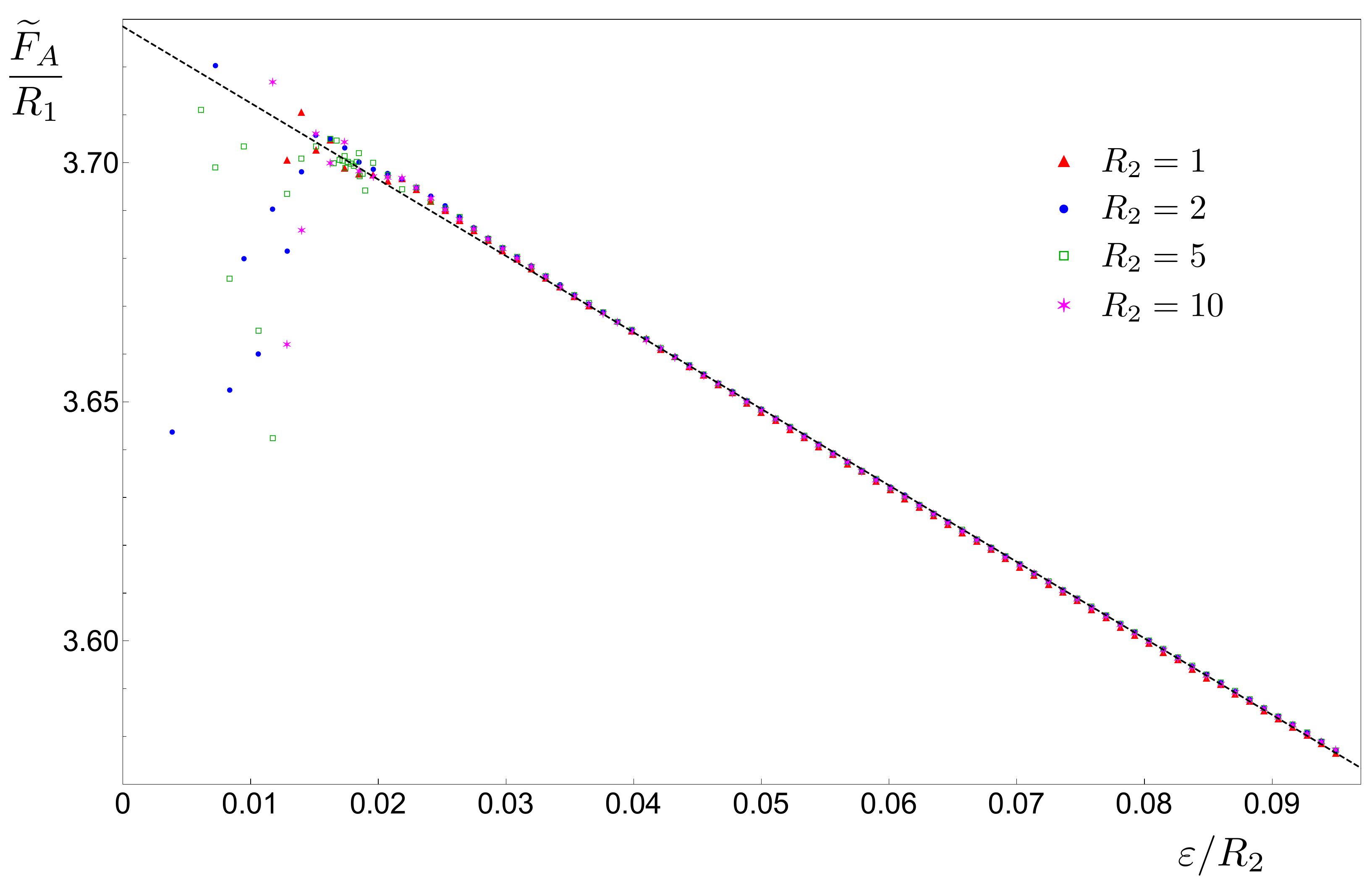}
\vspace{.1cm}
\caption{\label{fig:appepsfit}
The quantity $\widetilde{F}_A$ (see (\ref{FAtilde def})) computed with Surface Evolver for ellipses having $R_1=2R_2$ (see the bottom panel of Fig.\,\ref{fig:squircle min surfs}), for various $R_2$ and $\varepsilon$.
When $\varepsilon/R_2$ is too small, our numerical data are not stable.
The fitted value on the vertical axis is $3.728$.
}
\end{figure}

Due to the divergence of the area element $d\mathcal{A}=\sqrt{h}/z^{2}\, du^{1}du^{2} $ at $z=0$, the boundary curves used in the numerical work have been defined on the plane $z=\varepsilon$. In order to maximize the accuracy of the numerical solution, it is preferable to choose value of $\varepsilon$ that is much smaller than any other length scale in the problem and yet large enough to allow the convergence of the optimization steps. With this goal in mind, we have adopted an empirical selection criterion based on the following procedure. Let $\partial{\tilde{\gamma}}_{A}$ be an ellipse and let $R_{1}$ and $R_{2}=R_{1}/2$ be the semi-major and semi-minor axes. Using Surface Evolver we have calculated the finite part of the area $\widetilde{F}_{A}$ for various choices of $\varepsilon$ and $R_{1}$. In the limit of $\varepsilon\rightarrow 0$ the ratio $\widetilde{F}_{A}/R_{1}$ is expected to approach a finite value, but from the data shown in Fig.\,\ref{fig:appepsfit} we see that for $\varepsilon/R_{2}<0.02$, the accuracy of the numerical calculation starts to drop. Based on this numerical evidence we have set in most of our numerical calculations $\varepsilon/R=0.03$, where $R$ is the typical length scale of the boundary. 
It is worth remarking  that in our numerical computations it is easier (namely the evolution is more stable) to deal with smaller values of $\varepsilon/R$ by increasing $R$ than by decreasing $\varepsilon$. 
Smaller values of $\varepsilon/R$ obtained by decreasing $\varepsilon$ keeping $R$ fixed can be  achieved by setting up {\em ad hoc} evolutions, tailored for a specific type of boundary shape. This has been done only for the triangles in Fig.\,\ref{fig:polygons profiles}, while in the remaining figures we have increased $R$ keeping $\varepsilon=0.03$ fixed. 
Nevertheless, for $\varepsilon$ fixed, numerical instabilities are encountered when $R$ is too large as well.
The values of $\varepsilon/R$ adopted in our numerical calculations have been chosen to guarantee both stable evolutions and a satisfactory precision to compare the data with the analytic results, when they are available. 

Other alternative methods are available to construct minimal surfaces. A popular one by Chopp \cite{chopp} consists of evolving the surface level sets under the surface mean curvature flow. A variant of this method has been employed in \cite{Hubeny:2013gta} to study minimal surfaces in the Schwarzschild-AdS$_{D+2}$ background.

\section{Superellipse: a lower bound for $F_A$}
\label{app bounds}

In this appendix we provide a lower bound for the quantity $F_A$ (see (\ref{SA no corners})) associated with the entangling curves $\partial A$ given by the superellipses (\ref{eq superellipse}), that we have discussed in \S\ref{sec superellipse}.

If $A$ is a simply connected domain without corners in its boundary, let us consider a surface $\gamma_A^\ast$ anchored on $\partial A$, but different from $\tilde{\gamma}_A$, and such that $\mathcal{A}[\gamma_A^\ast] = P_A/\varepsilon - F_A^\ast + o(1)$ as $\varepsilon \to 0$.
Being $\tilde{\gamma}_A$ the minimal area surface anchored on $\partial A$, it is immediate to realize that $F_A^\ast < F_A$. 
Here we consider the superellipses (\ref{eq superellipse}), whose perimeter is given by
\be
P_A 
\label{perimetersuperellipse}
\,=\,
4 R_1 \int_{0}^{1}
\sqrt{1+  \left({R_2}/{R_1}\right)^2 h_n(\tilde{x})^2} \, d\tilde{x} \,,
 \qquad
 h_n(\tilde{x}) \equiv 
 \frac{\tilde{x}^{n-1}}{(1-\tilde{x}^n )^{1-1/n}} \,,
\ee
where the integration variable $\tilde{x}=x/R_1$ as been employed. 
Let us adapt to this case the choice of the trial surface suggested in \cite{Allais:2014ata} for the ellipse, namely we consider $\gamma_A^\ast$ such that any section along the $x$ direction provides the profile of the infinite strip whose width is given by $y(x)$ obtained from (\ref{eq superellipse}), i.e.
\be
\label{eq superellipse app}
y(\tilde{x})=R_2 \left(1-\tilde{x}^n\right)^{1/n} .
\ee
Given the symmetries of the superellipse, we are allowed to restrict ourselves to $x>0$ and $y>0$. 
From (\ref{strip profile ddim}) for $D=2$, 
we construct the trial surface $\gamma_A^\ast$ by requiring that
we have that any section at $x=\textrm{const}$ is given by
\be
\label{trial surf}
y(z,\tilde{x}) = z_\ast (\tilde{x}) \int_{z/z_\ast(\tilde{x})}^1  
\frac{Z^2}{\sqrt{1-Z^4}} \, d Z\,,
\qquad
z_\ast(\tilde{x}) \equiv   \frac{2\, y(\tilde{x})}{\sqrt{s_{\infty}}}\,,
\ee
where the integration variable $Z \equiv z/z_\ast$ has been employed
and $z_\ast(\tilde{x})$ has been introduced by taking $z_\ast$ in (\ref{profile for strip}) with $s_{\infty}$ defined in (\ref{area infinite strip}) and replacing $R_2$ with $y(\tilde{x})$ defined in (\ref{eq superellipse app}). 
From (\ref{trial surf}), it is straightforward to show that $y(0,\tilde{x})=y(\tilde{x})$ and this guarantees that the trial surface is anchored on the superellipse (\ref{eq superellipse app}).

The occurrence of the cutoff $\varepsilon$ in the holographic direction influences the integration domain along the $x$ direction. In particular, by employing (\ref{eq superellipse app}) and (\ref{trial surf}), the requirement $z_\ast(\tilde{x})\geqslant \varepsilon$
becomes $\tilde{x} \leqslant \tilde{x}_{\varepsilon} $, where 
\be
\label{xtilde def}
\tilde{x}_{\varepsilon} 
\equiv
\left[
1-\left(\frac{\sqrt{s_\infty}}{2 R_2} \, \varepsilon \right)^n
\,\right]^{1/n} .
\ee
Plugging (\ref{trial surf}) inside the area functional, being $y$ written in terms of $x$ and $z$, we get
\be
\label{Astar integ}
\mathcal{A}[\gamma_A^\ast]  
\,=\,
4 \int_{0}^{\tilde{x}_{\varepsilon} }d\tilde{x} \int_{\varepsilon}^{z_\ast(\tilde{x})} dz\, 
\frac{\sqrt{1+ (\partial_z y)^2+(\partial_x y)^2}}{z^2} 
\,=\,
\frac{2R_1 \sqrt{s_\infty} }{R_2}
\int_0^{\tilde{x}_{\varepsilon} }
\frac{M_{\varepsilon} (\tilde{x})}{ \left(1-\tilde{x}^n\right)^{1/n}}\, d\tilde{x}\,,
\ee
where 
\be
\label{eq mn cz}
M_{\varepsilon} (\tilde{x}) \equiv
\int_{\varepsilon/z_\ast(\tilde{x})}^{1}
 \frac{\sqrt{1+ \left({R_2}/{R_1}\right)^2h_n(\tilde{x})^2 \,C(Z)^2}	}{Z^2 \sqrt{1-Z^4}} \, dZ\,,
\qquad
C(Z) \equiv 
\frac{2}{\sqrt{s_\infty}}
\left(\, \int_Z^1  \sqrt{\frac{1-Z^4}{1-u^4}} \,u^2\, du - Z^3 \right).
\ee
Computing (\ref{Astar integ}) analytically is too hard, but one can check that the area law is satisfied. 
When $\varepsilon \to 0$, from (\ref{xtilde def}) we have that $\tilde{x}_{\varepsilon}  = 1+O(\varepsilon^n)$. 
In this limit, the most divergent term of $M_{\varepsilon} (\tilde{x}) $ comes from the limit of integration $\varepsilon/z_\ast(\tilde{x})$ and it can be found by considering an integration on the interval $[\varepsilon/z_\ast(\tilde{x}),a]$, where $Z$ is infinitesimal if $a \ll 1$. The remaining integral provides $O(1)$ terms. 
For $Z\to 0$ we have that $C(0)=1$ and therefore the leading term in (\ref{Astar integ}) is given by
\be
\label{Aast bis}
\mathcal{A}[\gamma_A^\ast]   
=
\frac{2R_1 \sqrt{s_\infty} }{R_2}
\int_0^1 d\tilde{x}\,
\frac{\sqrt{1+ \left({R_2}/{R_1}\right)^2h_n(\tilde{x})^2}
}{ \left(1-\tilde{x}^n\right)^{1/n}}
\int_{\varepsilon/z_\ast(\tilde{x})}^a
\frac{dZ}{Z^2} + O(1)
= 
\frac{P_A}{\varepsilon} + O(1)\,,
\ee
where $P_A$ given in (\ref{perimetersuperellipse}) can be recognized after (\ref{trial surf}) and (\ref{eq superellipse app}) have been employed.
We are not able to find $F^\ast_A$ analytically but it can be obtained numerically as $F^\ast_A=\lim_{\varepsilon\to 0}(P_A/\varepsilon - \mathcal{A}[\gamma_A^\ast]   )$, with $\mathcal{A}[\gamma_A^\ast] $ given by (\ref{Astar integ}), getting a lower bound for $F_A$ associated with the superellipse.

It is interesting to consider $F^\ast_A$ in the limit of a very elongated superellipses, namely when $R_1/R_2 \to \infty$. 
This means that (\ref{Astar integ}) must be studied in the double expansion $\varepsilon \to 0$ and $R_2/R_1 \to 0$.
Assuming that the order of this two limits does not matter, let us set $R_2/R_1 = 0$ in the expressions of $M_\varepsilon(\tilde{x})$ in (\ref{eq mn cz}) and expand it for small $\varepsilon$, finding
\be 
\label{Meps expanded}
M_{\varepsilon} (\tilde{x}) \big|_{\small{R_2/R_1=0}}
= 
\frac{z_\ast(\tilde{x})}{\varepsilon}-\frac{\sqrt{s_\infty}}{2} + O(\varepsilon^2)\,,
\ee
where $z_\ast(\tilde{x})$ is given in (\ref{trial surf}).
By plugging (\ref{Meps expanded}) into (\ref{Astar integ}) and expanding the resulting expression for $\varepsilon \to 0$, we have that
\be
\mathcal{A}[\gamma_A^\ast]  
\,=\,
\frac{4 R_1}{\varepsilon} 
- s_\infty  \,\frac{R_1}{R_2}\int_0^{\tilde{x}_\varepsilon} 
\frac{d\tilde{x}}{ \left(1-\tilde{x}^n\right)^{1/n}} 
+o(\varepsilon)
\,=\,
\frac{4 R_1}{\varepsilon} 
-  \frac{\pi s_\infty}{n \sin(\pi/n)} \, \frac{R_1}{R_2}
+o(\varepsilon)\,.
\ee
Notice that, from (\ref{perimetersuperellipse}), one can observe that 
$P_A=4 R_1\big[1+o(1)\big]$ when $R_1/R_2 \to \infty$.
We conclude that the leading term of $F^\ast_A $ as $R_1/R_2 \to \infty$ reads 
\be 
\label{Fa star expanded}
F^\ast_A 
\,=\,  
\frac{\pi s_\infty}{n \sin(\pi/n)} \, \frac{R_1}{R_2}
+ \dots\;.
\ee
When $n=2$, the result of \cite{Allais:2014ata} is recovered, as expected. 
Moreover, the expression (\ref{Fa star expanded}) in the special cases of $n=2$ and $n=3$ has been checked in Fig.\,\ref{fig:data squircles}  against the data obtained with Surface Evolver (see respectively the red and the blue dotted horizontal lines), finding a good agreement.
Notice that the expression in the r.h.s. of (\ref{Fa star expanded}) is strictly larger than the value of $F_A$ corresponding to the infinite strip (see (\ref{area infinite strip})), which is approached as $n \to \infty$.

\section{Some generalizations to AdS$_{D+2}$}
\label{app d-dims}

\subsection{Sections of the infinite strip}
\label{app caps}

In this section we discuss the computation of the area of the domain identified by an orthogonal  section of the minimal surfaces associated with the infinite strip.

The metric of AdS$_{D+2}$ in the Poincar\'e coordinates reads
\be
\label{ads metric ddim}
ds^2 = \frac{-\,dt^2+dz^2+dx_1^2+ \dots +dx_D^2}{z^2}\,.
\ee
Considering an infinite $D$-dimensional strip on the spatial slice $t=\textrm{const}$ extended along the $x_2, \dots , x_D$ directions whose width is given by $2R_2$, i.e. $|x_1| \leqslant R_2$, the minimal area surface associated with this domain is characterized by the profile $z=z(x_1)$.
Because of the symmetry of the problem, $z(x_1)$ is even and therefore we can restrict to $0\leqslant x_1 \leqslant R_2$. The profile is obtained by solving the following differential equation \cite{Ryu:2006bv, Ryu:2006ef}
\be
\label{strip profile ddim}
z' = -\,\frac{\sqrt{z_\ast^{2D} - z^{2D}}}{z^D} \,.
\ee
where $z_\ast$ is the maximum value of $z$, which is reached at $x_1=0$. 

A way to get an orthogonal section of the infinite strip is defined by $x_2 = \dots = x_D = \textrm{const}$. Then, one considers the two dimensional region enclosed by the profile $z(x_1)$ and the cutoff $z=\varepsilon$ in the plane $(x_1, z)$.
The domain along the $x_1$ axis is $|x_1| \leqslant R_2-a$, where $a$ is defined by $z(R_2-a) = \varepsilon$.
Its area reads
\be
\label{area cap ddim v1}
\hat{\mathcal{A}} 
\,= \,
2\int_{0}^{R_2-a} dx_1 \int_{\varepsilon}^{z(x_1)} \frac{dz}{z^2}
\,=\,
\frac{2(R_2-a)}{\varepsilon} - 
\frac{2}{D} \bigg[\, \frac{\pi}{2} - 
\arctan \bigg( \frac{\varepsilon^D}{\sqrt{z_\ast^{2D}} - \varepsilon^{2D}} \bigg)  \bigg]
\,=\,
\frac{2R_2}{\varepsilon} - 
\frac{\pi}{D} + o(1)\,,
\ee
where (\ref{strip profile ddim}) has been employed.

Another section of the infinite strip to study is defined by $x_i = \textrm{const}$ for some $2\leqslant i \leqslant D$ and $|x_j| \leqslant R_1$ for $j \neq i$.
In this case we are interested in the volume of the $D$ dimensional region enclosed by the profile $z(x_1)$ and $z=\varepsilon$, whose projection on the $z=0$ hyperplane is included within the section of the infinite strip we are dealing with.
It is given by
\be
\label{area cap ddim v2}
\hat{\mathcal{A}} 
\,= \,
2(2R_1)^{D-2}
\int_{0}^{R_2-a} dx_1 \int_{\varepsilon}^{z(x_1)} \frac{dz}{z^D}
\,=\,
(2R_1)^{D-2} (D-1)
\left[\,
\frac{2R_2}{\varepsilon^{D-1}} - 
\frac{\sqrt{\pi}\,\Gamma(1+1/D)}{z_\ast^{D-2}\,\Gamma(1/2 +1/D)} + o(1)
\,\right] .
\ee
Notice that for $D=2$ the expressions (\ref{area cap ddim v1}) and (\ref{area cap ddim v2}) coincide, as expected, and the result is employed in \S\ref{sec simply connected} to study the auxiliary surface, which corresponds to the dashed curve in Fig.\,\ref{fig:data squircles}.

\subsection{Annular domains}
\label{app annulus}

In this appendix we consider the surfaces anchored on the boundaries of annular domains which are local minima of the area functional because some analytic expressions can be found for them.

The metric of AdS$_{D+2}$ in Poincar\'e coordinates (\ref{ads metric xy}) written by employing spherical coordinates for the spatial part $\mathbb{R}^D$ of the boundary $z=0$ is
\be
ds^2 = \frac{dz^2-dt^2+d\rho^2 +\rho^2 d\Omega^2_{D-1}}{z^2}\,,
\ee
where $\rho \in [0,\infty)$ and the AdS radius has been set to one.

A spherically symmetric spatial region $A$ in the AdS boundary is completely specified by an interval in the radial direction.
Because of the symmetry of $A$, the minimal surface anchored on $\partial A$ is given by  $z=z(\rho)$ and, for a generic profile $z=z(\rho)$, the corresponding area of the two dimensional surface $\gamma_A$ reads
\begin{eqnarray}
\label{area z(rho)}
\mathcal{A}[\gamma_A] &=&  \textrm{Vol}(S^{D-1})  \, \mathcal{R}_D\,,
\qquad
\mathcal{R}_D
\,\equiv\, 
\int
\frac{\rho^{D-1}}{z^D} \sqrt{1+(z')^2} \, d\rho\,,
\end{eqnarray}
where $\textrm{Vol}(S^{D-1})$ is the volume of the $(D-1)$-dimensional unit sphere and $\mathcal{R}_D$ is the integral in the radial direction. 
We remark that the integration domain in $\mathcal{R}_D$ is not necessarily the interval defining $A$ in the radial direction, as it will be clear from the case discussed in the following. 
In order to find the minimal surface $\tilde{\gamma}_A$, one extremizes the area functional (\ref{area z(rho)}), obtaining 
\be
\label{eq sphere ddim}
z z'' +(1+z'^2)\left[D+(D-1) \frac{z z'}{\rho} \right] = 0 \,.
\ee
When $A$ is a sphere of radius $R$, we have that $0\leqslant \rho \leqslant R$ and it is well known that the corresponding minimal surface is a hemisphere \cite{Ryu:2006bv, Ryu:2006ef}.

Here we consider the region $A$ delimited by two concentric spheres, whose radii are $R_{\textrm{\tiny in}}$ and $R_{\textrm{\tiny out}}$, with $0<R_{\textrm{\tiny in}} < R_{\textrm{\tiny out}}$.
In this case $R_{\textrm{\tiny in}}\leqslant \rho \leqslant R_{\textrm{\tiny out}}$ and $A$ is not simply connected. 
For $D=2$ and $D=3$, the corresponding minimal surface extending in the bulk and anchored on $\partial A$ has been studied in \cite{Drukker:2005cu, Hirata:2006jx, Dekel:2013kwa}.
In order to solve (\ref{eq sphere ddim}) for this configuration, we find it convenient to introduce
\cite{Drukker:2005cu, Hirata:2006jx}
\be
\label{zrho subs}
z(\rho) \equiv \rho\, \tilde{z}(\rho)\,,
\qquad
u \equiv \log \rho\,,
\qquad
\tilde{z} _u \equiv \partial_u \tilde{z} \,.
\ee
Notice that $\tilde{z} = \tan \theta$ is the angular coefficient of the line connecting the origin to a point belonging to the surface. 
Given (\ref{zrho subs}), the differential equation (\ref{eq sphere ddim}) becomes
\be
\tilde{z} \, \tilde{z}_u\big(1+\partial_{\tilde{z}}\tilde{z}_u \big)
+
\big[1+(\tilde{z} + \tilde{z}_u)^2\big]
\big[D+(D-1) \tilde{z} (\tilde{z} + \tilde{z}_u)\big]
=0\,.
\ee
Integrating this equation, we find two solutions, namely
\be
\label{ztilde1 pm}
\tilde{z}_{u,\pm}(\tilde{z}) = - \frac{1+\tilde{z}^2}{\tilde{z}}
\left[1\pm \frac{\tilde{z}^{D-1}}{\sqrt{K (1+\tilde{z}^2)-\tilde{z}^{2D}}}\right]^{-1},
\qquad
K \,>\, 0\,,
\ee
which correspond to two different parts of the profile.
As for the integration constant $K$, it must be strictly positive because $\tilde{z} =0$ corresponds to the boundary $z=0$, which is included in the range of $z$.
The domain for $\tilde{z}$ is $0\leqslant \tilde{z} \leqslant \tilde{z}_m$, where $\tilde{z}_m$ is the first positive zero of the polynomial under the square root in (\ref{ztilde1 pm}). 
For $D=2$ we are lead to solve a biquadratic equation, which gives 
$\tilde{z}_m^2 = \big(K+\sqrt{K(K+4)}\,\big)/2$. Notice that $\tilde{z}_m \to 0$ when $K \to 0$. 

\begin{figure}[t] 
\vspace{-.4cm}
\hspace{.1cm}
\includegraphics[width=.98\textwidth]{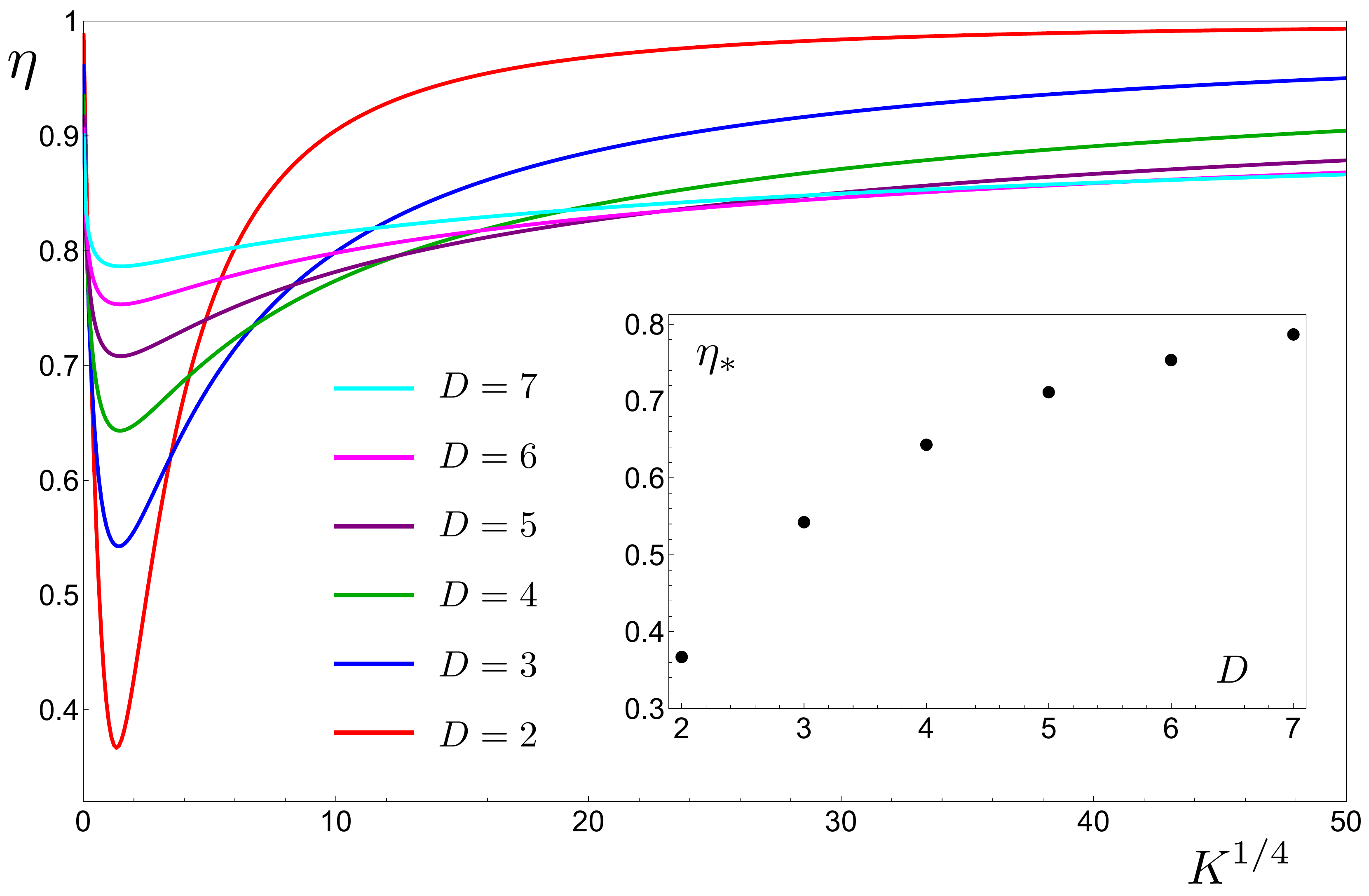}
\vspace{-.1cm}
\caption{\label{fig:appannulusk}
Curves for $\eta$ as function of $K$ obtained from the matching condition (\ref{integ ratio}) for various dimensions $2\leqslant D \leqslant 7$. 
For any $D$, a minimal value $\eta_\ast >1$ occurs, which is shown in the inset. 
Given a value $\eta \in (\eta_\ast,1)$, two values of $K$ correspond to it, providing two different radial profiles (see an example for $D=2$ in Fig.\,\ref{fig:appannulusprof}).
}
\end{figure}

\begin{figure}[t] 
\vspace{-.2cm}
\begin{center}
\includegraphics[width=.8\textwidth]{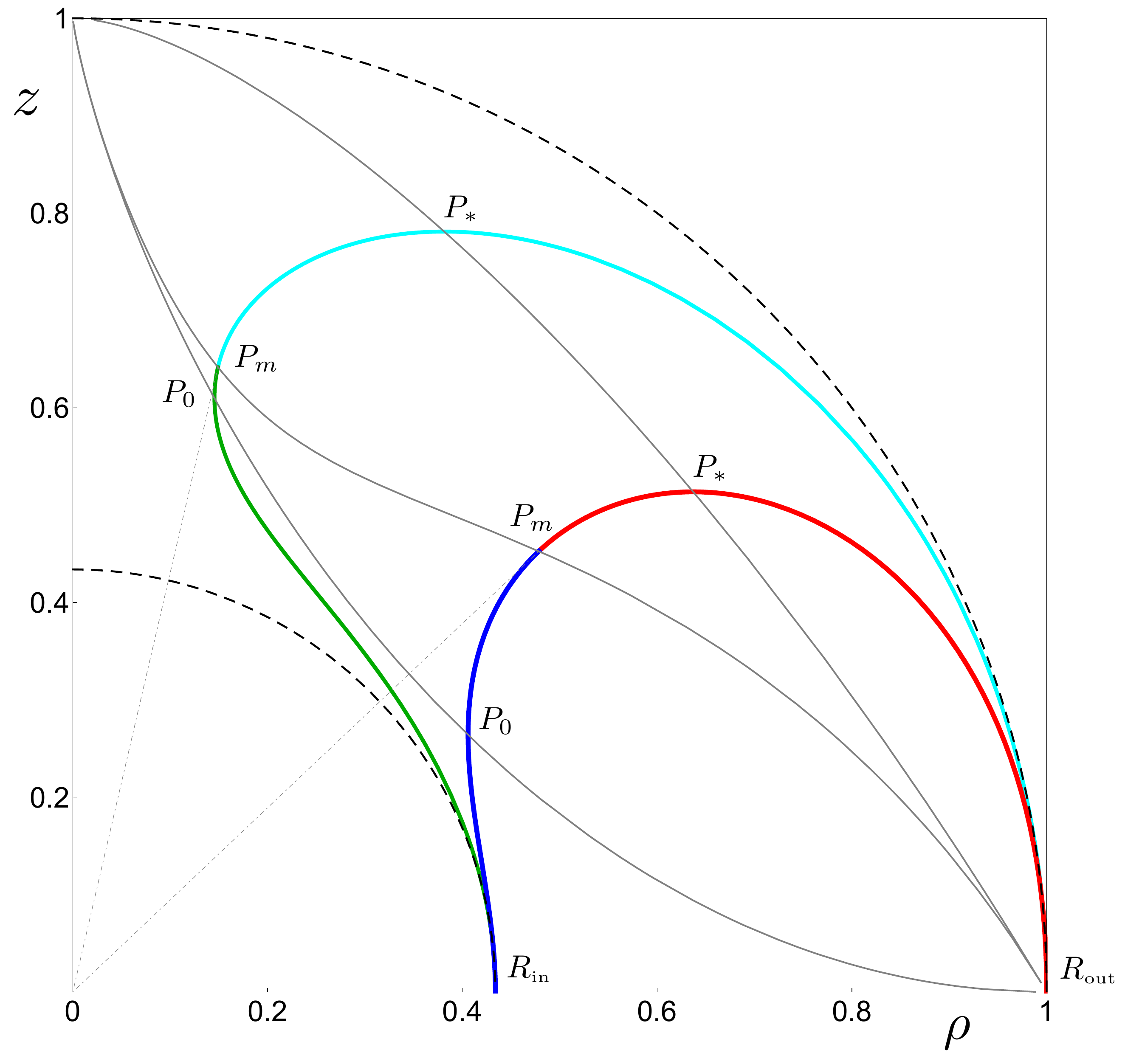}
\end{center}
\vspace{-.4cm}
\caption{\label{fig:appannulusprof}
Radial profiles in the $(\rho,z)$ plane for the connected surfaces anchored on the boundary of the same annulus $A$ having $R_{\textrm{\tiny in}} < R_{\textrm{\tiny out}}$. 
They correspond to local minima of the area functional and they are characterized by the two different values of $K$ associated with the same $\eta$.
These connected surfaces are obtained through (\ref{branches R1R2}) and (\ref{integ ratio}),
where different colours are used for the various branches.
The dashed curves represent the two concentric hemispheres anchored on $\partial A$ as well.
The continuous grey curves are the paths in the $(\rho,z)$ plane of the points $P_0$, $P_m$ and $P_\ast$ as $K\in (0,\infty)$.
Here $D=2$, $R_{\textrm{\tiny in}} =0.43$, $R_{\textrm{\tiny out}} =1$ and the values of $K$ are $K=0.81$ (global minimum) and $K=2.05$ (local minimum).
Comparing the area of the two connected surfaces, we find that the one having minimal area has $P_\ast$ closer to the boundary.
}
\end{figure}

The differential equation (\ref{ztilde1 pm}) can be solved through the separation of the variables. 
In particular, from the r.h.s. of (\ref{ztilde1 pm}), we find it convenient to introduce
\be
\label{fD def}
f^{(D)}_{\pm, K}(\tilde{z})
\equiv
\int_0^{\tilde{z}} 
\frac{\lambda}{1+\lambda^2} 
\left[1\pm \frac{\lambda^{D-1}}{\sqrt{K (1+\lambda^2)-\lambda^{2D}}}\right]
d\lambda\,.
\ee
Then, the profile of the radial section is given by the following two branches
\be
\label{branches R1R2}
\Bigg\{\begin{array}{l}
\rho = R_{\textrm{\tiny in}} \, e^{-f^{(D)}_{-, K}(\tilde{z})}\,,
\\
\rule{0pt}{.5cm}
\rho = R_{\textrm{\tiny out}} \, e^{-f^{(D)}_{+, K}(\tilde{z})}\,.
\end{array}
\ee
Imposing that these two branches match at the point $P_m$, whose $(\rho, z)$ coordinates are $(\rho_m , z_m \equiv z(\rho_m))$, where $z_m$ has been found above, we get the following relation
\be
\label{integ ratio}
-\, \log(\eta)
= f^{(D)}_{+, K}(\tilde{z}_m) - f^{(D)}_{-, K}(\tilde{z}_m) 
=
\int_0^{\tilde{z}_m} 
 \frac{2\,\lambda^{D}}{(1+\lambda^2)\sqrt{K (1+\lambda^2)-\lambda^{2D}}} \, 
d\lambda\,,
\qquad
\eta \equiv 
\frac{R_{\textrm{\tiny in}}}{R_{\textrm{\tiny out}}}\,.
\ee
Since $\tilde{z}_m$ depends on $K$, from (\ref{integ ratio}) we get a relation between $\eta$ and $K$, which is represented in Fig.\,\ref{fig:appannulusk} for $2\leqslant D \leqslant 7$.
The first feature to point out about (\ref{integ ratio}) is the existence of a minimal value for $\eta$ that will be denoted by $\eta_\ast>0$.
For instance, we find $\eta_\ast =0.367$, $\eta_\ast =0.542$ and $\eta_\ast =0.643$ for $D=2$, $D=3$ and $D=4$ respectively (see the inset in Fig.\,\ref{fig:appannulusk} for other $D$'s).
Then, for any $\eta_\ast < \eta <1$, there are two values of $K$ giving the same $\eta$,
while for $0< \eta <\eta_\ast $ connected solutions do not exist. 
The two different $K$'s associated with the same $\eta_\ast < \eta <1$ provide two different radial profiles and therefore two connected surfaces having the same $\partial A$. 
In order to find the global minimum of the area functional, we have to evaluate their area.
Through a numerical analysis, one observes that $z_m$ is an increasing function of $K$.

Beside $P_m$, another interesting point of the profile is $P_0=(\rho_0 , z_0 \equiv z(\rho_0))$, where $| z(\rho_0)' |$ diverges. From (\ref{ztilde1 pm}), this divergence occurs when
\be
\sqrt{K (1+\tilde{z}_0^2)-\tilde{z}_0^{2D}} \pm \tilde{z}_0^{D-1} = 0
\qquad
\Longrightarrow
\qquad
K = \tilde{z}_0^{2D-2} \equiv (\tan \theta_0)^{2D-2}\,.
\ee
This tells us that $K$ has a geometric meaning because it provides $\tilde{z}_0$.

Let us also introduce the point $P_\ast$, with coordinates $(\rho_\ast , z_\ast \equiv z(\rho_\ast))$ as the point having the maximum value of $z$, which corresponds to the maximal penetration of the minimal surface into the bulk.
The coordinate $z_\ast $ can be found by considering the branch $z(\rho)$ characterized by $f^{(D)}_{+, K}$ in (\ref{branches R1R2}) and then computing its derivative w.r.t. $\rho$, which is given by
\be 
\label{root P_ast}
\frac{d z}{d \rho}
= \frac{d ( \tilde{z} \rho)}{d \tilde{z}} \left( \frac{d \rho}{d \tilde{z}} \right)^{-1} 
= \tilde{z} -\left( \frac{d f^{(D)}_{+, K}(\tilde{z})}{d \tilde{z}}\right)^{-1} ,
\ee
where in the last step (\ref{branches R1R2}) has been used. 
When $D=2$ the root of (\ref{root P_ast}) can be found and it reads
\be
\tilde{z}_\ast = K^{1/4} \,.
\ee

An explicit example in $D=2$ is given in Fig.\,\ref{fig:appannulusprof}, where we have shown the two connected radial profiles having the same $\eta>\eta_\ast$ but different values of $K$. The two different branches in (\ref{branches R1R2}) at fixed $K$, supported by the matching condition (\ref{integ ratio}), have been denoted with different colours: the red and cyan curves are obtained through $f^{(2)}_{+, K}$ while the blue and the green ones through $f^{(2)}_{-, K}$. 
In Fig.\,\ref{fig:appannulusprof} the grey curves denote the paths described by the three points $P_m$, $P_0$ and $P_\ast$ introduced above as $K$ assumes all the positive real values.

We find it instructive to consider the limit $K \rightarrow + \infty$.
From (\ref{fD def}), in this limit one finds $f^{(D)}_{+, \infty}=f^{(D)}_{-, \infty}$ for any $D$, which reads
\be
\label{fpm K inf}
\lim_{K \rightarrow \infty} f^{(D)}_{\pm, K}(\tilde{z})
= \int_0^{\tilde{z}} 
\frac{\lambda}{1+\lambda^2} \, d\lambda = \frac{1}{2} \log(1+\tilde{z}^2 )\,,
\ee
and therefore $\eta \to 1$ from (\ref{integ ratio}), i.e. $R_{\textrm{\tiny in}} = R_{\textrm{\tiny out}} \equiv R$ (see also Fig.\,\ref{fig:appannulusk}).
From (\ref{fpm K inf}), both the branches in (\ref{branches R1R2}) become 
\be
\rho = \frac{R}{\sqrt{1+\tilde{z}^2}}\,,
\ee
which is the well known spherical solution $z^2 = R^2 - \rho^2$.
As for the points $P_m$, $P_0$ and $P_\ast$, they tend to the same point when $\eta \to 1$, as can be seen from Fig.\,\ref{fig:appannulusprof}, where the gray lines show the paths of these points in the $(\rho,z)$ plane as $K$ varies in $(0,\infty)$.

Given the radial profile (\ref{branches R1R2}), we can compute the area of the corresponding surface obtained by exploiting the rotational symmetry. 
From (\ref{zrho subs}), the radial integral in (\ref{area z(rho)}) can be written as
\be
\label{minarea tube pm}
\mathcal{R}_D^{\textrm{\tiny  con}}     
\,=\,
\int^{\tilde{\varepsilon}_{+} }_{\tilde{z}_m} 
\frac{\sqrt{1+(\tilde{z} + \tilde{z}_{u,+})^2}}{\tilde{z}^D \,\tilde{z}_{u,+}}
\,d\tilde{z}\,
+
\int^{\tilde{\varepsilon}_{-} }_{\tilde{z}_m} 
\frac{\sqrt{1+(\tilde{z} + \tilde{z}_{u,-})^2}}{\tilde{z}^D \,\tilde{z}_{u,-}}
\,d\tilde{z}\,,
\qquad
\hspace{.5cm}
\tilde{\varepsilon}_{+} \equiv \frac{\varepsilon}{R_{\textrm{\tiny out}}}\,,
\hspace{.5cm}
\tilde{\varepsilon}_{-} \equiv \frac{\varepsilon}{R_{\textrm{\tiny in}}}\,,
\ee
where $\tilde{z}_{u,\pm}$ have been defined in (\ref{ztilde1 pm}) and $0< \varepsilon \ll 1$ is the ultraviolet cutoff of the boundary theory. 
Notice that the domains of integration are different for the two branches of the profile.
Plugging (\ref{ztilde1 pm})  into (\ref{minarea tube pm}), the integrands become the same and, by 
splitting the first integral, (\ref{minarea tube pm}) becomes
\bea
\label{Rcon}
\mathcal{R}_D^{\textrm{\tiny  con}}   
&=&
\int_{\varepsilon/R_{\textrm{\tiny out}}}^{\tilde{z}_m}
\frac{\sqrt{K }\, d\tilde{z}}{\tilde{z}^D\sqrt{K (1+\tilde{z}^2)-\tilde{z}^{2D}}}
+
\int_{\varepsilon/R_{\textrm{\tiny in}}}^{\tilde{z}_m}
\frac{\sqrt{K }\, d\tilde{z}}{\tilde{z}^D\sqrt{K (1+\tilde{z}^2)-\tilde{z}^{2D}}}
\\
\rule{0pt}{.65cm}
\label{Rcon split}
&=&
2\int_{\varepsilon/R_{\textrm{\tiny in}}}^{\tilde{z}_m}
\frac{\sqrt{K }\, d\tilde{z}}{\tilde{z}^D\sqrt{K (1+\tilde{z}^2)-\tilde{z}^{2D}}}
+
\int_{\varepsilon/R_{\textrm{\tiny out}}}^{\varepsilon/R_{\textrm{\tiny in}}}
\frac{\sqrt{K }\, d\tilde{z}}{\tilde{z}^D\sqrt{K (1+\tilde{z}^2)-\tilde{z}^{2D}}}\,.
\eea
In the second integral of (\ref{Rcon split}), we can employ the expansion of the integrand for $\tilde{z} \sim 0$, which reads
\be
\frac{1}{\tilde{z}^D\sqrt{1+\tilde{z}^2-\tilde{z}^{2D}/K}}
=
\frac{1}{\tilde{z}^D} + \frac{\gamma_{D,D-1}}{\tilde{z}^{D-2}} + \frac{\gamma_{D,D-3}}{\tilde{z}^{D-4}}
+ \dots + 
\left\{ 
\begin{array}{ll}
\displaystyle 
\frac{\gamma_{D,\textrm{\tiny log}}}{\tilde{z}} + O(\tilde{z})
& \hspace{.6cm} \textrm{odd $D$,}
\\
\rule{0pt}{.5cm}
\gamma_{D,-1} + O(\tilde{z}^2)
& \hspace{.6cm}  \textrm{even $D$,}
\end{array}\right.
\ee
finding that it provides a non trivial contribution $\gamma_{D,\textrm{\tiny log}} \log(R_{\textrm{\tiny out}}/R_{\textrm{\tiny in}})$ to the finite term for odd $D$.

Given $R_{\textrm{\tiny in}}$ and $R_{\textrm{\tiny out}}$, besides the two connected surfaces having the same $\eta$ but different $K$, we have also another surface $\gamma_A$ which is a local minimum for the area functional (\ref{area z(rho)}) such that $\partial \gamma_A = \partial A$: it is made by two disjoint concentric hemispheres in the bulk with radii $R_{\textrm{\tiny in}}$ and $R_{\textrm{\tiny out}}$ which are anchored on the boundaries of the concentric spheres in  the boundary (see the dashed curves in Fig.\,\ref{fig:appannulusprof}). 
The area of a hemisphere  of radius $R$ in the bulk anchored on the boundary of a sphere with the same radius at $z=0$ can be found by integrating (\ref{area z(rho)}) for $0\leqslant \rho \leqslant R-a$, where $z(\varepsilon)\equiv a$, finding
\be
\label{Rsphere}
\mathcal{R}^{\textrm{\tiny sph}}_D(R)
=
\int_{\infty}^{\varepsilon/R}
\frac{\sqrt{1+(\tilde{z} + \tilde{z}_u)^2}}{\tilde{z}^D \tilde{z}_u}\, d\tilde{z}
\,=\,
\int^{\infty}_{\varepsilon/R}
\frac{d\tilde{z}}{\tilde{z}^D\, \sqrt{1+\tilde{z}^2}}\,,
\qquad
\varepsilon = \sqrt{R^2 -(R-a)^2} \,\ll\,1\,,
\ee
where  $\tilde{z}_u$ is (\ref{ztilde1 pm}) in the limit $K \rightarrow +\infty$, namely $\tilde{z}_u = -(1+\tilde{z}^2)/\tilde{z}$. 

Thus, the factor coming from the radial integration  in (\ref{area z(rho)}) for this configuration of two disjoint hemispheres is $\mathcal{R}^{\textrm{\tiny  dis}}_D  = \mathcal{R}^{\textrm{\tiny sph}}_D(R_{\textrm{\tiny out}}) 
+ \mathcal{R}^{\textrm{\tiny sph}}_D(R_{\textrm{\tiny in}}) $.

Having found three surfaces anchored on $\partial A$ for any given $R_{\textrm{\tiny in}} < R_{\textrm{\tiny out}}$ such that $\eta_\ast < \eta <1$ which are local minima of the area functional, the holographic entanglement entropy can be found by selecting the global minimum among them. \\
Considering a connected surface and the configuration made by the two disjoint hemispheres, we find it useful to introduce the following finite quantity 
\be
\label{Delta R def}
\Delta \mathcal{R}_D 
\equiv
\lim_{\varepsilon \to 0}
(  \mathcal{R}^{\textrm{\tiny  dis}}_D  - \mathcal{R}^{\textrm{\tiny con}}_D)\,.
\ee
From (\ref{Rcon}) and (\ref{Rsphere}), it can be written as
\be
\Delta \mathcal{R}_D  \,= \, 
\mathcal{J}^{(\textrm{\tiny in})}_D + \mathcal{J}^{(\textrm{\tiny out})}_D\,,
\ee
where we have introduced 
\be
\mathcal{J}^{(j)}_D
=
\lim_{\varepsilon \to 0}\left(\,
 \int^{\infty}_{\varepsilon/R_j}
\frac{d\tilde{z}}{\tilde{z}^D\, \sqrt{1+\tilde{z}^2}}
-
\int_{\varepsilon/R_j}^{\tilde{z}_m}
\frac{ d\tilde{z}}{\tilde{z}^D \sqrt{1+\tilde{z}^2-\tilde{z}^{2D}/K}}
\right) .
\ee
Splitting the second integral, we can take the limit, finding that $\mathcal{J}^{(\textrm{\tiny in})}_D = \mathcal{J}^{(\textrm{\tiny out})}_D$ and then
\be
\label{delta R finite}
\Delta \mathcal{R}_D = 2\left[\,
\int^{\infty}_{\tilde{z}_m}
\frac{d\tilde{z}}{\tilde{z}^D\, \sqrt{1+\tilde{z}^2}}
-
\int_{0}^{\tilde{z}_m}
\frac{1}{\tilde{z}^D \sqrt{1+\tilde{z}^2}}\left(
\frac{ 1}{\sqrt{1-\tilde{z}^{2D}/[K(1+\tilde{z}^2)]}}
- 1
\right) d\tilde{z}
\, \right] .
\ee
\\
Since $z_m=z_m(K)$ and $K$ depends on the ratio $\eta$ only, also $\Delta \mathcal{R}_D$ is a function of $\eta$.
Nevertheless, as discussed above, there are two values of $K$ associated with the same $\eta$ and, by computing $\Delta \mathcal{R}_D$ for both of them, we can easily find which surface has the minimal area between the two connected ones. 
It turns out that it is the one associated with the lowest value of $K$.
Since $z_m$ is an increasing function of $K$, the minimal area surface between the two connected ones has the lowest $z_m$.
In the example in Fig.\,\ref{fig:appannulusprof} for $D=2$, 
both the radial profiles of the two connected surfaces which are local minima of the area functional and which have the same $\eta$ are shown. 
The one described by the red and the blue curves characterizes the minimal area surface between the two connected ones.

Once the connected surface having minimal area has been found, the sign of the corresponding $\Delta \mathcal{R}_D $ determines the configuration with minimal area, providing therefore the global minimum of the area functional, and its root (which can be found numerically) gives the value of $\eta = \eta_c$ which characterizes the transition.
For $D=2$, $D=3$ and $D=4$ we get respectively $\eta_c = 0.419$ \cite{Olesen:2000ji, Drukker:2005cu}, $\eta_c = 0.562$ \cite{Hirata:2006jx} and $\eta_c=0.652$.
Thus, for any $\eta \in (\eta_\ast ,1)$, we have $\eta_c >\eta_\ast$ and  $\Delta \mathcal{R}_D >0$ when $\eta \in (\eta_c ,1)$.
This tells us that for $\eta < \eta_c$ the configuration occurring in the holographic entanglement entropy for the annular domains is the one made by two disjoint hemispheres.

\section{Elliptic integrals}
\label{sec:elliptic integrals}

When $D=2$, the integrals encountered in \S\ref{sec corners} and in \S\ref{app annulus} can be computed analytically in terms of elliptic integrals. 
Here we report their definitions for completeness, following \cite{abramowitz} (notice that Mathematica adopts the same notation).

The incomplete elliptic integrals of the first, second and third kind are defined respectively as follows
\bea
\mathbb{F}(x | m) &\equiv &
\int_0^x \frac{d\theta}{\sqrt{1-m \sin^2 \theta}} \,,
\\
\mathbb{E}(x | m) &\equiv &
\int_0^x  \sqrt{1-m \sin^2 \theta}\, d\theta \,,
\\
\Pi(n,x | m) &\equiv  &
\int_0^x \frac{d\theta}{(1-n \sin^2 \theta) \sqrt{1-m \sin^2 \theta}} \, .
\eea
Setting $x=\pi/2$ in these expressions, we have
\be 
\mathbb{K}(m)\,\equiv \,\mathbb{F}(\pi/2 | m)\,,
\qquad 
\mathbb{E}(m)\,\equiv \,\mathbb{E}(\pi/2 |m)\,,
\qquad
\Pi(n,m)\,\equiv \,\Pi(n,\pi/2 |m) \,,
\ee 
which are the complete elliptic integrals of the first, second and third kind respectively.

%%%%%%%%%%%%%%%%%%%%%%%%%%%%%%%%%%%%%%%%%%%%%


\begin{thebibliography}{99}



  
  %\cite{Srednicki:1993im}
\bibitem{Srednicki:1993im}
  M.~Srednicki,
  ``Entropy and area,''
  Phys.\ Rev.\ Lett.\  {\bf 71} (1993) 666
  [\hhref{hep-th/9303048}].
  %%CITATION = HEP-TH/9303048;%%
  %559 citations counted in INSPIRE as of 06 Jul 2014
  
  %\cite{Bombelli:1986rw}
\bibitem{Bombelli:1986rw}
  L.~Bombelli, R.~K.~Koul, J.~Lee and R.~D.~Sorkin,
  ``A Quantum Source of Entropy for Black Holes,''
  Phys.\ Rev.\ D {\bf 34} (1986) 373.
  %%CITATION = PHRVA,D34,373;%%
  %545 citations counted in INSPIRE as of 06 Jul 2014
  
  %\cite{Holzhey:1994we}
\bibitem{Holzhey:1994we}
  C.~Holzhey, F.~Larsen and F.~Wilczek,
  ``Geometric and renormalized entropy in conformal field theory,''
  Nucl.\ Phys.\ B {\bf 424} (1994) 443
  [\hhref{hep-th/9403108}].
  %%CITATION = HEP-TH/9403108;%%
  %339 citations counted in INSPIRE as of 24 Sep 2014
  
  %\cite{Calabrese:2004eu}
\bibitem{Calabrese:2004eu}
  P.~Calabrese and J.~L.~Cardy,
  ``Entanglement entropy and quantum field theory,''
  J.\ Stat.\ Mech.\  {\bf 0406} (2004) P06002
  [\hhref{hep-th/0405152}].
  %%CITATION = HEP-TH/0405152;%%
  %288 citations counted in INSPIRE as of 24 Sep 2014
  
  %\cite{Maldacena:1997re}
\bibitem{Maldacena:1997re}
  J.~M.~Maldacena,
  ``The Large N limit of superconformal field theories and supergravity,''
  Int.\ J.\ Theor.\ Phys.\  {\bf 38} (1999) 1113
   [Adv.\ Theor.\ Math.\ Phys.\  {\bf 2} (1998) 231]
  [\hhref{hep-th/9711200}].
  %%CITATION = HEP-TH/9711200;%%
  %10132 citations counted in INSPIRE as of 24 Sep 2014
  
  %\cite{Witten:1998qj}
\bibitem{Witten:1998qj}
  E.~Witten,
  ``Anti-de Sitter space and holography,''
  Adv.\ Theor.\ Math.\ Phys.\  {\bf 2} (1998) 253
  [\hhref{hep-th/9802150}].
  %%CITATION = HEP-TH/9802150;%%
  %6760 citations counted in INSPIRE as of 24 Sep 2014
  
  %\cite{Gubser:1998bc}
\bibitem{Gubser:1998bc}
  S.~S.~Gubser, I.~R.~Klebanov and A.~M.~Polyakov,
  ``Gauge theory correlators from noncritical string theory,''
  Phys.\ Lett.\ B {\bf 428} (1998) 105
  [\hhref{hep-th/9802109}].
  %%CITATION = HEP-TH/9802109;%%
  %5910 citations counted in INSPIRE as of 24 Sep 2014
  
    %\cite{Aharony:1999ti}
\bibitem{Aharony:1999ti}
  O.~Aharony, S.~S.~Gubser, J.~M.~Maldacena, H.~Ooguri and Y.~Oz,
  ``Large N field theories, string theory and gravity,''
  Phys.\ Rept.\  {\bf 323} (2000) 183
  [\hhref{hep-th/9905111}].
  %%CITATION = HEP-TH/9905111;%%
  %3266 citations counted in INSPIRE as of 18 Sep 2014

%\cite{Ryu:2006bv}
\bibitem{Ryu:2006bv}
  S.~Ryu and T.~Takayanagi,
  ``Holographic derivation of entanglement entropy from AdS/CFT,''
  Phys.\ Rev.\ Lett.\  {\bf 96} (2006) 181602
  [\hhref{hep-th/0603001}].
  %%CITATION = HEP-TH/0603001;%%

%\cite{Ryu:2006ef}
\bibitem{Ryu:2006ef}
  S.~Ryu and T.~Takayanagi,
  ``Aspects of Holographic Entanglement Entropy,''
  JHEP {\bf 0608} (2006) 045
   [\hhref{hep-th/0605073}].
  %%CITATION = HEP-TH/0605073;%%
  
  %\cite{Hubeny:2007xt}
\bibitem{Hubeny:2007xt}
  V.~E.~Hubeny, M.~Rangamani and T.~Takayanagi,
  ``A Covariant holographic entanglement entropy proposal,''
  JHEP {\bf 0707} (2007) 062
   [\hhref{0705.0016} [hep-th]].
  %%CITATION = ARXIV:0705.0016;%%
 
  
  %\cite{Calabrese:2009qy}
\bibitem{Calabrese:2009qy}
  P.~Calabrese and J.~Cardy,
  ``Entanglement entropy and conformal field theory,''
  J.\ Phys.\ A {\bf 42} (2009) 504005
  [\hhref{0905.4013} [cond-mat.stat-mech]].
  %%CITATION = ARXIV:0905.4013;%%
  %144 citations counted in INSPIRE as of 24 Sep 2014
  
  %\cite{Casini:2009sr}
\bibitem{Casini:2009sr}
  H.~Casini and M.~Huerta,
  ``Entanglement entropy in free quantum field theory,''
  J.\ Phys.\ A {\bf 42} (2009) 504007
  [\hhref{0905.2562} [hep-th]].
  %%CITATION = ARXIV:0905.2562;%%
  %112 citations counted in INSPIRE as of 24 Sep 2014
 
  
  %\cite{Takayanagi:2012kg}
\bibitem{Takayanagi:2012kg}
  T.~Takayanagi,
  ``Entanglement Entropy from a Holographic Viewpoint,''
  Class.\ Quant.\ Grav.\  {\bf 29} (2012) 153001
  [\hhref{1204.2450} [gr-qc]].
  %%CITATION = ARXIV:1204.2450;%%
  %73 citations counted in INSPIRE as of 01 Sep 2014
  
      %\cite{Maldacena:1998im}
\bibitem{Maldacena:1998im}
  J.~M.~Maldacena,
  ``Wilson loops in large N field theories,''
  Phys.\ Rev.\ Lett.\  {\bf 80} (1998) 4859
  [\hhref{hep-th/9803002}].
  %%CITATION = HEP-TH/9803002;%%
  %1145 citations counted in INSPIRE as of 05 Sep 2014
  
  %\cite{Rey:1998ik}
\bibitem{Rey:1998ik}
  S.~J.~Rey and J.~T.~Yee,
  ``Macroscopic strings as heavy quarks in large N gauge theory and anti-de Sitter supergravity,''
  Eur.\ Phys.\ J.\ C {\bf 22} (2001) 379
  [\hhref{hep-th/9803001}].
  %%CITATION = HEP-TH/9803001;%%
  %959 citations counted in INSPIRE as of 05 Sep 2014
  
    %\cite{Casini:2011kv}
\bibitem{Casini:2011kv}
  H.~Casini, M.~Huerta and R.~C.~Myers,
  ``Towards a derivation of holographic entanglement entropy,''
  JHEP {\bf 1105} (2011) 036
  [\hhref{1102.0440} [hep-th]].
  %%CITATION = ARXIV:1102.0440;%%
  %209 citations counted in INSPIRE as of 22 Sep 2014
  
  
    %\cite{Myers:2010tj}
\bibitem{Myers:2010tj}
  R.~C.~Myers and A.~Sinha,
  ``Holographic c-theorems in arbitrary dimensions,''
  JHEP {\bf 1101} (2011) 125
  [\hhref{1011.5819} [hep-th]].
  %%CITATION = ARXIV:1011.5819;%%
  %166 citations counted in INSPIRE as of 06 Jul 2014
  
  %\cite{Klebanov:2011gs}
\bibitem{Klebanov:2011gs}
  I.~R.~Klebanov, S.~S.~Pufu and B.~R.~Safdi,
  ``F-Theorem without Supersymmetry,''
  JHEP {\bf 1110} (2011) 038
  [\hhref{1105.4598} [hep-th]].
  %%CITATION = ARXIV:1105.4598;%%
  %70 citations counted in INSPIRE as of 06 Jul 2014
  
  %\cite{Casini:2012ei}
\bibitem{Casini:2012ei}
  H.~Casini and M.~Huerta,
  ``On the RG running of the entanglement entropy of a circle,''
  Phys.\ Rev.\ D {\bf 85} (2012) 125016
  [\hhref{1202.5650} [hep-th]].
  %%CITATION = ARXIV:1202.5650;%%
  %68 citations counted in INSPIRE as of 06 Jul 2014
  
 %\cite{Solodukhin:2008dh}
\bibitem{Solodukhin:2008dh}
  S.~N.~Solodukhin,
  ``Entanglement entropy, conformal invariance and extrinsic geometry,''
  Phys.\ Lett.\ B {\bf 665} (2008) 305
  [\hhref{0802.3117} [hep-th]].
  %%CITATION = ARXIV:0802.3117;%%
  %90 citations counted in INSPIRE as of 18 Oct 2014
  
    %\cite{Hubeny:2012ry}
\bibitem{Hubeny:2012ry}
  V.~E.~Hubeny,
  ``Extremal surfaces as bulk probes in AdS/CFT,''
  JHEP {\bf 1207} (2012) 093
  [\hhref{1203.1044} [hep-th]].
  %%CITATION = ARXIV:1203.1044;%%
  
    %\cite{Klebanov:2012yf}
\bibitem{Klebanov:2012yf}
  I.~R.~Klebanov, T.~Nishioka, S.~S.~Pufu and B.~R.~Safdi,
  ``On Shape Dependence and RG Flow of Entanglement Entropy,''
  JHEP {\bf 1207} (2012) 001
  [\hhref{1204.4160} [hep-th]].
  %%CITATION = ARXIV:1204.4160;%%
  
    %\cite{Myers:2013lva}
\bibitem{Myers:2013lva}
  R.~C.~Myers, R.~Pourhasan and M.~Smolkin,
  ``On Spacetime Entanglement,''
  JHEP {\bf 1306} (2013) 013
  [\hhref{1304.2030} [hep-th]].
  %%CITATION = ARXIV:1304.2030;%%
  %21 citations counted in INSPIRE as of 20 Sep 2014
  
      %\cite{Papadimitriou:2010as}
\bibitem{Papadimitriou:2010as}
  I.~Papadimitriou,
  ``Holographic renormalization as a canonical transformation,''
  JHEP {\bf 1011} (2010) 014
  [\hhref{1007.4592} [hep-th]].
  %%CITATION = ARXIV:1007.4592;%%
  
    %\cite{Hung:2011ta}
\bibitem{Hung:2011ta}
  L.~-Y.~Hung, R.~C.~Myers and M.~Smolkin,
  ``Some Calculable Contributions to Holographic Entanglement Entropy,''
  JHEP {\bf 1108} (2011) 039
  [\hhref{1105.6055} [hep-th]].
  %%CITATION = ARXIV:1105.6055;%%
  %31 citations counted in INSPIRE as of 06 Jul 2014
  
    %\cite{Astaneh:2014uba}
\bibitem{Astaneh:2014uba}
  A.~F.~Astaneh, G.~Gibbons and S.~N.~Solodukhin,
  ``What surface maximizes entanglement entropy?,''
  \hhref{1407.4719} [hep-th].
  %%CITATION = ARXIV:1407.4719;%%
  %3 citations counted in INSPIRE as of 16 Sep 2014
  
    %\cite{Allais:2014ata}
\bibitem{Allais:2014ata}
  A.~Allais and M.~Mezei,
  ``Some results on the shape dependence of entanglement and R\'enyi entropies,''
  \hhref{1407.7249} [hep-th].
  %%CITATION = ARXIV:1407.7249;%%
  %1 citations counted in INSPIRE as of 01 Sep 2014
  
  %\cite{Vidal:2002zz}
\bibitem{Vidal:2002zz}
  G.~Vidal and R.~F.~Werner,
  ``Computable measure of entanglement,''
  Phys.\ Rev.\ A {\bf 65} (2002) 032314
  [\hhref{quant-ph/0102117}].
  %%CITATION = PHRVA,A65,032314;%%
  %114 citations counted in INSPIRE as of 21 Oct 2014
  
  %\cite{Calabrese:2012ew}
\bibitem{Calabrese:2012ew}
  P.~Calabrese, J.~Cardy and E.~Tonni,
  ``Entanglement negativity in quantum field theory,''
  Phys.\ Rev.\ Lett.\  {\bf 109} (2012) 130502
  [\hhref{1206.3092} [cond-mat.stat-mech]].
  %%CITATION = ARXIV:1206.3092;%%
  %12 citations counted in INSPIRE as of 21 Oct 2014
  
  %\cite{Calabrese:2012nk}
\bibitem{Calabrese:2012nk}
  P.~Calabrese, J.~Cardy and E.~Tonni,
  ``Entanglement negativity in extended systems: A field theoretical approach,''
  J.\ Stat.\ Mech.\  {\bf 1302} (2013) P02008
  [\hhref{1210.5359} [cond-mat.stat-mech]].
  %%CITATION = ARXIV:1210.5359;%%
  %9 citations counted in INSPIRE as of 21 Oct 2014
  
  %\cite{Calabrese:2014yza}
\bibitem{Calabrese:2014yza}
  P.~Calabrese, J.~Cardy and E.~Tonni,
  ``Finite temperature entanglement negativity in conformal field theory,''
  \hhref{1408.3043} [cond-mat.stat-mech].
  %%CITATION = ARXIV:1408.3043;%%
  %1 citations counted in INSPIRE as of 21 Oct 2014
  
    %\cite{Caraglio:2008pk}
\bibitem{Caraglio:2008pk}
  M.~Caraglio and F.~Gliozzi,
  ``Entanglement Entropy and Twist Fields,''
  JHEP {\bf 0811} (2008) 076
  [\hhref{0808.4094} [hep-th]].
  %%CITATION = ARXIV:0808.4094;%%
  %34 citations counted in INSPIRE as of 24 Sep 2014
  
  %\cite{Furukawa:2008uk}
\bibitem{Furukawa:2008uk}
  S.~Furukawa, V.~Pasquier and J.~Shiraishi,
  ``Mutual Information and Compactification Radius in a c=1 Critical Phase in One Dimension,''
  Phys.\ Rev.\ Lett.\  {\bf 102} (2009) 170602
  [\hhref{0809.5113} [cond-mat.stat-mech]].
  %%CITATION = ARXIV:0809.5113;%%
  %40 citations counted in INSPIRE as of 24 Sep 2014
  
    %\cite{Calabrese:2009ez}
\bibitem{Calabrese:2009ez}
  P.~Calabrese, J.~Cardy and E.~Tonni,
  ``Entanglement entropy of two disjoint intervals in conformal field theory,''
  J.\ Stat.\ Mech.\  {\bf 0911} (2009) P11001
  [\hhref{0905.2069} [hep-th]].
  %%CITATION = ARXIV:0905.2069;%%
  %46 citations counted in INSPIRE as of 02 Jul 2014
  
  %\cite{Calabrese:2010he}
\bibitem{Calabrese:2010he}
  P.~Calabrese, J.~Cardy and E.~Tonni,
  ``Entanglement entropy of two disjoint intervals in conformal field theory II,''
  J.\ Stat.\ Mech.\  {\bf 1101} (2011) P01021
  [\hhref{1011.5482} [hep-th]].
  %%CITATION = ARXIV:1011.5482;%%
  %30 citations counted in INSPIRE as of 02 Jul 2014
 
  
%\cite{Cardy:2013nua}
\bibitem{Cardy:2013nua}
  J.~Cardy,
  ``Some results on the mutual information of disjoint regions in higher dimensions,''
  J.\ Phys.\ A {\bf 46} (2013) 285402
  [\hhref{1304.7985} [hep-th]].
  %%CITATION = ARXIV:1304.7985;%%
  
        %\cite{Casini:2008wt}
\bibitem{Casini:2008wt} 
  H.~Casini and M.~Huerta,
  ``Remarks on the entanglement entropy for disconnected regions,''
  JHEP {\bf 0903}, 048 (2009)
  [\hhref{0812.1773} [hep-th]].
  %%CITATION = ARXIV:0812.1773;%%
  
  %\cite{Hung:2014npa}
\bibitem{Hung:2014npa}
  L.~Y.~Hung, R.~C.~Myers and M.~Smolkin,
  ``Twist operators in higher dimensions,''
  \hhref{1407.6429} [hep-th].
  %%CITATION = ARXIV:1407.6429;%%
  %7 citations counted in INSPIRE as of 10 Oct 2014
  
  
    %\cite{Hubeny:2007re}
\bibitem{Hubeny:2007re}
  V.~E.~Hubeny and M.~Rangamani,
  ``Holographic entanglement entropy for disconnected regions,''
  JHEP {\bf 0803} (2008) 006
  [\hhref{0711.4118} [hep-th]].
  %%CITATION = ARXIV:0711.4118;%%
  %35 citations counted in INSPIRE as of 24 Sep 2014
  
  %\cite{Headrick:2010zt}
\bibitem{Headrick:2010zt}
  M.~Headrick,
  ``Entanglement Renyi entropies in holographic theories,''
  Phys.\ Rev.\ D {\bf 82} (2010) 126010
  [\hhref{1006.0047} [hep-th]].
  %%CITATION = ARXIV:1006.0047;%%
  
    %\cite{Tonni:2010pv}
\bibitem{Tonni:2010pv}
  E.~Tonni,
  ``Holographic entanglement entropy: near horizon geometry and disconnected regions,''
  JHEP {\bf 1105} (2011) 004
  [\hhref{1011.0166} [hep-th]].
  %%CITATION = ARXIV:1011.0166;%%
  
      %\cite{Gross:1998gk}
\bibitem{Gross:1998gk}
  D.~J.~Gross and H.~Ooguri,
  ``Aspects of large N gauge theory dynamics as seen by string theory,''
  Phys.\ Rev.\ D {\bf 58} (1998) 106002
  [\hhref{hep-th/9805129}].
  %%CITATION = HEP-TH/9805129;%%
  %303 citations counted in INSPIRE as of 05 Sep 2014
  
  %\cite{Zarembo:1999bu}
\bibitem{Zarembo:1999bu}
  K.~Zarembo,
  ``Wilson loop correlator in the AdS / CFT correspondence,''
  Phys.\ Lett.\ B {\bf 459} (1999) 527
  [\hhref{hep-th/9904149}].
  %%CITATION = HEP-TH/9904149;%%
  %53 citations counted in INSPIRE as of 05 Sep 2014
  
  %\cite{Olesen:2000ji}
\bibitem{Olesen:2000ji}
  P.~Olesen and K.~Zarembo,
  ``Phase transition in Wilson loop correlator from AdS / CFT correspondence,''
  \hhref{hep-th/0009210}.
  %%CITATION = HEP-TH/0009210;%%
  %32 citations counted in INSPIRE as of 05 Sep 2014
  
  %\cite{Kim:2001td}
\bibitem{Kim:2001td}
  H.~Kim, D.~K.~Park, S.~Tamarian and H.~J.~W.~Muller-Kirsten,
  ``Gross-Ooguri phase transition at zero and finite temperature: Two circular Wilson loop case,''
  JHEP {\bf 0103} (2001) 003
  [\hhref{hep-th/0101235}].
  %%CITATION = HEP-TH/0101235;%%
  %21 citations counted in INSPIRE as of 05 Sep 2014
  
  
        %\cite{Faulkner:2013ana}
\bibitem{Faulkner:2013ana}
  T.~Faulkner, A.~Lewkowycz and J.~Maldacena,
  ``Quantum corrections to holographic entanglement entropy,''
  JHEP {\bf 1311} (2013) 074
  [\hhref{1307.2892}].
  %%CITATION = ARXIV:1307.2892;%%
  %30 citations counted in INSPIRE as of 06 Jul 2014
  
  %\cite{Headrick:2007km}
\bibitem{Headrick:2007km}
  M.~Headrick and T.~Takayanagi,
  ``A Holographic proof of the strong subadditivity of entanglement entropy,''
  Phys.\ Rev.\ D {\bf 76} (2007) 106013
  [\hhref{0704.3719} [hep-th]].
  %%CITATION = ARXIV:0704.3719;%%
  %91 citations counted in INSPIRE as of 26 Oct 2014
  
  %\cite{Azeyanagi:2007bj}
\bibitem{Azeyanagi:2007bj}
  T.~Azeyanagi, T.~Nishioka and T.~Takayanagi,
  ``Near Extremal Black Hole Entropy as Entanglement Entropy via AdS(2)/CFT(1),''
  Phys.\ Rev.\ D {\bf 77} (2008) 064005
  [\hhref{0710.2956} [hep-th]].
  %%CITATION = ARXIV:0710.2956;%%
  %60 citations counted in INSPIRE as of 26 Oct 2014
  
  %\cite{Hubeny:2013gta}
\bibitem{Hubeny:2013gta}
  V.~E.~Hubeny, H.~Maxfield, M.~Rangamani and E.~Tonni,
  ``Holographic entanglement plateaux,''
  JHEP {\bf 1308} (2013) 092
  [\hhref{1306.4004} [hep-th]].
  %%CITATION = ARXIV:1306.4004,;%%
  %22 citations counted in INSPIRE as of 26 Oct 2014
  
        %\cite{Drukker:2005cu}
\bibitem{Drukker:2005cu} 
  N.~Drukker and B.~Fiol,
  ``On the integrability of Wilson loops in $AdS_5 \times S^5$: Some periodic ansatze,''
  JHEP {\bf 0601}, 056 (2006)
  [\hhref{hep-th/0506058}].
  %%CITATION = HEP-TH/0506058;%%
  
        %\cite{Hirata:2006jx}
\bibitem{Hirata:2006jx} 
  T.~Hirata and T.~Takayanagi,
  ``AdS/CFT and strong subadditivity of entanglement entropy,''
  JHEP {\bf 0702}, 042 (2007)
  [\hhref{hep-th/0608213}].
  %%CITATION = HEP-TH/0608213;%%
 
  
    %\cite{Dekel:2013kwa}
\bibitem{Dekel:2013kwa}
  A.~Dekel and T.~Klose,
  ``Correlation Function of Circular Wilson Loops at Strong Coupling,''
  JHEP {\bf 1311} (2013) 117
  [\hhref{1309.3203} [hep-th]].
  %%CITATION = ARXIV:1309.3203;%%
  %5 citations counted in INSPIRE as of 02 Sep 2014
  
  %\cite{Krtous:2013vha}
\bibitem{Krtous:2013vha}
  P.~Krtous and A.~Zelnikov,
  ``Entanglement entropy of spherical domains in anti-de Sitter space,''
  Phys.\ Rev.\ D {\bf 89} (2014) 104058
  [\hhref{1311.1685} [hep-th]].
  %%CITATION = ARXIV:1311.1685;%%
  %1 citations counted in INSPIRE as of 24 Sep 2014
  
%\cite{Krtous:2014pva}
\bibitem{Krtous:2014pva}
  P.~Krtous and A.~Zelnikov,
  ``Minimal surfaces and entanglement entropy in anti-de Sitter space,''
  JHEP {\bf 1410} (2014) 77
  [\hhref{1406.7659} [hep-th]].
  %%CITATION = ARXIV:1406.7659;%%
  
  
  \bibitem{evolverpaper}
  K.~Brakke,
  ``The Surface Evolver,''
  Experimental Mathematics {\bf 1} (2): 141 (1992).

\bibitem{evolverlink}
{\it Surface Evolver} program:
\href{http://www.susqu.edu/brakke/evolver/evolver.html}{http://www.susqu.edu/brakke/evolver/evolver.html}
    
  
  %%%%%  CORNERS  %%%%%%%%
  
  
  %\cite{Fradkin:2006mb}
\bibitem{Fradkin:2006mb}
  E.~Fradkin and J.~E.~Moore,
  ``Entanglement entropy of 2D conformal quantum critical points: hearing the shape of a quantum drum,''
  Phys.\ Rev.\ Lett.\  {\bf 97} (2006) 050404
  [\hhref{cond-mat/0605683} [cond-mat.str-el]].
  %%CITATION = COND-MAT/0605683;%%
  %35 citations counted in INSPIRE as of 02 Sep 2014
  
    %\cite{Casini:2006hu}
\bibitem{Casini:2006hu}
  H.~Casini and M.~Huerta,
  ``Universal terms for the entanglement entropy in 2+1 dimensions,''
  Nucl.\ Phys.\ B {\bf 764} (2007) 183
  [\hhref{hep-th/0606256}].
  %%CITATION = HEP-TH/0606256;%%
  %39 citations counted in INSPIRE as of 10 Sep 2014
 
  
  
  %\cite{Kallin:2014oka}
\bibitem{Kallin:2014oka}
  A.~B.~Kallin, E.~M.~Stoudenmire, P.~Fendley, R.~R.~P.~Singh and R.~G.~Melko,
  ``Corner contribution to the entanglement entropy of an O(3) quantum critical point in 2+1 dimensions,''
  J. Stat. Mech. (2014) P06009
  [\hhref{1401.3504} [cond-mat.str-el]].
  %%CITATION = ARXIV:1401.3504;%%
 
    
  %\cite{Drukker:1999zq}
\bibitem{Drukker:1999zq}
  N.~Drukker, D.~J.~Gross and H.~Ooguri,
  ``Wilson loops and minimal surfaces,''
  Phys.\ Rev.\ D {\bf 60} (1999) 125006
  [\hhref{hep-th/9904191}].
  %%CITATION = HEP-TH/9904191;%%
  %303 citations counted in INSPIRE as of 03 Jul 2014
  
  
    %\cite{Myers:2012vs}
\bibitem{Myers:2012vs}
  R.~C.~Myers and A.~Singh,
  ``Entanglement Entropy for Singular Surfaces,''
  JHEP {\bf 1209} (2012) 013
  [\hhref{1206.5225} [hep-th]].
  %%CITATION = ARXIV:1206.5225;%%
  %4 citations counted in INSPIRE as of 01 Dec 2013
  
      %\cite{Klebanov:2007ws}
\bibitem{Klebanov:2007ws}
  I.~R.~Klebanov, D.~Kutasov and A.~Murugan,
  ``Entanglement as a probe of confinement,''
  Nucl.\ Phys.\ B {\bf 796} (2008) 274
  [\hhref{0709.2140} [hep-th]].
  %%CITATION = ARXIV:0709.2140;%%
  %130 citations counted in INSPIRE as of 11 Nov 2014
  
    %\cite{Berenstein:1998ij}
\bibitem{Berenstein:1998ij}
  D.~E.~Berenstein, R.~Corrado, W.~Fischler and J.~M.~Maldacena,
  ``The Operator product expansion for Wilson loops and surfaces in the large N limit,''
  Phys.\ Rev.\ D {\bf 59} (1999) 105023
  [\hhref{hep-th/9809188}].
  %%CITATION = HEP-TH/9809188;%%
  %224 citations counted in INSPIRE as of 10 Jul 2014
  

    %\cite{Coser:2013qda}
\bibitem{Coser:2013qda}
  A.~Coser, L.~Tagliacozzo and E.~Tonni,
  ``On R\'enyi entropies of disjoint intervals in conformal field theory,''
  J.\ Stat.\ Mech.\  {\bf 2014} (2014) P01008
  [\hhref{1309.2189} [hep-th]].
  %%CITATION = ARXIV:1309.2189;%%
  %2 citations counted in INSPIRE as of 02 Jul 2014
  
      %\cite{Hayden:2011ag}
\bibitem{Hayden:2011ag}
  P.~Hayden, M.~Headrick and A.~Maloney,
  ``Holographic Mutual Information is Monogamous,''
  Phys.\ Rev.\ D {\bf 87} (2013) 4,  046003
  [\hhref{1107.2940} [hep-th]].
  %%CITATION = ARXIV:1107.2940;%%
  %29 citations counted in INSPIRE as of 02 Jul 2014
 
  
        %\cite{Headrick:2013zda}
\bibitem{Headrick:2013zda}
  M.~Headrick,
  ``General properties of holographic entanglement entropy,''
  JHEP {\bf 1403} (2014) 085
  [\hhref{1312.6717} [hep-th]].
  %%CITATION = ARXIV:1312.6717;%%
  %8 citations counted in INSPIRE as of 06 Jul 2014
  
    %\cite{Balasubramanian:2014hda}
\bibitem{Balasubramanian:2014hda}
  V.~Balasubramanian, P.~Hayden, A.~Maloney, D.~Marolf and S.~F.~Ross,
  ``Multiboundary Wormholes and Holographic Entanglement,''
  Class.\ Quant.\ Grav.\  {\bf 31} (2014) 185015
  [\hhref{1406.2663} [hep-th]].
  %%CITATION = ARXIV:1406.2663;%%
  %5 citations counted in INSPIRE as of 05 Nov 2014
 
  
  %\cite{AbajoArrastia:2010yt}
\bibitem{AbajoArrastia:2010yt}
  J.~Abajo-Arrastia, J.~Aparicio and E.~Lopez,
  ``Holographic Evolution of Entanglement Entropy,''
  JHEP {\bf 1011} (2010) 149
  [\hhref{1006.4090} [hep-th]].
  %%CITATION = ARXIV:1006.4090;%%
  %92 citations counted in INSPIRE as of 24 Sep 2014
 
  
  %\cite{Balasubramanian:2011ur}
\bibitem{Balasubramanian:2011ur}
  V.~Balasubramanian, A.~Bernamonti, J.~de Boer, N.~Copland, B.~Craps, E.~Keski-Vakkuri, B.~Muller and A.~Schafer {\it et al.},
  ``Holographic Thermalization,''
  Phys.\ Rev.\ D {\bf 84} (2011) 026010
  [\hhref{1103.2683} [hep-th]].
  %%CITATION = ARXIV:1103.2683;%%
  %108 citations counted in INSPIRE as of 24 Sep 2014
  
  %\cite{Balasubramanian:2011at}
\bibitem{Balasubramanian:2011at}
  V.~Balasubramanian, A.~Bernamonti, N.~Copland, B.~Craps and F.~Galli,
  ``Thermalization of mutual and tripartite information in strongly coupled two dimensional conformal field theories,''
  Phys.\ Rev.\ D {\bf 84} (2011) 105017
  [\hhref{1110.0488} [hep-th]].
  %%CITATION = ARXIV:1110.0488;%%
  %31 citations counted in INSPIRE as of 07 Jul 2014
  
    %\cite{Allais:2011ys}
\bibitem{Allais:2011ys}
  A.~Allais and E.~Tonni,
  ``Holographic evolution of the mutual information,''
  JHEP {\bf 1201} (2012) 102
  [\hhref{1110.1607} [hep-th]].
  %%CITATION = ARXIV:1110.1607;%%
  %36 citations counted in INSPIRE as of 07 Jul 2014
  
  %\cite{Callan:2012ip}
\bibitem{Callan:2012ip}
  R.~Callan, J.~Y.~He and M.~Headrick,
  ``Strong subadditivity and the covariant holographic entanglement entropy formula,''
  JHEP {\bf 1206} (2012) 081
  [\hhref{1204.2309} [hep-th]].
  %%CITATION = ARXIV:1204.2309;%%
  %30 citations counted in INSPIRE as of 24 Sep 2014
 
  
  %\cite{Liu:2013qca}
\bibitem{Liu:2013qca}
  H.~Liu and S.~J.~Suh,
  ``Entanglement growth during thermalization in holographic systems,''
  Phys.\ Rev.\ D {\bf 89} (2014) 066012
  [\hhref{1311.1200} [hep-th]].
  %%CITATION = ARXIV:1311.1200;%%
  %17 citations counted in INSPIRE as of 24 Sep 2014
    
 \bibitem{Kreyszig:1991} 
E. Kreyszig, ``Differential geometry'' (Dover Publications, New York, 1991).

 \bibitem{chopp} 
D.~L.~Chopp, ``Computing Minimal Surfaces via Level Set Curvature Flow,''
J.\ Comput.\ Phys. {\bf 106} (1993) 77-91.

 \bibitem{abramowitz} 
  M.~Abramowitz and I.~Stegun,
  ``Handbook of Mathematical Functions with Formulas, Graphs, and Mathematical Tables'' (Dover Publications, New York, 1964).
  


  

  

  








\end{thebibliography}
\end{document}